\documentclass[10pt,a4paper]{article}
\pdfoutput=1
\usepackage{jheppub}
\usepackage{arydshln}
\usepackage{multirow}
\usepackage{tikz}
\usetikzlibrary{positioning}
\usepackage{lipsum}
\usepackage{parskip}
\usepackage{tabularray}
\usepackage[all]{xy}
\usepackage{amsfonts}
\usepackage{amscd}
\usepackage{amssymb}
\usepackage{amsmath,bbm}
\usepackage{graphicx}
\usepackage{epsfig}
\usepackage{latexsym}
\usepackage{mathtools}
\usepackage{hyperref}
\usepackage{amsthm}
\usepackage[vcentermath]{youngtab}
\usepackage{accents}
\usepackage{bm}
    
\usepackage{calligra}

\DeclareMathAlphabet{\mathcalligra}{T1}{calligra}{m}{n}
\DeclareFontShape{T1}{calligra}{m}{n}{<->s*[2.2]callig15}{}

\def \be  {\begin{equation}}
\def \ee  {\end{equation}}
\def \bea {\begin{equation}\begin{aligned}}
\def \eea {\end{aligned}\end{equation}}
\def \ba  {\begin{eqnarray}}
\def \ea  {\end{eqnarray}}
\def \bb  {\begin{thebibliography}}
\def \eb  {\end{thebibliography}}
\def \lab #1 {\label{#1}}

\newcommand\cN{\mathcal{N}}

\definecolor{cardinal}{rgb}{0.6,0,0}
\definecolor{darkgreen}{rgb}{0,0.5,0}
\definecolor{golden}{rgb}{0.92, 0.7, 0}
\definecolor{midnight}{rgb}{0, 0, 0.5}
\definecolor{darkblue}{rgb}{0.2, 0, 0.8}

\theoremstyle{definition}

\def\CB{{\cal B}}
\def\CF{{\cal F}}
\def\CH{{\cal H}}\def\CI{{\cal I}}
\def\CM{{\cal M}}
\def\CN{{\cal N}}

\def\CT{{\cal T}}
\def\CW{{\cal W}}

\def\a{\alpha}\def\b{\beta}

\usepackage{tikz}

\newcommand\Tr{{\mathrm{Tr}}\,}

\newif\ifniklas\niklastrue
\RequirePackage{verbatim}

\makeatletter
\newwrite\bibinl@out
{\immediate\closeout\bibinl@out\@esphack}
\makeatother

\usetikzlibrary{arrows}
\usetikzlibrary{positioning}
\usetikzlibrary{decorations.pathmorphing}
\usetikzlibrary{decorations.markings}
\def\arrowhead{angle 90}
\tikzset{>=\arrowhead}
\pgfarrowsdeclaredouble{<<}{>>}{\arrowhead}{\arrowhead}
\pgfarrowsdeclaretriple{<<<}{>>>}{\arrowhead}{\arrowhead}
\pgfarrowsdeclaredouble{<<<<}{>>>>}{<<}{>>}
\tikzstyle{W}=[minimum size=2em, scale=1, inner sep=2pt]
\tikzstyle{B}=[draw,circle,fill=black,scale=1]
\tikzstyle{D}= [circle, minimum size=1em]
\tikzstyle{H}=[draw,circle,fill=gray,scale=1]
\tikzstyle{R}=[draw, inner sep=4pt, minimum size=1.8em]
\tikzstyle{every picture}=[scale=1,baseline=(current bounding box.south)]

\title{From BPS Spectra of Argyres-Douglas Theories\\to Families of 3d TFTs}
\author[1]{Byeonggi Go,}
\author[1]{Qiang Jia,}
\author[1]{Heeyeon Kim,}
\author[2]{and Sungjoon Kim}
\affiliation[1]{Department of Physics, Korea Advanced Institute of Science and Technology,Daejeon 34141, Republic of Korea}
\affiliation[2]{Korea Institute for Advanced Study, 85 Hoegiro, Dongdaemun-Gu, Seoul 02455, Korea}
\emailAdd{wjdzm14@kaist.ac.kr}
\emailAdd{qjia1993@kaist.ac.kr}
\emailAdd{heeyeon.kim@kaist.ac.kr}
\emailAdd{sungjoon@kias.re.kr}

\abstract{Vertex operator algebras (VOAs) arise in protected subsectors of supersymmetric quantum field theories, notably in 4d $\CN=2$ superconformal field theories (SCFT) via the Schur sector and in twisted 3d $\CN=4$ theories via boundary algebras. These constructions are connected through twisted circle compactifications, which can be best understood from the dynamics of BPS particles in the Coulomb branch of the 4d SCFT. This data is encoded in an operator $\hat\Phi$ acting on the Hilbert space of an auxiliary quantum mechanics of BPS particles, whose trace yields the partition functions of a 3d topological field theory (TFT) bounding the VOA.
We generalize this trace formula by considering higher powers of $\hat\Phi$, leading to a finite family of VOAs associated with a given 4d SCFT. Applying this framework to Argyres-Douglas theories labeled by $(A_1, G)$, where $G$ is an ADE-type group of rank up to 8, we extract the modular data of the family of boundary VOAs via TFT partition function calculations on Seifert manifolds. Our results suggest that the modular data obtained from different powers of $\hat\Phi$ are related by Galois transformations.}
\begin{document}
	
\maketitle

\section{Introduction}

A vertex operator algebra (VOA) is a mathematical framework that encapsulates the algebraic structure of two-dimensional conformal field theories, which often emerges as a subsector of higher-dimensional quantum field theories. A prominent example is the 4d/2d correspondence \cite{Beem:2013sza}, which associates a 4d $\CN=2$ superconformal field theory (SCFT) with a subsector of local operators -- known as Schur operators -- that form a VOA. This is a protected subsector which exhibits a rich structure that significantly constrains the dynamics of underlying 4d SCFT.

A closely related recent development is the appearance of VOAs on the boundaries of 3d supersymmetric field theories. \cite{Gaiotto:2016wcv,Costello:2018fnz,Costello:2018swh,Creutzig:2021ext,Garner:2022rwe,Coman:2023xcq} A 3d $\CN=4$ theory can be topologically twisted to generate a pair of topological field theories (TFTs) which admit holomorphic boundary conditions. The boundary local operators then form a VOA structure, leading to a broad generalization of the bulk-boundary correspondence originally formulated for 3d Chern-Simons theories and 2d WZW models \cite{Moore:1988qv,Moore:1989vd}, extending to non-unitary and non-rational cases. 

In both of these constructions, non-Lagrangian SCFTs play a central role. Notably, it has been conjectured that rational VOAs can arise from 3d $\CN=4$ rank-zero SCFTs \cite{Gang:2021hrd,Gang:2023rei,Dedushenko:2023cvd,Ferrari:2023fez,Creutzig:2024ljv,ArabiArdehali:2024ysy,ArabiArdehali:2024vli} and 4d $\CN=2$ Argyres-Dogulas type theories \cite{Cordova:2015nma,Buican:2015ina,Beem:2017ooy,Beem:2014rza,Creutzig:2017qyf}, both of which are believed to be inherently non-Lagrangian.

The two constructions -- VOAs from 4d $\CN=2$ and 3d $\CN=4$ non-Lagrangian theories -- are interconnected via a twisted circle compactification. See \cite{Dedushenko:2018bpp,Dedushenko:2023cvd} for recent discussions. The precise relation is most naturally understood from the perspective of the Coulomb branch of the original 4d SCFT. \cite{Cecotti:2010fi,Cecotti:2011iy,Gaiotto:2024ioj} In this setup, we perform a twist by the $U(1)_r$ symmetry along the compactification circle, which induces a closed Janus-like trajectory in the Coulomb branch, along which the central charges undergo smooth $2\pi$ phase rotations. As a result, the 3d BPS particles become trapped at specific loci along the loop determined by their central charges. This leads to an effective 3d $\cN=2$ abelian Chern-Simons matter theory (CSM) description with various superpotential couplings. 

These abelian 3d $\CN=2$ theories are expected to flow to a non-Lagrangian SCFT with enhanced $\CN=4$ supersymmetry, which, upon twisting, supports the corresponding VOA on its boundary. This picture gives rise to the following wall-crossing invariant trace formula that computes the vacuum character of the VOA \cite{Cecotti:2010fi,Cordova:2015nma,Cordova:2017mhb,Cordova:2017ohl}
\be
I_{\text{Schur}} =  (q)_\infty^{2r} \, \Tr \, \hat M^{-1}\ ,
\ee
where $\hat M$ is an operator constructed out of the quantum torus algebra variables for each Coulomb branch BPS particle. This expression can also be interpreted as a half-index of the 3d SCFT in the infrared.

Motivated by this, one can construct other wall-crossing invariant quantities that calculate various supersymmetric observables of 3d SCFTs. In particular, it is proposed in \cite{Gaiotto:2024ioj} that the ellipsoid partition function of the twisted SCFT can be computed by the trace formula
\be\label{Sb phi trace}
S_b = \Tr \,\hat \Phi\ ,
\ee
where now $\hat\Phi$ is an operator acting on the Hilbert space of an auxiliary quantum mechanics of BPS particles, encoding the data of the IR physics in a similar manner. 

A natural generalization of these formulas is to consider higher powers of the operators inside the trace, which gives rise to a family of distinct wall-crossing invariant quantities. A generalization of the Schur index involving the trace of higher powers of $\hat M$, i.e.
\be\label{family schur}
    I_N (q) = (q)_\infty^{2r} \, \Tr\, \hat M^N \ , 
\ee
has been considered in \cite{Kim:2024dxu}. (See also \cite{Cecotti:2015lab,Cecotti:2010fi} for earlier discussions.) Physically, this construction corresponds to a twisted circle compactification along a closed path that wraps around the Coulomb branch Janus loop multiple times. For a given 4d SCFT, the trace formula \eqref{family schur} generates a set of $q$-series, which can be identified with the vacuum characters of a family of VOAs.

In this paper, we consider the higher powers of the monodromy operator that produces a family of the ellipsoid partition functions of the 3d topological field theories
\be
S_b^{(M)}= \Tr \,\hat\Phi^M\ ,
\ee
which generalizes \eqref{Sb phi trace}. The operator $\hat\Phi$ exhibits significantly better convergence properties than $\hat M$, providing a powerful tool for extracting 3d gauge theory descriptions that flow to either a topological theory or an $\CN=4$ SCFT (admitting topological twists) that support the family of VOAs on their boundary.

As discussed in \cite{Dedushenko:2023cvd}, the $U(1)_r$ twisting is non-trivial for the non-Lagrangian theories whose Coulomb branch operators carry fractional $r$-charges. Let $n$ be the least common multiple (lcm) of the denominators of these charges. Then we expect that $n$-th wrapping of the Janus trajectory would trivialize the twisting. Indeed, one can explicitly verify that
\be
\hat\Phi^n \sim {\bf 1}\ ,
\ee
where $\bf 1$ is the identity operator acting on the Hilbert space of the auxiliary quantum mechanics. This identity allows us to consider at most $n-1$ distinct trace formulas, from which we can extract the 3d theories that bound a finite family of VOAs.

Once we have a 3d $\CN=2$ gauge theory description of the theory, one can calculate the partition functions of the associated topological theories on Seifert manifolds, utilizing the method developed in \cite{Closset:2018ghr}, which allows us to compute the modular data of the boundary VOAs. We carry out this computation for a large class of Argyres-Douglas theories labeled by $(A_1, G)$ where $G$ is an ADE group with rank up to 8. For simplicity, we refer to these as $G$-type AD theories in the main text.

From these results, we identify the candidate boundary VOAs compatible with the modular data computed from the partition functions, whenever possible. For a given 4d SCFT with fixed $G$, we find that the modular data for different $M$'s are related by Galois transformations. Note that a similar observation was made in recent works \cite{ArabiArdehali:2024ysy,ArabiArdehali:2024vli} by considering a high-temperature limit of the 4d $\CN=1$ Lagrangian descriptions. A concrete physical explanation of this phenomenon from the 4d perspective remains unclear, and we hope to return to this question in the future.

This paper is organized as follows. In section \ref{sec: review}, we review the twisted circle compactification of Argyres-Douglas theories and introduce the trace formula. From section \ref{sec: A2N} to section \ref{sec:EN}, we evaluate the trace formula for all $G=A,D,E$ of rank up to 8 and extract the gauge theory description that gives the modular data out of the partition function computation. We summarize our conventions for the partition function computation in appendix \ref{app: conventions}. In appendix \ref{app:numerical evaluataion}, we summarize the results for the numerical evaluations of the ellipsoid partition functions. In appendix \ref{app: identity}, we summarize the identities of the Faddeev quantum dilogarithm, including a proof of the periodicity relation that we use extensively in the main text. In appendix \ref{app:modular data}, we summarize the modular data of various VOAs that appear in the main text.

\newpage

\section{Twisted compactifications of Argyres-Douglas theories}\label{sec: review}

The relevant setup is the holomorphic-topological twist of 4d $\CN=2$ SCFTs of Argyres-Douglas type on $C_q\times \mathbb{C}$, where $C_q$ is a topologically twisted cigar \cite{Dedushenko:2023cvd}. The compactification along the cigar circle $S^1$ can be understood from the Coulomb branch effective theory, which is described by an abelian gauge theory. The dynamics of BPS particles at a point $u$ in the Coulomb branch are encoded in the set of central charge functions
\be
\{Z_\gamma(u)\},\quad \gamma\in \Gamma\ ,
\ee
where $\Gamma$ is the electromagnetic charge lattice with the Dirac pairing $\langle~,~ \rangle$ and the flavor lattice $\Gamma_f$ is defined to be the lattice of central elements in $\Gamma$. Let us focus on a particular chamber of the Coulomb branch which contains only a finitely many hypermultiplets.

Due to the topological twist on the cigar, the compactification along $S^1$ involves an extra twisting by the $U(1)_r$ symmetry, which induces the uniform rotation of the phase of the central charges in the Coulomb branch effective theory
\be
Z_\gamma(u) \rightarrow e^{2\pi i}Z_\gamma(u)\ ,
\ee
along a closed path in the Coulomb branch. As explained in \cite{Gaiotto:2024ioj,Cecotti:2011iy}, this gives rise to the Janus-like configuration on the Coulomb branch which leads to an effective description of 3d $\CN=2$ abelian Chern-Simons matter (CSM) theory coupled to chiral multiplets of charge $\pm \gamma$, subject to a potential superpotential deformation. The class of 3d CSM theories obtained in this way are expected to flow to a superconformal fixed point with an $\CN=4$ supersymmetry enhancement. It is proposed in \cite{Gaiotto:2024ioj} that the ellipsoid partition function of the topologically twisted theory can be computed by the universal formula 
\be
S_b = \text{Tr}~ \hat\Phi\ ,
\ee
where $\hat\Phi$ is an operator acting on the auxiliary Hilbert space $\CH=L^2(\mathbb{R}^{\text{rk}(\Gamma/\Gamma_f)/2})$ constructed out of the BPS particles
\be\label{Sb proposal}
\hat\Phi =\prod^{\curvearrowright}_\gamma \Phi_b (x_\gamma) \ , 
\ee
where 
\be
\Phi_b(x) = \exp\left[\frac14\int_{\mathbb{R}+i\epsilon}\frac{e^{-2ixw}}{\sinh(wb)\sinh(w/b)}\frac{dw}{w}\right]
\ee
is the \emph{Faddeev quantum dilogarithm} and $x_\gamma$'s are the generators of the Weyl algebra satisfying
\be
[x_\gamma, x_{\gamma'}] = \frac{1}{2\pi i}\ , 
\ee
for $\langle \gamma, \gamma'\rangle=1$. The order of product in \eqref{Sb proposal} must be chosen according to the phase of the central charges at given $u$. The product is designed in the way that $S_b$ is invariant under wall-crossings, as we discuss below.
\subsection{Wall-crossing identities}

The formula \eqref{Sb proposal} is by construction a wall-crossing invariant quantity. This follows from the pentagon relation that the quantum dilogarithm satisfies
\be
\Phi_b (x_\gamma) \Phi_b(x_{\gamma'}) =  \Phi_b(x_{\gamma'})  \Phi_b(x_{\gamma}+x_{\gamma'})\Phi_b(x_{\gamma})\ ,
\label{eq: pentagon identity}
\ee
if $\langle \gamma,\gamma'\rangle=1$. Another useful identity is
\be
\Phi_b(x)\Phi_b(-x)  = \Phi_b(0)^2 e^{i\pi x^2}\ ,\quad \Phi_b(0) = e^{\frac{\pi i}{24}(b^2+b^{-2})}:=C_b\ .
\label{eq: inversion relation}
\ee
In the following, we utilize these identities to simplify the trace formula to obtain simple 3d gauge theory descriptions. At the level of 3d theories, these movements correspond to applying basic abelian dualities. \cite{Ekholm:2019lmb,Gaiotto:2024ioj} To avoid clutter, we often use the following abbreviation
\be\label{notation (a)}
(a) := \Phi_b(x_{\gamma_a})\ ,\quad e^a :=e^{i\pi x_{\gamma_a}^2} \ .
\ee

\subsection{Higher powers of monodromy operators}

It is natural to consider the trace of a higher power of the operator $\hat\Phi$, which corresponds to the twisted compactification along the Janus configuration which wraps the $r$-flow loop multiple times. We propose that the ellipsoid partition functions of the topological theory in the IR can be computed by the formula
\be\label{multi trace}
S^{(M)}_b =\text{Tr}~ \hat\Phi^M\ . 
\ee
For the Argyres-Douglas type theories where the Coulomb branch operators have fractional R-charges, there exists a positive integer $n$ such that the $n$-th wrapping of the Janus loop trivializes the $U(1)_r$ twisting. Therefore, we expect that if $n$ is the least common multiple (lcm) of the denominators of the Coulomb branch operators, $\hat\Phi^n$ is central in the operator algebra, which can be written as
\be\label{identity}
\hat\Phi^n = C_b^{s}~ f(m_\gamma)\cdot {\bf 1}\ .
\ee
Here $\bf 1$ is the identity operator acting on the Hilbert space $\CH$ defined above, and $f(m_\gamma)$ is a function of mass parameters which can be a non-trivial function in the presence of flavor symmetry. To illustrate how this relation arises, let us consider the simplest example, the $A_2$ AD theory with $n=5$. In this case, we have
\bea
\hat\Phi &= \Phi_b(x_{\gamma_1})\Phi_b(x_{\gamma_2})\Phi_b(-x_{\gamma_1})\Phi_b(-x_{\gamma_2}) \\
& =C_b^2 \Phi_b(x_{\gamma_2})\Phi_b(x_{\gamma_1+\gamma_2})e^{i\pi x_{\gamma_1}^2}\Phi_b(-x_{\gamma_2}) \\
&=C_b^4\Phi_b(x_{\gamma_2})e^{i\pi x_{\gamma_1}^2}e^{i\pi x_{\gamma_2}^2}\ .
\eea
Then we can compute
\bea
\hat\Phi^5 &=\left[C^4_b \Phi_b(x_{\gamma_2})e^{i\pi x_{\gamma_1}^2}e^{i\pi x_{\gamma_2}^2}\right]^5 \\
& = C_b^{24}(e^{i\pi x_{\gamma_1}^2}e^{i\pi x_{\gamma_2}^2})^6\ .
\eea
Identifying the generators with the canonical variables for the auxiliary one-particle quantum mechanics, i.e., $x_{\gamma_1} = p$ and $x_{\gamma_2}=q$, the integral kernel for this operator can be written as
\be
\langle q | \hat\Phi^5 |q'\rangle =  C_b^{24}\langle q | (e^{i\pi p^2}e^{i\pi q^2})^6 |q'\rangle\ . 
\ee
It is straightforward to check that this is proportional to the delta function $\delta(q-q')$, which implies that $\hat\Phi^5=C_b^{24}\cdot{\bf 1}$.\footnote{These operator identities in this paper hold up to a constant that is independent of $b$ and mass parameters.}

For general $G$ AD theories, we have
\begin{align}
\begin{array}{c|c|c}
G & ~~n & s\\
\hline
A_{2N} &~~ 2N+3 ~~& 8N(2N+1) \\
A_{2N+1} & ~~N+2 ~~& 4(N+1)(2N+1) \\
D_{2N} & ~~N ~~& 4N(2N-1) \\
D_{2N+1} & ~~2N+1 ~~& 8N(2N+1) \\
E_6 & 7 & 144 \\
E_7 & 5 & 126 \\
E_8 & 8 & 240
\end{array}
\end{align}

Below we show that \eqref{identity} holds for all $G$ AD theories of rank up to $8$ by explicit computations of corresponding integral kernels.

\subsection{Gauge theory descriptions and the modular data}
To evaluate the trace, we choose the basis of the generators $\{x_\gamma\}$'s to be 
\be
\{p_i,q_i\}_{i=1,\cdots, \frac12\text{rk}(\Gamma/\Gamma_f)}\ ,
\ee
which satisfies the canonical commutation relations
\be
[p_i,q_j] = \frac{1}{2\pi i}\delta_{ij}\ .
\ee
Inserting the unity in complete basis
\be
{\bf 1} = \int \prod_i dp_i~|{ p_i}\rangle \langle { p_i}| = \int \prod_i dq_i~|{ q_i}\rangle \langle { q_i}|\ ,
\ee
the partition function can be written as an integral formula
\be\label{integral formula}
S_b^{(N)} = C_b^{\kappa}\int \prod_{a=1}^m dx_a~ e^{i\pi{\bf x}^t A{\bf x}} \prod_{c}\Phi_b(x_c)^{n_c}\ ,
\ee
for some positive integers $m$, $n_a$'s, and an $m\times m$ integer matrix $A$. This expression can be interpreted as the ellipsoid partition function of an $\CN=2$ $U(1)^m$ Chern-Simons matter theory, where the effective mixed CS level is given by
\be
K = A + Q Q^T\ ,
\label{eq: A to K}
\ee
where $Q$ is the matrix of charges of chiral multiplets. The overall $b$-dependent phase factor $C_b^{\kappa}$ can be understood as the contributions from the supersymmetric gravitational Chern-Simons term
\be
S_{\text{grav}} = \frac{i k_g}{192\pi}\int \text{tr}\left(w \wedge dw + \frac23 w\wedge w \wedge w\right) + \cdots
\ee
and the background CS level for the $U(1)_R$ symmetry
\be
S_{\text{RR}} = \frac{ik_{RR}}{4\pi}\int  A_{(R)}\wedge dA_{(R)}+ \cdots\ ,
\ee
which evaluates on the squashed sphere background to \cite{Closset:2012vp,Closset:2019hyt,Cassani:2021fyv}
\be\label{kg and krr}
\frac{\pi i k_g}{24} (b^2+b^{-2}+f)\ ,\quad \text{and} \quad -\frac{\pi i k_{RR}}{4}(b^2+b^{-2}+2)\ ,
\ee
respectively.
Here $f\in \mathbb{Z}$ corresponds to the choice of framing.
In this paper, we do not keep track of the overall $b$- and mass-independent phase factors of the partition function, which are closely related to the choice of framing in our supersymmetric partition function calculation. 

The ellipsoid partition function is insensitive to the superpotential deformations, which are expected to be generated during the compactification process through non-perturbative effects. The 3d $\CN=2$ gauge theories derived from the trace formula allow for various half-BPS gauge invariant operators built from dressed monopole operators, which can contribute to the superpotential to deform the theory. For each description, we identify the collection of such operators and determine the flavor symmetries that are lifted by these deformations.

The existence of such $\CN=2$ Lagrangian descriptions allows us to explicitly calculate various supersymmetric observables to probe the IR phases of the compactified theory. In particular, we calculate the superconformal index of the IR theory and the partition functions of associated topological field theories on Seifert three-manifolds. For example, the partition functions on a degree-$p$ circle bundle over a closed Riemann surface of genus $g$ can be calculated by the formula (See e.g., \cite{Closset:2019hyt} and references therein)
\be\label{Mgp}
Z(M_{g,p}) = \sum_{P_a(x_a^*)=1} \CH^{g-1}(x_*)\CF^p(x_*) = \sum_\alpha S_{0\alpha}^{2-2g} T_{\alpha\alpha}^{-p}\ ,
\ee
from which we extract the modular data of boundary vertex operator algebra. See appendix \ref{app: conventions} for more details of the computation of \eqref{Mgp} and the conventions. 

Finally, we remark that a pair of theories extracted from $S_b^{(N)}$ and $S_b^{(-N)}$ are related by the orientation reversal, up to a background gravitational CS coupling. We can write
\bea
S_b^{(-N)} &= \Tr \left(\hat\Phi^{-1}\right)^N \\
& = C_b^{-\kappa}\int \prod_{a=1}^m dx_a~ e^{-i\pi{\bf x}^t A{\bf x}} \prod_{c}\Phi_b(x_c)^{-n_c}\ ,
\eea
where $\kappa, A$ and $n_c$ are defined in \eqref{integral formula}. Using the identity \eqref{eq: inversion relation}, we can write
\be
S_b^{(-N)} = C_b^{-\kappa -2\sum_c n_c} \int \prod_{a=1}^m dx_a~ e^{-i\pi{\bf x}^t A{\bf x}-i\pi\sum_c n_c x_c^2} \prod_{c}\Phi_b(-x_c)^{n_c}\ .
\ee
This is the partition function of the orientation reversed description of the theory derived from $S_b^{(N)}$. Together with the periodicity relation \eqref{identity}, we also conclude that the two theories obtained from $S_b^{(N)}$ and $S_b^{(n-N)}$ are related by the orientation reversal up to a gravitation CS couplings and possibly the background CS coupling for the flavor symmetries.

\newpage

\section{$A_{2N}$} \label{sec: A2N}
\begin{figure}[h]
  \centering
  \begin{tikzpicture}
			\node[W] (1) at (0,0){$\gamma_1$};
		 	\node[W] (2) at (2,0) {$\gamma_2$};
            \node[W] (3) at (4,0) {$\gamma_3$};
            \node[W] (4) at (6,0) {$\cdots$};
            \node[W] (5) at (8,0) {$\gamma_{2N}$};
			\draw[->] (1)--(2);
            \draw[<-] (2)--(3);
            \draw[->] (3)--(4);
            \draw[->] (4)--(5);
  \end{tikzpicture}
  \caption{ \label{A2N quiver}
  BPS quiver for the $A_{2N}$ theory in the canonical chamber}
  \end{figure}

Let us first consider the trace formula for the $A_{2N}$ theory whose BPS spectrum in the canonical chamber is described by the quiver in Figure \ref{A2N quiver}. We have
\be\label{ellipsoid A2N}
S_b^{(M)} = \text{Tr}~ \hat\Phi^M\ ,
\ee
where
\be
\hat\Phi =\prod_{i:\text{odd}}\Phi_b(x_{\gamma_i})\prod_{i:\text{even}}\Phi_b(x_{\gamma_i})\prod_{i:\text{odd}}\Phi_b(-x_{\gamma_i})\prod_{i:\text{even}}\Phi_b(-x_{\gamma_i})\ .
\label{eq: A2n monodromy}
\ee

In this section, we evaluate this formula for $N\leq 4$ and find that the monodromy operator satisfies the relation 
\be
\hat\Phi^{2N+3} = C_b^{8N(2N+1)}{\bf 1}\ .
\ee
For each $N$, we write down a 3d $\CN=2$ gauge theory description for $M=1,\cdots, 2N+2$, extracted from the trace formula. Performing the F-maximization for each example, we compute the superconformal index at the fixed point to probe the IR phases of the theory. We find that, if $M$ and $2N+3$ are co-prime, the theory either flows to
\begin{itemize}
\item[$\bullet$] a superconformal fixed point with $\CN=4$ supersymmetry enhancement, with the zero-dimensional Higgs and Coulomb branch. The IR theory can be topologically twisted to produce a pair of non-unitary semi-simple TFTs, which can support rational VOAs on their holomorphic boundary.
\item[$\bullet$] a unitary TFT, which admits a boundary condition that supports a unitary rational VOA on the boundary. 
\end{itemize}

The partition functions of the topological field theories on a Seifert manifold can be computed as summarized in appendix \ref{app: conventions}. We evaluate this when the three-manifold is a degree $p$ circle fibration over the genus $g$ Riemann surface, from which we extract the modular data of the boundary algebra, that are expected to be rational.

Note that the partition function computations \eqref{Mgp} do not uniquely fix the modular data of the boundary VOA, and we can only extract the following \emph{partial} information
\be
\{|S_{0\alpha}|\}\ ,\quad \{T_{\alpha\alpha}\}\ , 
\ee
as an ordered set, up to an overall phase factor for $T$. For $N=1,2,3,4$, we calculate these data and write down candidate boundary vertex operator algebras compatible with this result. 

When the theory flows to a rank-zero SCFT, it is natural to expect that the ellipsoid formula \eqref{ellipsoid A2N} calculates the partition function of the $A$-twisted theory, which can be obtained by the dimensional reduction of Donaldson's twist in 4d $\CN=2$ theory. 

If we normalize the $T$ matrix so that $T_{00}=1$, we will see that the entries of modular matrices lie in the field $\mathbb{Q}[{\rm e}_{2N+3}]$, a cyclotomic extension of rationals by ${\rm e}_n :=e^{2\pi i/n}$. For fixed $N$, we find that the modular data we obtain for different choices of $M$'s are related by Galois transformations.\footnote{See \cite{Harvey:2019qzs} and references therein for more discussions of the Galois actions on RCFTs and their modular data. For small values of $N$, such as $N=1$ and $N=2$, we explicitly checked that the modular data for different choices of $M$'s can be obtained by solving the pentagon and hexagon equations of simple lines, which satisfy the same fusion rules as $M(2,5)$ and $M(2,7)$ cases. } The Galois group, $\text{Gal}(\mathbb{Q}[{\rm e}_{2N+3}]/\mathbb{Q})$, is defined as the automorphism group of $\mathbb{Q}[{\rm e}_{2N+3}]$ that fixes $\mathbb{Q}$. This is isomorphic to the multiplicative group of integers co-prime to $2N+3$ (mod $2N+3$). For an integer $M$ co-prime to $2N+3$, the action of the Galois group element $\sigma_M$ on the modular $T$-matrix is
\be 
\sigma_M: T \rightarrow  T^M\ .
\label{eq: Gal T}
\ee
And on the $S$-matrix, it acts as permutations of entries with signs
\be
\sigma_M: S_{\a\b}\rightarrow \epsilon_M(\a,\b) S_{\a,\pi_M(\b)} \ ,
\ee
where $\epsilon_M$ is a sign and $\pi_M$ is a permutation.

If $M$ and $2N+3$ are not co-prime, we find that the resulting theory has runaway vacua with no supersymmetric vacuum, and the supersymmetric indices and partition functions are ill-defined. The first example can be found from the $A_6$ theory whose monodromy operator satisfies $\hat\Phi^9\sim {\bf 1}$, with $M=3,6$. See section \ref{sec: degenerate} for more detail.

\subsection{$A_2$}

Let us first consider the simplest Argyres-Douglas theory, and examine the formula
\be
S_b^{(M)} = \text{Tr}\left[\Phi_b(x_{\gamma_1})\Phi_b(x_{\gamma_2})\Phi_b(-x_{\gamma_1})\Phi_b(-x_{\gamma_2})\right]^M\ .
\ee

\subsubsection{$M=1$}

This is the case studied in \cite{Gaiotto:2024ioj}. We can write 
\bea\label{M1 A2 integral}
S_b^{(1)} &= \text{Tr}\left[\Phi_b(x_{\gamma_1})\Phi_b(x_{\gamma_2})\Phi_b(-x_{\gamma_1})\Phi_b(-x_{\gamma_2})\right] \\
& = C_b^4\text{Tr}\left[e^{i\pi x_{\gamma_1}^2}\Phi_b(x_{\gamma_2})e^{i\pi x_{\gamma_2}^2}\right]\\
& = C_b^4 \int dx ~\Phi_b(x)e^{i\pi x^2}\ ,
\eea
up to an overall $b$-independent phase factor. The final expression is the ellipsoid partition function for the $\CN=2$ $U(1)$ gauge theory with
\be\label{Tmin gauge}
K=(2)\ , Q=(1)\ ,
\ee
together with the gravitational CS level $k_g=4$ and $k_{RR}=0$. The theory has a $U(1)$ global symmetry 
\be\label{choice A A11}
A = U(1)_{\text{top}}\ ,
\ee
which is the topological symmetry for the gauge group.

\paragraph{Superconformal index} This is the first example of the gauge theory which is expected to flow to the simplest $\CN=4$ rank-zero theory, called ${\CT}_\text{min}$. \cite{Gang:2018huc, Gang:2021hrd} The superconformal index at the IR fixed point reads
\be
I_{\text{SCI}} = 1-q-\left(\eta +\frac{1}{\eta }\right) q^{3/2}-2 q^2-\left(\eta +\frac{1}{\eta }\right) q^{5/2}-2 q^3-\left(\eta +\frac{1}{\eta }\right) q^{7/2}-2 q^4+ \cdots\ ,
\ee
where $\eta$ is the fugacity for the global symmetry $A$.
This expression is a strong signal that the theory flows to a conformal fixed point with a supersymmetry enhancement.\footnote{See appendix \ref{app: conventions} for more details.} Being an $\CN=4$ theory, the IR theory can be topologically twisted to produce a pair of topological field theories. The corresponding choice of R-symmetry is parametrized by
\be
R_\nu = R_{0} + \nu A\ ,
\ee
where $A$ is the $U(1)_A = U(1)_C- U(1)_H$ symmetry in the IR superconformal algebra and $R_0=U(1)_C+U(1)_H$ is the superconformal $R$ symmetry. The choices $\nu=-1$ and $\nu=1$ correspond to the topological A- and B-twist respectively in this convention.

\paragraph{Modular data ($\nu=-1$ twist)} 

The modular data we extract from the supersymmetric partition function is
\be\label{data Ia1}
\left\{|S_{0\alpha}|\right\} = \left\{\frac{2}{\sqrt 5}\sin(2\pi/5), ~\frac{2}{\sqrt 5}\sin (\pi/5)\right\}\ ,\quad \left\{T_{\alpha\alpha}\right\} = \left\{ 1, e^{2\pi i(-1/5)}\right\}\ ,
\ee
up to an overall phase factor for $T$. This data is compatible with the modular data of Virasoro minimal model $M(2,5)$ where
\be
S = \frac{2}{\sqrt 5} \left(\begin{array}{cc}-\sin (2\pi/5) & \sin (\pi/5) \\ \sin(\pi/5) &  \sin(2\pi/5)\end{array}\right)\ ,\quad T = \text{diag}\left( e^{2\pi i(11/60)}, e^{2\pi i(-1/60)}\right)\ ,
\ee
with $c=-22/5$.
It is argued in \cite{Ferrari:2023fez,Creutzig:2024ljv} that this SCFT indeed admits a boundary condition that supports $M(2,5)$, by explicit computations of the boundary OPEs. 

We can also check, by numerically evaluating the integral \eqref{M1 A2 integral} that 
\be
|S_b^{(1)}| \approx 0.851397 \approx \frac{2}{\sqrt 5}\sin (2\pi/5)\ 
\ee
where the RHS coincides with $|S_{00}|$, which is compatible with the claim that the trace formula computes the partition function of the topologically twisted theory. See appendix \ref{app:numerical evaluataion} for more detail.

It is interesting to note that, the partial modular data \eqref{data Ia1} are also compatible with the modular data of an intermediate algebra called affine $E_{7\frac12}$ at level 1
\be\label{modular E7.5}
S = \frac{2}{\sqrt 5} \left(\begin{array}{cc}\sin (2\pi/5) & \sin (\pi/5) \\ \sin(\pi/5) & - \sin(2\pi/5)\end{array}\right)\ ,\quad T = \text{diag}\left( e^{2\pi i(-19/60)}, e^{2\pi i(29/60)}\right)
\ee
with $c=38/5$.
The corresponding two modular invariant modules for \eqref{modular E7.5} can be constructed by considering the half-index of a gauge theory which is conjecturally dual to $\CT_{\text{min}}$. \cite{Kim:2024dxu}
However, as first noted in \cite{Mathur:1988na}, the fusion rule contains negative coefficients and therefore cannot define a consistent rational VOA.

\paragraph{Modular data ($\nu=1$ twist)} The modular data we extract for the other twist is 
\be
\{|S_{0\alpha}|\} = \left\{\frac{2}{\sqrt 5}\sin(2\pi/5), ~\frac{2}{\sqrt 5}\sin (\pi/5)\right\}\ ,\quad \left\{T_{\alpha\alpha}\right\} = \left\{1, ~e^{2\pi i(1/5)}\right\}\ ,
\ee
up to an overall phase factor for $T$.
Notice that the partition functions on Seifert manifolds are complex conjugate to that of the $\nu=1$ twist computed from \eqref{data Ia2}, which is consistent with the conjecture that the rank-zero theory $\CT_{\text{min}}$ is mirror to its orientation reversal $\overline{\CT}_{\text{min}}$. This data is compatible with the modular data of two irreducible modules in affine $\mathrm{osp}(1|2)$ at level 1 \cite{Ferrari:2023fez}
\be
S = \frac{2}{\sqrt 5} \left(\begin{array}{cc}\sin (2\pi/5) & -\sin (\pi/5) \\ -\sin(\pi/5) & - \sin(2\pi/5)\end{array}\right)\ ,\quad T = \text{diag}\left( e^{2\pi i(-1/60)}, e^{2\pi i(11/60)}\right)\ ,
\ee
with $c=2/5$.
It is argued in \cite{Ferrari:2023fez} that this SCFT indeed admits a boundary condition that supports this VOA, by an explicit computation of the boundary OPEs.

\subsubsection{$M=2$} The trace of the second power of the monodromy operator is 
\bea\label{311 M=2}
S_b^{(2)} &= \text{Tr}\left[\Phi_b(x_{\gamma_1})\Phi_b(x_{\gamma_2})\Phi_b(-x_{\gamma_1})\Phi_b(-x_{\gamma_2})\right]^2 \\
&=C_b^8\text{Tr}\left[e^{i\pi x_{\gamma_1}^2}\Phi_b(x_{\gamma_2})e^{i\pi x_{\gamma_2}^2}\right]^2 \\
&= C_b^8\int d x_1dx_2dy_1dy_2 ~ e^{i\pi(x_1^2+x_2^2+y_1^2+y_2^2)-2\pi i(x_1-x_2)(y_1-y_2)}\Phi_b(x_1)\Phi_b(x_2)\\
&= C_b^8 \int dx_1 dx_2~e^{-i \pi x_1^2+ 4 i \pi x_1 x_2-i \pi x_2^2}\Phi_b(x_1)\Phi_b(x_2)\ , 
\eea
where we integrated out $y_1$ and $y_2$ for the last equality. 
The final expression is the partition function of an $\CN=2$ $U(1)^2$ gauge theory coupled to two chiral multiplets with the CS level and the charge matrix
\be\label{M2 matrices}
K =     \begin{pmatrix} 0 & 2 \cr 2 & 0 \end{pmatrix}\ ,\quad Q =  \begin{pmatrix} 1 & 0 \cr 0 & 1 \end{pmatrix}\ ,
\ee
with the gravitational CS level $k_g-6k_{RR}=8$.
The gauge theory admits two gauge invariant half-BPS monopole operators 
\be\label{monopole 1}
\phi_2^2 V_{(-1,0)}\ ,\quad \phi_1^2 V_{(0,-1)}\ ,
\ee
which we can use to deform the theory. 

\paragraph{Superconformal index} The theory deformed by the two monopole superpotentials \eqref{monopole 1} does not have any global symmetry to exhibit supersymmetry enhancement to $\CN=4$ in the infrared. Instead, we find that, upon a proper choice of background couplings ($R_\phi=1$ and $k_{RR}=1$)
\be
I_{\text{SCI}} = 1\ ,
\ee
which suggests that the theory directly flows to a unitary topological field theory.

\paragraph{Modular data} The modular data we extract from the partition function of the topological field theory is
\be\label{data F4}
\left\{|S_{0\alpha}|\right\} = \left\{\frac{2}{\sqrt 5}\sin(\pi/5), ~\frac{2}{\sqrt 5}\sin (2\pi/5)\right\}\ ,\quad \left\{T_{\alpha\alpha}\right\} = \left\{1, ~e^{2\pi i(3/5)}\right\}\ ,
\ee
again up to an overall phase factor for $T$. This data is compatible with the modular data of affine Lie algebra $F_4$ at level one, where 
\be 
S=\frac{2}{\sqrt{5}}\begin{pmatrix} \text{sin}(\pi/5) & \text{sin}(2\pi/5) \cr \text{sin}(2\pi/5) & -\text{sin}(\pi/5) \end{pmatrix}\ , \qquad T= e^{2\pi i(\frac{7}{60})}\begin{pmatrix} 1 & 0 \cr 0 & e^{2\pi i(\frac{3}{5})} \end{pmatrix}\ ,
\ee
with $c=26/5$, and is sometimes called the modular representation of conjugate Fibonacci MTC.

\subsubsection{$M=3$}
 The trace of the third power of the monodromy operator is 
\bea
S_b^{(3)} &= \text{Tr}\left[\Phi_b(x_{\gamma_1})\Phi_b(x_{\gamma_2})\Phi_b(-x_{\gamma_1})\Phi_b(-x_{\gamma_2})\right]^3 \\
&=C_b^{12}\text{Tr}\left[e^{i\pi x_{\gamma_1}^2}\Phi_b(x_{\gamma_2})e^{i\pi x_{\gamma_2}^2}\right]^3 \\
&=C_b^{12}\text{Tr}\left[ \Phi_b(x_{\gamma_1})\Phi_b(x_{\gamma_2}) \left(e^{i\pi x_{\gamma_1}^2}e^{i\pi x_{\gamma_2}^2}\right)^3\right] \\
&= C_b^{12}\int dx_1 dx_2~    e^{-4\pi i x_1 x_2} \Phi_b(x_1) \Phi_b(x_2)\ .
\eea
The final expression is the partition function of an $\CN=2$ $U(1)^2$ gauge theory coupled to two chiral multiplets with the CS level and the charge matrix
\be
K =     \begin{pmatrix} 1 & -2 \cr -2 & 1 \end{pmatrix}\ ,\quad Q =  \begin{pmatrix} 1 & 0 \cr 0 & 1 \end{pmatrix}\ .
\ee
We find that this gauge theory is precisely the orientation reversal of the theory \eqref{M2 matrices} up to gravitational CS terms, which is expected from the relation $\hat\Phi^5=C_b^{24}{\bf 1}$. This description also admits the two gauge invariant half-BPS monopole operators 
\be\label{monopole 2}
\phi_2^2 V_{(1,0)}\ ,\quad \phi_1^2 V_{(0,1)}\ .
\ee
\paragraph{Superconformal index} From the fact that this theory is the orientation reversal of \eqref{M2 matrices}, we immediately have
\be
I_{\text{SCI}} = 1\ .
\ee

\paragraph{Modular data} The modular data we extract from the partition function of the topological field theory is
\be
\left\{|S_{0\alpha}|\right\} = \left\{\frac{2}{\sqrt 5}\sin(\pi/5), ~\frac{2}{\sqrt 5}\sin (2\pi/5)\right\}\ ,\quad \left\{T_{\alpha\alpha}\right\} = \left\{1, ~e^{2\pi i(2/5)}\right\}\ ,
\ee
and therefore the partition function is the complex conjugate of that of \eqref{M2 matrices}. This data is compatible with the modular data of affine Lie algebra $G_2$ at level one, where 
\be \label{Modular matrices of Fibonacci alg_bar}
S=\frac{2}{\sqrt{5}}\begin{pmatrix} \text{sin}(\pi/5) & \text{sin}(2\pi/5) \cr \text{sin}(2\pi/5) & -\text{sin}(\pi/5) \end{pmatrix}\ , \qquad T= e^{2\pi i(\frac{53}{60})}\begin{pmatrix} 1 & 0 \cr 0 & e^{2\pi i(\frac{2}{5})} \end{pmatrix}\ ,
\ee
with $c=14/5$, and is sometimes called the modular representation of Fibonacci MTC.

\subsubsection{$M=4$}

The trace of the fourth power gives
\bea
S_b^{(4)} &= \text{Tr}\left[\Phi_b(x_{\gamma_1})\Phi_b(x_{\gamma_2})\Phi_b(-x_{\gamma_1})\Phi_b(-x_{\gamma_2})\right]^4 \\
& = C_b^{18}\text{Tr}~\Phi_b(x_{\gamma_1}) \left(e^{i\pi x_{\gamma_2}^2}e^{i\pi x_{\gamma_1}^2}\right)^4 e^{i\pi x_{\gamma_2}^2} \\
& = C_b^{18}\int dx ~e^{-2\pi i x^2}\Phi_b(x)\ .
\eea
This is the ellipsoid partition function of an $\CN=2$ $U(1)$ gauge theory with
\be
K=(-1)\ , Q=(1)\ ,
\ee
with  $k_g-6k_{RR}=18$. Up to a background gravitational Chern-Simons couplings, this theory describes the orientation reversal of the gauge theory \eqref{Tmin gauge}. Therefore we expect that the theory flows to an $\CN=4$ superconformal fixed point $\overline{\CT}_{\text{min}}$, which is the orientation reversal of the minimal rank-zero SCFT. We can choose
\be
A=-U(1)_{\text{top}}\ .
\ee

\paragraph{Superconformal index} The superconformal index coincides with that of $\CT_{\text{min}}$
\be
I_{\text{SCI}} = 1-q-\left(\eta +\frac{1}{\eta }\right) q^{3/2}-2 q^2-\left(\eta +\frac{1}{\eta }\right) q^{5/2}-2 q^3-\left(\eta +\frac{1}{\eta }\right) q^{7/2}-2 q^4+ \cdots\ .
\ee

\paragraph{Modular data ($\nu=-1$ twist)} 
The modular data we extract from the partition function coincides with the $M=1$ case:
\be
\{|S_{0\alpha}|\} = \left\{\frac{2}{\sqrt 5}\sin(2\pi/5), ~\frac{2}{\sqrt 5}\sin (\pi/5)\right\}\ ,\quad \left\{T_{\alpha\alpha}\right\} = \left\{1, ~e^{2\pi i(1/5)}\right\}\ ,
\ee
which is compatible with the affine $\mathrm{osp}(1|2)$ at level one. 

\paragraph{Modular data ($\nu=1$ twist)}

The modular data we extract from the partition function again coincides with the $M=1$ case:
\be\label{data Ia2}
\left\{|S_{0\alpha}|\right\} = \left\{\frac{2}{\sqrt 5}\sin(2\pi/5), ~\frac{2}{\sqrt 5}\sin (\pi/5)\right\}\ ,\quad \left\{T_{\alpha\alpha}\right\} = \left\{ 1, e^{2\pi i(-1/5)}\right\}\ ,
\ee
up to an overall phase, which is compatible with $M(2,5)$ or $(E_{7\frac12})_1$ as discussed before.

\subsection{$A_4$}

Let us consider the next simplest example, the $A_4$ theory. The trace formula reads
\be
S_b^{(M)} = \text{Tr}\left[\Phi_b(x_{\gamma_1})\Phi_b(x_{\gamma_3})\Phi_b(x_{\gamma_2})\Phi_b(x_{\gamma_4})\Phi_b(-x_{\gamma_1})\Phi_b(-x_{\gamma_3})\Phi_b(-x_{\gamma_2})\Phi_b(-x_{\gamma_4})\right]^M\ .
\ee
In section \ref{subsec: A4 periodicity} we show that 
\be
\hat\Phi^7=C_b^{80} \bf{1}\ ,
\ee
as expected,
which implies that the theories extracted from $S_b^{(M)}$ and $S_b^{(7-M)}$ are related by the orientation reversal. Therefore it is enough to consider the cases $M=1,2,3$. 

We will use the abbreviation \eqref{notation (a)} throughout the rest of the paper.

\subsubsection{$M=1$}
The monodromy operator for $A_4$ theory reads
\begin{align}
    \hat\Phi = (1)(3)(2)(4)(-1)(-3)(-2)(-4)\ ,
\end{align}
which can be simplified to
\begin{align}
    \hat\Phi 
    &=C_b^2(3)(2) e^1 (2)(4)(-3)(-2)(-4)
    \nonumber\\
    &=C_b^4(3)(4)(2)(-3) e^1 e^2 (-3)(-4)\ .
\end{align}
The ellipsoid partition function of the theory is
\begin{align}\label{evaluation A41}
    S_b^{(1)} &= \Tr \hat\Phi
    \nonumber\\
    &= C_b^{8} \Tr [ e^1 e^2 (-4) e^3 e^4 (2)(-3) ]
    \nonumber\\
    &= C_b^{10} \Tr [ e^1 e^2 e^3 e^4 (2) e^3 (2) ]
    \nonumber\\
    & = C_b^{10} \int dx_1 dx_2\;
    e^{\pi i (-2x_1^2 + 4 x_1 x_2 -  x_2^2)} \Phi_b(x_1)\Phi_b(x_2) 
    \end{align}
which is the integral representation of the ellipsoid partition function of an $\CN=2$ $U(1)^2$ gauge theory coupled to two chiral multiplets with
\begin{align}
    K = \left(
    \begin{array}{cc}
        -1 & 2 \\
        2 & 0
    \end{array}
    \right)
    \;,\;\;
    Q = \left(
    \begin{array}{cc}
        1 & 0 \\
        0 & 1
    \end{array}
    \right)\ .
\end{align}
This description has a single gauge invariant half-BPS monopole operator
\begin{align}
    \phi_1^2 V_{(0,-1)}.
    \label{eq: A4^1 monopole}
\end{align}
Deforming by the monopole superpotential, the remaining flavor symmetry is
\be
A = U(1)_{\text{top}}^{(1)}\ ,
\ee
which is the topological symmetry that corresponds to the first $U(1)$ factors of the gauge group.

\paragraph{Superconformal index}
Based on the F-maximization computation, we claim that the gauge theory flows to the $\CN=4$ rank-zero theory in the IR, upon the superpotential deformation \eqref{eq: A4^1 monopole}. The superconformal index at the IR fixed point is 
\begin{align}
    I_{\text{SCI}} = 1 - q -\left(\eta + \frac{1}{\eta}\right)q^{3/2} - 2 q^2 - \eta\, q^{5/2} - \left(1-\frac{1}{\eta^2}\right)q^3+\left(\frac{1}{\eta}-\eta\right)q^{7/2}+\frac{1}{\eta^2}q^4 +\cdots
\end{align}
which is strong evidence for the supersymmetry enhancement to $\CN=4$ at the IR fixed point. This computation also suggests that the IR rank-zero theory is dual to the theory $T_2$ discussed in \cite{Gang:2023rei}.
\paragraph{Modular data ($\nu=-1$ twist)}
The modular data for $\nu=-1$ twist extracted from the supersymmetric partition function is 
\begin{align}
    \{ |S_{0\a}| \} = \left\{
    \frac{2}{\sqrt{7}} \cos(3\pi/14)
    ,
    \frac{2}{\sqrt{7}} \sin(\pi/7)
    ,
    \frac{2}{\sqrt{7}} \cos(\pi/14)
    \right\}
    \;,\;\;
    \{T_{\a\a}\}
    =
    \left\{
    1,e^{2\pi i (4/7)},e^{2\pi i (5/7)}
    \right\}\ .
\end{align}
These are compatible with
\be \label{Modular data of M(2,7)}
S =\frac{2}{\sqrt{7}}\left(
\begin{array}{ccc}
 \cos \left(\frac{3 \pi }{14}\right) & \sin \left(\frac{\pi }{7}\right) & -\cos \left(\frac{\pi }{14}\right) \\
 \sin \left(\frac{\pi }{7}\right) & \cos \left(\frac{\pi }{14}\right) & \cos \left(\frac{3 \pi }{14}\right) \\
 -\cos \left(\frac{\pi }{14}\right) & \cos \left(\frac{3 \pi }{14}\right) & -\sin \left(\frac{\pi }{7}\right) \\
\end{array}
\right) \ , T =\left(
\begin{array}{ccc}
 e^{2\pi i (17/42)} & 0 & 0 \\
 0 & e^{2\pi i (-1/42)} & 0 \\
 0 & 0 & e^{2\pi i (5/42)} \\
\end{array}
\right) \ ,
\ee
which are the modular data of $M(2,7)$ with central charge $c=-68/7$.

\paragraph{Modular data ($\nu=1$ twist)}
The modular data for $\nu=1$ twist extracted from the supersymmetric partition function are
\begin{align}
    \{ |S_{0\a}| \} = \left\{
    \frac{2}{\sqrt{7}} \cos(\pi/14)
    ,
    \frac{2}{\sqrt{7}} \cos(3\pi/14)
    ,
    \frac{2}{\sqrt{7}} \sin(\pi/7) 
    \right\}
    \;,\;\;
    \{T_{\a\a}\}
    =
    \left\{
    1,e^{2\pi i (1/7)},e^{2\pi i (3/7)}
    \right\}\ .
\end{align}
These are compatible with
\be
S =\frac{2}{\sqrt{7}}\left(
\begin{array}{ccc}
 \cos \left(\frac{\pi }{14}\right) & -\cos \left(\frac{3 \pi }{14}\right) & \sin \left(\frac{\pi }{7}\right) \\
 -\cos \left(\frac{3 \pi }{14}\right) & -\sin \left(\frac{\pi }{7}\right) & \cos \left(\frac{\pi }{14}\right) \\
 \sin \left(\frac{\pi }{7}\right) & \cos \left(\frac{\pi }{14}\right) & \cos \left(\frac{3 \pi }{14}\right) \\
\end{array}
\right) \ , T =\left(
\begin{array}{ccc}
 e^{2\pi i (-1/42)} & 0 & 0 \\
 0 & e^{2\pi i (5/42)} & 0 \\
 0 & 0 & e^{2\pi i (17/42)} \\
\end{array}
\right) \ ,
\ee
which are the modular data of the affine $\mathrm{osp}(1|2)$ at level 2 with central charge $c=4/7$. \cite{Ferrari:2023fez}

\subsubsection{$M=2$}
Consider the trace of the second power of the monodromy operator
\begin{align}
    S_b^{(2)} &= \Tr \hat\Phi^2
    \nonumber\\
    &= C_b^{14} \Tr [(3)(4)(2)(-3) e^1 e^2 e^3 e^4 (2) e^3 (2) e^1 e^2 (-3)(-4)]
    \nonumber\\
    &= C_b^{18} \Tr[ (3)(4) e^3 e^1 e^2 e^3 e^4 e^3 e^1 e^2 (-1) e^3 (-4) ]
    \nonumber\\
    &= C_b^{20} \Tr[ (3) e^4 e^3 e^1 e^2 e^3 e^4 e^3 e^1 e^2 e^3 (-4) ]
    \nonumber\\
    &= C_b^{20} \Tr [ e^1 e^2 e^3 e^4 e^1 e^2 e^3 e^1 e^2 e^1 (-2)(-4) ]
    \nonumber\\
    &= C_b^{20} \int dx_1 dx_2\; e^{\pi i ( 2 x_1^2 - 4 x_1 x_2)} \Phi_b(x_1) \Phi_b(x_2)
\end{align}
The integral is the partition function of an $\CN=2$ $U(1)^2$ gauge theory coupled to 2 chiral multiplets with
\begin{align}
    K = \left(
    \begin{array}{cc}
        3 & -2 \\
        -2 & 1
    \end{array}
    \right)
    \;,\;\;
    Q = \left(
    \begin{array}{cc}
        1 & 0 \\
        0 & 1
    \end{array}
    \right)\ .
\end{align}
In this description, there is a single gauge invariant half-BPS monopole operator
\begin{align}
    \phi_1^2 V_{(0,1)}\ .
    \label{eq: A4^2 monopole}
\end{align}
Deforming by the monopole potential, the remaining flavor symmetry can be identified with
\be
A = -U(1)_{\text{top}}^{(1)}\ .
\ee

\paragraph{Superconformal index}
After the superpotential deformation \eqref{eq: A4^2 monopole}, the gauge theory is expected to flow to an IR fixed point with supersymmetry enhancement. The superconformal index at the IR fixed point reads
\begin{align}
    I_{\text{SCI}} = 1-q-\left(\eta +\frac{1}{\eta }\right) q^{3/2}-2 q^2-\eta\,  q^{5/2}+\left(\frac{1}{\eta^2}-1\right) q^3-\left(\eta-\frac{1}{\eta } \right) q^{7/2}+ \frac{1}{\eta^2}\, q^4 +\cdots\ ,
\end{align}
which is a strong signal that the IR fixed point is a rank-zero SCFT. This result also suggests that the theory is mirror dual to the one obtained in the $M=1$ case.

\paragraph{Modular data ($\nu=-1$ twist)}
The modular data for $\nu=-1$ twist extracted from the supersymmetric partition function is
\begin{align}
    \{ |S_{0\a}| \} = \left\{
    \frac{2}{\sqrt{7}} \cos(\pi/14)
    ,
    \frac{2}{\sqrt{7}} \cos(3\pi/14)
    ,
    \frac{2}{\sqrt{7}} \sin(\pi/7) 
    \right\}
    \;,\;\;
    \{T_{\a\a}\}
    =
    \left\{
    1,e^{2\pi i (1/7)},e^{2\pi i (3/7)}
    \right\}\ ,
\end{align}
which agrees with the affine $\mathrm{osp}(1|2)$ at level 2, as discussed in the previous example.

\paragraph{Modular data ($\nu=1$ twist)}

The modular data for $\nu=1$ twist extracted from the supersymmetric partition function is
\begin{align}
    \{ |S_{0\a}| \} = \left\{
    \frac{2}{\sqrt{7}} \cos(3\pi/14)
    ,
    \frac{2}{\sqrt{7}} \sin(\pi/7)
    ,
    \frac{2}{\sqrt{7}} \cos(\pi/14)
    \right\}
    \;,\;\;
    \{T_{\a\a}\}
    =
    \left\{
    1,e^{2\pi i (4/7)},e^{2\pi i (5/7)}
    \right\}\ .
\end{align}
which agrees with the modular data of $M(2,7)$ as discussed in the previous example.

\subsubsection{$M=3$}
The trace of the third power of the monodromy operator reads
\begin{align}
    S_b^{(3)} &= \Tr \hat\Phi^3
    \nonumber\\
    &= C_b^{28} \Tr [e^1 e^2 e^3 e^4 e^1 e^2 e^3 e^1 e^2 e^1 (-2)(3) e^4 (2) e^3 (2) e^1 e^2 (-3)(-4)]
    \nonumber\\
    &= C_b^{32} \Tr [e^1 e^2 e^3 e^4 e^1 e^2 e^3 e^1 e^2 (3) e^4 e^1 e^2 e^3 (2) e^1 e^2 (-3) e^4 (-3)]
    \nonumber\\
    &= C_b^{32} \Tr [ (e^1 e^2 e^3 e^4)^3 (-2)(-1)e^2 e^3 (-2)(-4) e^3 e^4 ]
    \nonumber\\
    &= C_b^{32} \int dx_1 dx_2 dx_3 dx_4\;
    e^{\pi i ( -x_1^2 -x_2^2 -x_3^2 -x_4^2 +4 x_1 x_2 -4 x_2 x_3 + 4 x_3 x_4 )} \Phi_b(x_1)\Phi_b(x_2)\Phi_b(x_3)\Phi_b(x_4)\ .
\end{align}
The integral expression can be interpreted as a $\CN=2$ $U(1)^4$ gauge theory coupled to four chiral multiplets with
\begin{align}
    K = \left(
    \begin{array}{cccc}
        0 & 2 & 0 & 0 \\
        2 & 0 & -2 & 0 \\
        0 & -2 & 0 & 2 \\
        0 & 0 & 2 & 0
    \end{array}
    \right)
    \;,\;\;
    Q = \left(
    \begin{array}{cccc}
        1 & 0 & 0 & 0 \\
        0 & 1 & 0 & 0 \\
        0 & 0 & 1 & 0 \\
        0 & 0 & 0 & 1
    \end{array}
    \right)
\end{align}
This description has four gauge invariant half-BPS monopole operators
\begin{align}
    \phi_1^2 V_{(0,-1,0,-1)}
    \;,\;\;
    \phi_2^2 V_{(-1,0,0,0)}
    \;,\;\;
    \phi_3^2 V_{(0,0,0,-1)}
    \;,\;\;
    \phi_4^2 V_{(-1,0,-1,0)}\ .
    \label{eq: A4^3 monopole}
\end{align}
\paragraph{Superconformal index}
Deforming the theory by all four monopole operators breaks all flavor symmetry in the UV gauge theory. The superconformal index in this setting gives
\begin{align}
    I_{\text{SCI}} = 1\ ,
\end{align}
which is a strong signal that the theory flows to a unitary topological field theory in the IR.

\paragraph{Modular data}
The modular data of the IR topological theory extracted from the supersymmetric partition functions are
\begin{align}
    \{ |S_{0\a}| \} = \left\{
    \frac{2}{\sqrt{7}} \sin(\pi/7)
    ,
    \frac{2}{\sqrt{7}} \cos(\pi/14)
    ,
    \frac{2}{\sqrt{7}} \cos(3\pi/14)
    \right\}
    \;,\;\;
    \{T_{\a\a}\}
    =
    \left\{
    1,e^{2\pi i (5/7)},e^{2\pi i (1/7)}
    \right\}\ .
\end{align}
These are compatible with
\be
S =\frac{2}{\sqrt{7}}\left(
\begin{array}{ccc}
 \sin \left(\frac{\pi }{7}\right) & \cos \left(\frac{\pi }{14}\right) & \cos \left(\frac{3 \pi }{14}\right) \\
 \cos \left(\frac{\pi }{14}\right) & -\cos \left(\frac{3 \pi }{14}\right) & \sin \left(\frac{\pi }{7}\right) \\
 \cos \left(\frac{3 \pi }{14}\right) & \sin \left(\frac{\pi }{7}\right) & -\cos \left(\frac{\pi }{14}\right) \\
\end{array}
\right) \ , T =\left(
\begin{array}{ccc}
 e^{2\pi i (2/42)} & 0 & 0 \\
 0 & e^{2\pi i (-10/42)} & 0 \\
 0 & 0 & e^{2\pi i (8/42)} \\
\end{array}
\right) \ ,
\ee
which are the complex conjugate of the modular data of $(A_1,5)_{\frac12}$. See appendix \ref{app:modular data} for the definitions and conventions. 
The bulk topological theory is expected to be dual to the orientation reversal of the pure CS theory 
$(SU(2)_{5}\times U(1)_{-2})/\mathbb{Z}^{[1]}_2$, where we denote the gauging by the non-anomalous $\mathbb{Z}_2$ one-form symmetry by $/\mathbb{Z}^{[1]}_2$.

\subsubsection{$M=7$}\label{subsec: A4 periodicity}
The seventh power of the monodromy operator simplifies to
\begin{align}
    \hat\Phi^7 = C_b^{80} (e^1 e^2 e^3 e^4)^{10}\ .
\end{align}
It is straightforward to check by an explicit but tedious computation that this operator is proportional to the identity operator in the two-particle Hilbert space
    \begin{equation}
        \hat\Phi^7 =C_b^{80} \mathbf{1} \ .
    \end{equation}
In general, we show in appendix \ref{app: identity} that for $n$-particle quantum mechanics, the following identity holds
\be
\left(\prod_{i=1}^{2n} e^{i\pi x_{\gamma_i}^2} \right)^{4n+2} = {\bf 1} \ ,
\ee
for general $(A_1,A_{2n})$ theories.

\subsection{$A_6$}
Next, we consider $A_6$ theory which gives
\begin{align}
    S_b^{(M)} = \Tr[(1)(3)(5)(2)(4)(6)(-1)(-3)(-5)(-2)(-4)(-6)]^M\ .
\end{align}
In section \ref{sec: A6^9}, we show that
\begin{align}
    \hat\Phi^9 = C_b^{168} \bf{1}\ ,
\end{align}
on a three-particle quantum mechanics Hilbert space. This implies that the modular data extracted from $S_b^{(M)}$ and $S_b^{(9-M)}$ are related by complex conjugation, up to an overall phase for the $T$-matrices.

\subsubsection{$M=1$}
The trace of the first power of the monodromy operator gives
\begin{align}
    S_b^{(1)} &= \Tr \hat\Phi
    \nonumber\\
    &= C_b^{16} \Tr[ e^1 e^2 e^3 e^4 e^5 e^6 e^4 (2) e^3 (2) (-4) ]
    \nonumber\\
    &= C_b^{16} \int dx_1 dx_2 dx_3 \;
    e^{\pi i ( 2 x_2^2 + x_3^2 - 4 x_1 x_2 + 4 x_2 x_3 )} \Phi_b(x_1) \Phi_b(x_2) \Phi_b(x_3)\ ,
\end{align}
where the integral expression corresponds to an $\CN=2$ $U(1)^3$ gauge theory coupled to three chiral multiplets with
\begin{align}
    K = \left(
    \begin{array}{ccc}
        1 & -2 & 0 \\
        -2 & 3 & 2 \\
        0 & 2 & 2
    \end{array}
    \right)
    \;,\;\;
    Q = \left(
    \begin{array}{ccc}
        1 & 0 & 0 \\
        0 & 1 & 0 \\
        0 & 0 & 1
    \end{array}
    \right)\ .
\end{align}
This gauge theory admits two gauge invariant half-BPS monopole operators
\begin{align}
    \phi_1^2 V_{(0,1,-1)}
    \;\;,\;\;
    \phi_2^2 V_{(1,0,0)}\;\;.
    \label{eq: A6^1 monopole}
\end{align}
Deforming by these monopole operators, the remaining flavor symmetry can be identified with
\be
A = U(1)_{\text{top}}^{(2)} + U(1)_{\text{top}}^{(3)}\ .
\ee
\paragraph{Superconformal index}
Upon a superpotential deformation by the two monopole operators above, the superconformal index at the fixed point reads
\bea
    I_{\text{SCI}} =& 1\!-q\!-\!\left(\eta +\frac{1}{\eta }\right) q^{3/2}\!-2 q^2\!-\eta\, q^{5/2}+\left(\frac{1}{\eta ^2}-1\right) q^3\!+\left(\frac{1}{\eta }\!-\!\eta \right) q^{7/2}\!+\left(\eta \!+\!\frac{2}{\eta } \!-\!\frac{1}{\eta ^3}\right) q^{9/2} \!+\cdots\ ,
\eea
which is a strong signal that the theory flows to an $\CN=4$ rank-zero fixed point.
\paragraph{Modular data ($\nu=-1$ twist)}
The modular data extracted from the twisted partition functions are
\begin{align}
    \{|S_{0\a}|\}
    =
    \frac{2}{3}\sin(\pi/9)\times
    \left\{
    d,d^2-1,1,d+1
    \right\}
    \;\;,\;\;
    \left\{T_{\a\a}\right\}
    =
    \left\{
    1,
    e^{2\pi i (4/9)},
    e^{2\pi i (3/9)},
    e^{2\pi i (6/9)}
    \right\}
\end{align}
with $d=2\cos(\pi/9)$. They are compatible with
\be
S =\frac{2}{3}\sin(\pi/9)\begin{pmatrix} -d  & 1\!-\!d^2 & 1 & d+1 \\ 
1\!-\!d^2 & 0 & d^2 \!-\! 1 & 1\!-\!d^2 \\ 
1 & d^2 \!-\! 1 & d+1 & d \\ 
d+1 & 1\!-\!d^2 & d & -1
\end{pmatrix} \ , T =\left(
\begin{array}{cccc}
 e^{2\pi i (-13/36)} & 0 & 0 & 0 \\
 0 & e^{2\pi i (3/36)} & 0 & 0 \\
 0 & 0 & e^{2\pi i (-1/36)} & 0 \\
 0 & 0 & 0 & e^{2\pi i (11/36)} \\
\end{array}
\right) \ ,
\ee
which are the modular data of $M(2,9)$ with central charge $c=-46/3$.

\paragraph{Modular data ($\nu=1$ twist)}
The modular data extracted from the twisted partition functions are
\begin{align}
    \{|S_{0\a}|\}
    =
    \frac{2}{3}\sin(\pi/9)\times
    \left\{
    d+1,d^2-1,d,1
    \right\}
    \;\;,\;\;
    \left\{T_{\a\a}\right\}
    =
    \left\{
    1,
    e^{2\pi i (1/9)},
    e^{2\pi i (3/9)},
    e^{2\pi i (6/9)}
    \right\}
\end{align}
with $d=2\cos(\pi/9)$. They are compatible with
\be
S =\frac{2}{3}\sin(\pi/9)\begin{pmatrix} d+1 & 1\!-\!d^2 & d & -1 \\ 1\!-\!d^2 & 0 & d^2 \!-\! 1 & 1\!-\!d^2 \\ d & d^2 \!-\! 1 & -1 & -d-1 \\ -1 & 1\!-\!d^2 & -d-1 & -d 
\end{pmatrix} \ , T =\left(
\begin{array}{cccc}
 e^{2\pi i (-1/36)} & 0 & 0 & 0 \\
 0 & e^{2\pi i (3/36)} & 0 & 0 \\
 0 & 0 & e^{2\pi i (11/36)} & 0 \\
 0 & 0 & 0 & e^{2\pi i (-13/36)} \\
\end{array}
\right) \ ,
\ee
which are the modular data of the affine $\mathrm{osp}(1|2)$ at level 3 with central charge $c=2/3$.

\subsubsection{$M=2$}

The trace of the second power gives
\begin{align}
    S_b^{(2)} &= \Tr \hat\Phi^2
    \nonumber\\
    &= C_b^{36} \Tr [ (e^1 e^2 e^3 e^4 e^5 e^6)^2 e^1 e^2 e^3 e^4 (2)(1) e^2 e^3 (2) ]
    \nonumber\\
    &=C_b^{36}\int dx_1 dx_2 dx_3\;
    e^{\pi i ( x_1^2 + x_2^2 - x_3^2 - 4 x_1 x_2 + 4 x_2 x_3 )} \Phi_b(x_1) \Phi_b(x_2) \Phi_b(x_3)
\end{align}
where the integral is the partition function of an $\CN=2$ $U(1)^3$ gauge theory coupled to three chiral multiplets with
\begin{align}
    K = \left(
    \begin{array}{ccc}
        2 & -2 & 0 \\
        -2 & 2 & 2 \\
        0 & 2 & 0
    \end{array}
    \right)
    \;,\;\;
    Q = \left(
    \begin{array}{ccc}
        1 & 0 & 0 \\
        0 & 1 & 0 \\
        0 & 0 & 1
    \end{array}
    \right)\ .
\end{align}
In this description, there are two gauge invariant half-BPS monopole operators
\begin{align}
    \phi_2^2 V_{(0,0,-1)}
    \;\;,\;\;
    \phi_3^2 V_{(-1,-1,0)}
    \;\;.
    \label{eq: A6^2 monopole}
\end{align}
Deforming by the two monopole superpotentials, we are left with
\be
A = U(1)_{\text{top}}^{(1)} - U(1)_{\text{top}}^{(2)}\ .
\ee

\paragraph{Superconformal index}
Performing the F-maximization, we find that the superconformal index at the fixed point reads
\bea
    I_{\text{SCI}} =& 1\!-\!q\!-\!\left(\eta \!+\!\frac{1}{\eta }\right) q^{3/2}\!-2 q^2-\frac{q^{5/2}}{\eta }+\left(\eta ^2\!-\!1\right) q^3+\left(\eta \!-\!\frac{1}{\eta }\right) q^{7/2}+\left(\frac{1}{\eta } \!+\! 2 \eta\!-\!\eta ^3\right) q^{9/2} + \cdots
\eea
which is strong evidence that the theory flows to a $\CN=4$ rank-zero fixed point in the infrared. 
\paragraph{Modular data ($\nu=-1$ twist)}
The modular data extracted from the supersymmetric partition function are 
\begin{align}
    \left\{|S_{0\a}|\right\}
    =
    \frac{2}{3}\sin(\pi/9)\times
    \left\{
    d+1,d^2-1,d,1
    \right\}
    \;\;,\;\;
    \left\{T_{\a\a}\right\}
    =
    \left\{
    1,
    e^{2\pi i (8/9)},
    e^{2\pi i (6/9)},
    e^{2\pi i (3/9)}
    \right\}
\end{align}
with $d=2\cos(\pi/9)$. The result agrees with the complex conjugate of the modular data of the affine $\mathrm{osp}(1|2)$ at level 3 discussed above.

\paragraph{Modular data ($\nu=1$ twist)}
The modular data extracted from the supersymmetric partition function are
\begin{align}
    \{|S_{0\a}|\}
    =
    \frac{2}{3}\sin(\pi/9)\times
    \left\{
    d,d^2-1,1,d+1
    \right\}
    \;\;,\;\;
    \left\{T_{\a\a}\right\}
    =
    \left\{
    1,
    e^{2\pi i (5/9)},
    e^{2\pi i (6/9)},
    e^{2\pi i (3/9)}
    \right\}
\end{align}
with $d=2\cos(\pi/9)$. The result agrees with the complex conjugate of the modular data of $M(2,9)$ discussed above.

\subsubsection{$M=3$} \label{sec: degenerate}
The trace of the third power gives
\begin{align}
    S_b^{(3)} &= \Tr \hat\Phi^3
    \nonumber\\
    &= C_b^{56} \Tr [(e^1 e^2 e^3 e^4 e^5 e^6)^4 e^1 e^2 e^3 e^4]
    \nonumber\\
    &= \frac{1}{2} C_b^{56} \int dx_1 dx_2 dx_3 \;
    e^{\pi i ( x_1^2 + x_2^2 + x_3^2 - 2 x_1 x_2 - 2 x_2 x_3 + 2 x_1 x_3 )}\ .
\end{align}
The integral is the partition function of an $\CN=2$ $U(1)^3$ pure CS theory with
\begin{align}
    K=\left(
    \begin{array}{ccc}
        1 & -1 & 1 \\
        -1 & 1 & -1 \\
        1 & -1 & 1
    \end{array}
    \right)\,.
\end{align}
The integral can be further simplified by a change of variables
\begin{align}
    S_b^{(3)} = \frac{1}{2} C_b^{56} \int dx_1 dx_2 dx_3 e^{\pi i x_3^2}\ ,
\end{align}
which can be interpreted as a $U(1)^3$ pure $\CN=2$ CS theory with 
\begin{align}
    K=\left(
    \begin{array}{ccc}
        0 & 0 & 0 \\
        0 & 0 & 0 \\
        0 & 0 & 1
    \end{array}
    \right)\,.
\end{align}
This theory has runaway vacua with no supersymmetric vacuum, which makes the supersymmetric indices and the partition functions ill-defined. Note that this is an example where $M$ is not coprime to $2N+3=9$ with no corresponding Galois group element.

Similarly, if we consider the sixth power of the monodromy operator, we find
\begin{align}
    S_b^{(6)} = C_b^{112} \int dx_1 \cdots dx_6\,
    e^{\pi i {\bf{x}}^t K {\bf{x}} }
\end{align}
where $K$ is
\begin{align}
    K = \left(
    \begin{array}{cccccc}
        0 & 0 & 0 & -2 & 2 & 0 \\
        0 & 0 & 1 & 1 & -1 & 1 \\
        0 & 1 & -2 & -1 & 1 & -3 \\
        -2 & 1 & -1 & -2 & 3 & -1 \\
        2 & -1 & 1 & 3 & -4 & 1 \\
        0 & 1 & -3 & -1 & 1 & -4
    \end{array}
    \right)\ .
\end{align}
This expression can be interpreted as a partition function of an $\CN=2$ $U(1)^6$ pure CS theory with $\det K=0$, which has runaway vacua. In fact, we can use a change of variable to rewrite
\be
S_b^{(6)} = C_b^{112}\int dx_1\cdots dx_6 ~ e^{2\pi i (x_2x_3+x_4x_5)}\ .
\ee

\subsubsection{$M=4$}
The trace of the fourth power gives
\begin{align}
    S_b^{(4)} &= \Tr \hat\Phi^4
    \nonumber\\
    &= C_b^{72}\Tr [ (e^1 e^2 e^3 e^4 e^5 e^6)^3 e^2 e^1 e^2 e^2 e^3 e^2 e^2 e^3 e^4 e^5 e^3 (2)(4)(6) e^1 e^2 e^3 e^4 e^5 e^6 (2)(4)(6) e^3 ]
    \nonumber\\
    &= C_b^{72} \int 
    \left(\prod_{i=1}^6 dx_i \Phi_b(x_i)\right)
    e^{\pi i ( -x_1^2 - x_2^2 -x_3^2 -x_4^2 -x_5^2 -x_6^2 + 4 x_1 x_2 - 4 x_2 x_3 + 4 x_2 x_4 + 4 x_3 x_5 - 4 x_4 x_5 + 4 x_4 x_6 )} 
\end{align}
which corresponds to an $\CN=2$ $U(1)^6$ gauge theory coupled to six chiral multiplets with
\begin{align}
    K = \left(
    \begin{array}{cccccc}
        0 & 2 & 0 & 0 & 0 & 0 \\
        2 & 0 & -2 & 2 & 0 & 0 \\
        0 & -2 & 0 & 0 & 2 & 0 \\
        0 & 2 & 0 & 0 & -2 & 2 \\
        0 & 0 & 2 & -2 & 0 & 0 \\
        0 & 0 & 0 & 2 & 0 & 0
    \end{array}
    \right)
    \;\;,\;\;
    Q = \left(
    \begin{array}{cccccc}
        1 & 0 & 0 & 0 & 0 & 0 \\
        0 & 1 & 0 & 0 & 0 & 0 \\
        0 & 0 & 1 & 0 & 0 & 0 \\
        0 & 0 & 0 & 1 & 0 & 0 \\
        0 & 0 & 0 & 0 & 1 & 0 \\
        0 & 0 & 0 & 0 & 0 & 1
    \end{array}
    \right)\ .
\end{align}
In this description, we have six gauge invariant half-BPS monopole operators
\begin{align}
    &\phi_1^2 V_{(0,-1,0,0,-1,0)}
    \;\;,\;\;
    \phi_2^2 V_{(-1,0,0,0,0,0)}
    \;\;,\;\;
    \phi_3^2 V_{(0,0,0,0,-1,-1)}
    \;,
    \nonumber\\
    &\phi_4^2 V_{(0,0,0,0,0,-1)}
    \;\;,\;\;
    \phi_5^2 V_{(-1,0,-1,0,0,0)}
    \;\;,\;\;
    \phi_6^2 V_{(0,0,-1,-1,0,0)}
    \;\;.
    \label{eq: A6^4 monopole}
\end{align}

\paragraph{Superconformal index}
By deforming the theory with all of the monopole operators above, we find that the superconformal index reduces to
\begin{align}
    I_{\text{SCI}} = 1\ ,
\end{align}
which implies that the theory flows directly to a unitary topological field theory in the infrared. 

\paragraph{Modular data}
The modular data extracted from the supersymmetric partition function is \footnote{The computation of partition function via the Bethe vacua formalism is numerically unstable for this example, particularly for the vacuum labeled by $\alpha=1$ (The corresponding modular matrix entries are marked with $\star$). To obtain a consistent result, we first disabled the second monopole operator $\phi_2^2V_{(-1,0,0,0,0,0)}$ and set the corresponding fugacity to approach $1$ from the right.}
\begin{align}
    \left\{ |S_{0\a}| \right\} = \frac{2}{3} \sin(\pi/9)\times\left\{
    1,d^2-1^\star,d+1,d
    \right\}
    \;,\;\;
    \{T_{\a\a}\}
    =
    \left\{
    1,{e^{2\pi i (7/9)}}^\star,e^{2\pi i (3/9)},e^{2\pi i (6/9)}
    \right\}
\end{align}
with $d = 2 \cos(\pi/9)$. This data is compatible with 
\be
S =\frac{2}{3}\sin(\pi/9)\begin{pmatrix} 1 & d^2 \!-\! 1 & d+1 & d \\ d^2 \!-\! 1 & 0 & 1\!-\!d^2 & d^2 \!-\! 1 \\ d+1 & 1\!-\!d^2 & d & -1 \\ d & d^2 \!-\! 1 & -1 & -d-1
\end{pmatrix} \ , T =\left(
\begin{array}{cccc}
 e^{2\pi i (5/36)} & 0 & 0 & 0 \\
 0 & e^{2\pi i (-3/36)} & 0 & 0 \\
 0 & 0 & e^{2\pi i (17/36)} & 0 \\
 0 & 0 & 0 & e^{2\pi i (-7/36)} \\
\end{array}
\right) \ ,
\ee
which are the complex conjugate of the modular data of $(A_1,7)_{\frac12}$. See appendix \ref{app:modular data} for the definitions and conventions. 
The bulk topological theory is dual to the orientation reversal of the
 pure CS theory $(SU(2)_{7}\times U(1)_{2})/\mathbb{Z}_2^{[1]}$.

\subsubsection{$M=9$}\label{sec: A6^9}
The ninth power of the monodromy operator simplifies to
\begin{align}
    \hat\Phi^9 = C_b^{168} (e^1 e^2 e^3 e^4 e^5 e^6)^{14}.
\end{align}
In Appendix \ref{app: identity}, we show the operator $(e^1 e^2 e^3 e^4 e^5 e^6)^{14}$ is the identity operator in the three-particle Hilbert space
    \begin{equation}
        (e^1 e^2 e^3 e^4 e^5 e^6)^{14} = \mathbf{1}.
    \end{equation}
It then implies $\hat\Phi^9 = C_b^{168}\bf{1}$ as expected. 

\subsection{$A_8$}
Let us consider the $A_8$ theory with the trace formula
\begin{align}
    S_b^{(M)} = \Tr [ (1)(3)(5)(7)(2)(4)(6)(8)(-1)(-3)(-5)(-7)(-2)(-4)(-6)(-8) ]^M
\end{align}
we will see in section \ref{sec: A8^11} that
\begin{align}
    \hat\Phi^{11} = C_b^{288} \bf{1}\;,
\end{align}
which implies that the theories from $S_b^{(M)}$ and $S_b^{(11-M)}$ are related by the orientation reversal. This allows us to consider the cases $M=1,2,3,4,$ and $5$.
\subsubsection{$M=1$}
The trace of the first power of the monodromy operator gives
\begin{align}
   S_b^{(1)} &= \Tr \hat\Phi
   \nonumber\\
   &= C_b^{22} \Tr [ e^1 e^2 e^3 e^4 e^5 e^6 e^7 e^8 (-4)e^5 (6) e^4 (-8) e^7 (-8) ]
   \nonumber\\
   &= C_b^{22} \int \left(
   \prod_{i=1}^4 dx_i \Phi_b(x_i)
   \right)
   \;
   e^{\pi i ( -2x_1^2 - 2 x_2^2 - x_3^2 + 4 x_1 x_2 -4 x_1 x_3 + 4 x_2 x_3 - 4 x_1 x_4 )}\ .
\end{align}
The integral is the partition function of an $\CN=2$ $U(1)^4$ gauge theory coupled to four chiral multiplets with 
\begin{align}
    K = \left(
    \begin{array}{cccc}
        -1 & 2 & -2 & -2 \\
        2 & -1 & 2 & 0 \\
        -2 & 2 & 0 & 0 \\
        -2 & 0 & 0 & 1
    \end{array}
    \right)
    \;\;,\;\;
    Q = \left(
    \begin{array}{cccc}
        1 & 0 & 0 & 0 \\
        0 & 1 & 0 & 0 \\
        0 & 0 & 1 & 0 \\
        0 & 0 & 0 & 1
    \end{array}
    \right)\ .
\end{align}
This description has three gauge invariant half-BPS monopole operators
\begin{align}
    \phi_1^2 V_{(0,0,0,1)},\quad \phi_2^2 V_{(0,0,-1,1)},\quad \phi_4^2 V_{(1,1,0,0)}\ .
    \label{eq: A8^1 monopole}
\end{align}
Deforming by the three monopole operators, we are left with 
\be
A = -U(1)^{(1)}_{\text{top}} + U(1)^{(2)}_{\text{top}}\ .
\ee

\paragraph{Superconformal index}
Performing the F-maximization, we find that the superconformal index at the fixed point reads
\bea
    I_{\text{SCI}} =& 1\!-\!q\!-\!\left(\eta \!+\!\frac{1}{\eta }\right) q^{3/2}\!-2 q^2-\eta {q^{5/2}}+\left(\frac{1}{\eta ^2}\!-\!1\right) q^3+\left(-\eta \!+\!\frac{1}{\eta }\right) q^{7/2}+\left( \eta \!+\!\frac{2}{\eta}\!-\!\frac{1}{\eta ^3}\right) q^{9/2} +\cdots
\eea
which is a strong signal that the theory flows to an $\CN=4$ rank-zero fixed point.

\paragraph{Modular data ($\nu=-1$ twist)}

The modular data extracted from the supersymmetric partition function are
\begin{align}
    \left\{ | S_{0\a} | \right\} &= 
    \frac{2}{\sqrt{11}}\times
    \left\{
    \sin(2\pi/11),
    \cos(3\pi/22),
    \cos(\pi/22),
    \cos(5\pi/22),
    \sin(\pi/11)
    \right\}
    \nonumber\\
    \left\{T_{\a\a}\right\} &= 
    \left\{ 
    1,
    e^{2\pi i \left(\frac{7}{11}\right)},
    e^{2\pi i \left(\frac{4}{11}\right)},
    e^{2\pi i \left(\frac{2}{11}\right)},
    e^{2\pi i \left(\frac{1}{11}\right)}
   \right\}\ .
\end{align}
They are compatible with
\bea
S &=\frac{2}{\sqrt{11}}\left(
\begin{array}{ccccc}
 \sin \left(\frac{2 \pi }{11}\right) & -\cos \left(\frac{3 \pi }{22}\right) & \cos \left(\frac{\pi }{22}\right) & -\cos \left(\frac{5 \pi }{22}\right) & \sin \left(\frac{\pi }{11}\right) \\
 -\cos \left(\frac{3 \pi }{22}\right) & \cos \left(\frac{5 \pi }{22}\right) & \sin \left(\frac{\pi }{11}\right) & -\cos \left(\frac{\pi }{22}\right) & \sin \left(\frac{2 \pi }{11}\right) \\
 \cos \left(\frac{\pi }{22}\right) & \sin \left(\frac{\pi }{11}\right) & -\cos \left(\frac{3 \pi }{22}\right) & -\sin \left(\frac{2 \pi }{11}\right) & \cos \left(\frac{5 \pi }{22}\right) \\
 -\cos \left(\frac{5 \pi }{22}\right) & -\cos \left(\frac{\pi }{22}\right) & -\sin \left(\frac{2 \pi }{11}\right) & \sin \left(\frac{\pi }{11}\right) & \cos \left(\frac{3 \pi }{22}\right) \\
 \sin \left(\frac{\pi }{11}\right) & \sin \left(\frac{2 \pi }{11}\right) & \cos \left(\frac{5 \pi }{22}\right) & \cos \left(\frac{3 \pi }{22}\right) & \cos \left(\frac{\pi }{22}\right) \\
\end{array}
\right) \ ,\\ 
T &=e^{2\pi i (-4/33)}\left(
\begin{array}{ccccc}
 1 & 0 & 0 & 0 & 0 \\
 0 & e^{2\pi i (7/11)} & 0 & 0 & 0 \\
 0 & 0 & e^{2\pi i (4/11)} & 0 & 0 \\
 0 & 0 & 0 & e^{2\pi i (2/11)} & 0 \\
 0 & 0 & 0 & 0 & e^{2\pi i (1/11)} \\
\end{array}
\right) \ ,
\eea
which are the modular data of $M(2,11)$ with central charge $c=-232/11$.

\paragraph{Modular data ($\nu=1$ twist)}

The modular data extracted from the supersymmetric partition function are
\begin{align}
    \left\{ | S_{0\a} | \right\} &= 
    \frac{2}{\sqrt{11}}\times
    \left\{
    \cos(\pi/22),
    \sin(\pi/11),
    \cos(3\pi/22),
    \sin(2\pi/11),
    \cos(5\pi/22)
    \right\}
    \nonumber\\
    \left\{T_{\a\a}\right\} &= 
    \left\{ 
    1,
    e^{2\pi i (\frac{10}{11})},
    e^{2\pi i (\frac{1}{11})},
    e^{2\pi i (\frac{6}{11})},
    e^{2\pi i (\frac{3}{11})}
    \right\}
\end{align}
They are compatible with
\bea
S &=\frac{2}{\sqrt{11}}\left(
\begin{array}{ccccc}
 \cos \left(\frac{\pi }{22}\right) & \sin \left(\frac{\pi }{11}\right) & -\cos \left(\frac{3 \pi }{22}\right) & -\sin \left(\frac{2 \pi }{11}\right) & \cos \left(\frac{5 \pi }{22}\right) \\
 \sin \left(\frac{\pi }{11}\right) & \sin \left(\frac{2\pi }{11}\right) & \cos \left(\frac{5 \pi }{22}\right) & \cos \left(\frac{3 \pi }{22}\right) & \cos \left(\frac{\pi }{22}\right) \\
 -\cos \left(\frac{3 \pi }{22}\right) & \cos \left(\frac{5 \pi }{22}\right) & \sin \left(\frac{\pi }{11}\right) & -\cos \left(\frac{\pi }{22}\right) & \sin \left(\frac{2 \pi }{11}\right) \\
 -\sin \left(\frac{2 \pi }{11}\right) & \cos \left(\frac{3 \pi }{22}\right) & -\cos \left(\frac{\pi }{22}\right) & \cos \left(\frac{5 \pi }{22}\right) & -\sin \left(\frac{\pi }{11}\right) \\
 \cos \left(\frac{5 \pi }{22}\right) & \cos \left(\frac{\pi }{22}\right) & \sin \left(\frac{2 \pi }{11}\right) & -\sin \left(\frac{\pi }{11}\right) & -\cos \left(\frac{3 \pi }{22}\right) \\
\end{array}
\right) \ ,\\ 
T &=e^{2\pi i (-1/33)}\left(
\begin{array}{ccccc}
 1 & 0 & 0 & 0 & 0 \\
 0 & e^{2\pi i (10/11)} & 0 & 0 & 0 \\
 0 & 0 & e^{2\pi i (1/11)} & 0 & 0 \\
 0 & 0 & 0 & e^{2\pi i (6/11)} & 0 \\
 0 & 0 & 0 & 0 & e^{2\pi i (3/11)} \\
\end{array}
\right) \ ,
\eea
which are the modular data of affine $\mathrm{osp}(1|2)$ at level 4 with central charge $c=8/11$.

\subsubsection{$M=2$}
The trace of the second power gives
\begin{align}
   S_b^{(2)} &= \Tr \hat\Phi^2
   \nonumber\\
   &= C_b^{48} \Tr [ e^1 e^2 e^3 e^4 e^5 e^6 e^7 e^8 (6) e^5 e^6 e^7 e^8 e^5 e^6 e^7 e^3 e^4 e^5 e^6 (4) e^2 e^3 e^4 e^5 (2) e^3 (2) ]
   \nonumber\\
   &= C_b^{48} \int \left(
   \prod_{i=1}^4 dx_i \Phi_b(x_i)
   \right)
   \;
   e^{\pi i ( 3 x_1^2 + x_2^2 - x_3^2 - x_4^2 + 4 x_1 x_2 + 2 x_1 x_3 + 2 x_1 x_4 + 2 x_3 x_4 )}
\end{align}
where the integral is the partition function of an $\CN=2$ $U(1)^4$ gauge theory coupled to four chiral multiplets with
\begin{align}
    K = \left(
    \begin{array}{cccc}
        4 & 2 & 1 & 1 \\
        2 & 2 & 0 & 0 \\
        1 & 0 & 0 & 1 \\
        1 & 0 & 1 & 0
    \end{array}
    \right)
    \;\;,\;\;
    Q = \left(
    \begin{array}{cccc}
        1 & 0 & 0 & 0 \\
        0 & 1 & 0 & 0 \\
        0 & 0 & 1 & 0 \\
        0 & 0 & 0 & 1
    \end{array}
    \right)\ .
\end{align}
In this description, we have three gauge invariant half-BPS monopole operators,
\begin{align}
    \phi_1 \phi_3 V_{(0,0,0,-1)},\quad \phi_1 \phi_4 V_{(0,0,-1,0)},\quad \phi_3\phi_4 V_{(-1,2,0,0)}
    \label{eq: A8^2 monopole}
\end{align}
Deforming the theory by the three monopole operators, we are left with
\be
A = 2U(1)_{\text{top}}^{(1)} + U(1)_{\text{top}}^{(2)}\ .
\ee

\paragraph{Superconformal index}
The superconformal index at the fixed point reads
\begin{align}
    I_{\text{SCI}} = 1 - q - \left(\eta+\frac{1}{\eta}\right) q^{3/2} - 2 q^2 + \left(1+\eta^2 + \frac{1}{\eta^2}\right) q^3 + 2 \left(\eta+\frac{1}{\eta}\right)q^{7/2} + \left(3+\frac{1}{\eta^2}\right)q^4 + \cdots
\end{align}
which is a strong signal that the theory flows to an $\CN=4$ rank-zero theory in the infrared.

\paragraph{Modular data ($\nu=-1$ twist)}
The modular data extracted from the partition functions are
\begin{align}
    \left\{ | S_{0\a} | \right\} &= 
    \frac{2}{\sqrt{11}}
    \times
    \left\{
    \cos(3\pi/22),
    \cos(5\pi/22),
    \sin(\pi/11),
    \cos(\pi/22),
    \sin(2\pi/11)
    \right\}
    \nonumber\\
    \left\{T_{\a\a}\right\} &= 
    \left\{ 
    1,
    e^{2\pi i (\frac{3}{11})},
    e^{2\pi i (\frac{8}{11})},
    e^{2\pi i (\frac{4}{11})},
    e^{2\pi i (\frac{2}{11})}
    \right\}
\end{align}
We were not able to identify these with any modular data of rational VOAs known in the literature. From the $SL(2,\mathbb{Z})$ relations $S^2=C$ and $(ST)^3=C$, the expected central charge is $20/11$ (mod 4).

\paragraph{Modular data ($\nu=1$ twist)}
The modular data extracted from the partition functions are
\begin{align}
    \left\{ | S_{0\a} | \right\} &= 
    \frac{2}{\sqrt{11}}
    \times
    \left\{
    \cos(5\pi/22),
    \cos(\pi/22),
    \sin(2\pi/11),
    \sin(\pi/11),
    \cos(3\pi/22)
    \right\}
    \nonumber\\
    \left\{T_{\a\a}\right\} &= 
    \left\{ 
    1,
    e^{2\pi i (\frac{5}{11})},
    e^{2\pi i (\frac{6}{11})},
    e^{2\pi i (\frac{3}{11})},
    e^{2\pi i (\frac{7}{11})}
    \right\}
\end{align}
From the $SL(2,\mathbb{Z})$ relations, the expected central charge is $92/11$ (mod 4). We were not able to identify these with any modular data of rational VOAs known in the literature, but the $T$-matrix coincides with the modular data of a vector-valued modular form which appears as solutions to a fifth-order modular linear differential equation, which corresponds to characters with central charge $224/11$. \cite{Kaidi:2021ent} See Table 7 in \emph{loc.cit.} The corresponding S-matrix gives rise to negative fusion rule coefficients.

\subsubsection{$M=3$}
The trace of the third power of the monodromy operator gives
\begin{align}
   S_b^{(3)} &= \Tr \hat\Phi^3
   \nonumber\\
   &= C_b^{72} \Tr [ e^1 e^2 e^3 e^4 e^5 e^6 e^7 e^8 e^1 e^2 e^3 e^4 e^5 e^6 e^7 e^1 e^2 e^3 e^4 e^5 e^6 
   \nonumber\\
   &\qquad\qquad\qquad\qquad  \times e^1 e^2 e^3 e^4 e^5 e^1 e^2 e^3 e^4 e^1 e^2 e^3 e^1 e^2 e^1 (-2)(-4)(-6)(-8)]
   \nonumber\\
   &= C_b^{72} \int \left(
   \prod_{i=1}^4 dx_i \Phi_b(x_i)
   \right)
   \;
   e^{\pi i ( 2 x_1^2 + 2 x_2^2 - 4 x_1 x_2 - 4 x_1 x_3 + 4 x_1 x_4 - 4 x_2 x_4 )}
\end{align}
which corresponds to an $\CN=2$ $U(1)^4$ gauge theory coupled to four chiral multiplets with
\begin{align}
    K = \left(
    \begin{array}{cccc}
        3 & -2 & -2 & 2 \\
        -2 & 3 & 0 & -2 \\
        -2 & 0 & 1 & 0 \\
        2 & -2 & 0 & 1
    \end{array}
    \right)
    \;\;,\;\;
    Q = \left(
    \begin{array}{cccc}
        1 & 0 & 0 & 0 \\
        0 & 1 & 0 & 0 \\
        0 & 0 & 1 & 0 \\
        0 & 0 & 0 & 1
    \end{array}
    \right)
\end{align}
This description has three gauge invariant half-BPS monopole operators
\begin{align}
    \phi_1^2 V_{(0,0,1,0)},\quad \phi_2^2 V_{(0,0,1,1)},\quad 
    \phi_3^2  V_{(1,1,0,0)}\ .
    \label{eq: A8^3 monopole}
\end{align}
Deforming by the monopole operators leaves 
\be
A = U(1)_{\text{top}}^{(1)}- U(1)_{\text{top}}^{(2)}\ .
\ee

\paragraph{Superconformal index}
Performing the F-maximization, the superconformal index at the fixed point is
\bea
    I_{\text{SCI}} = 1\!-\!q-\left(\eta \!+\!\frac{1}{\eta }\right) q^{3/2}\!-2 q^2-\frac{q^{5/2}}{\eta }+\left(\eta ^2\!-\!1\right) q^3+\left(\eta \!-\!\frac{1}{\eta }\right) q^{7/2}+\left(\frac{1}{\eta }\!+\!2 \eta\!-\!\eta ^3\right) q^{9/2} +\cdots
\eea
which is a strong signal that the theory flows to an $\CN=4$ rank-zero theory.

\paragraph{Modular data ($\nu=-1$ twist)}
The modular data extracted from the partition function are
\begin{align}
    \left\{ | S_{0\a} | \right\} &= 
    \frac{2}{\sqrt{11}}\times
    \left\{
    \cos(\pi/22),
    \sin(\pi/11),
    \cos(3\pi/22),
    \sin(2\pi/11),
    \cos(5\pi/22)
    \right\}
    \nonumber\\
    \left\{T_{\a\a}\right\} &= 
    \left\{ 
    1,
    e^{2\pi i (\frac{10}{11})},
    e^{2\pi i (\frac{1}{11})},
    e^{2\pi i (\frac{6}{11})},
    e^{2\pi i (\frac{3}{11})}
    \right\}
\end{align}
which are compatible with the modular data of affine $\mathrm{osp}(1|2)$ at level 4.

\paragraph{Modular data ($\nu=1$ twist)}
The modular data extracted from the partition function are
\begin{align}
    \left\{ | S_{0\a} | \right\} &= 
    \frac{2}{\sqrt{11}}\times
    \left\{
    \sin(2\pi/11),
    \cos(3\pi/22),
    \cos(\pi/22),
    \cos(5\pi/22),
    \sin(\pi/11)
    \right\}
    \nonumber\\
    \left\{T_{\a\a}\right\} &= 
    \left\{ 
    1,
    e^{2\pi i (\frac{7}{11})},
    e^{2\pi i (\frac{4}{11})},
    e^{2\pi i (\frac{2}{11})},
    e^{2\pi i (\frac{1}{11})}
    \right\}
\end{align}
which are compatible with the modular data of $M(2,11)$.

\subsubsection{$M=4$}
The trace of the fourth power of the monodromy operator gives
\begin{align}
   S_b^{(4)} &= \Tr \hat\Phi^4
   \nonumber\\
   &= C_b^{102} \Tr [ (e^1 e^2 e^3 e^4 e^5 e^6 e^7 e^8)^2 e^1 e^2 e^3 e^4 e^5 e^6 e^1 e^2 e^3 e^4 e^5 e^6 e^7 e^8 
   \nonumber\\
   &\qquad\qquad\qquad\qquad \times  e^1 e^2 e^3 e^4 e^1 e^2 e^3 e^1 e^2 e^1 e^2 e^3 e^4 e^5 e^6 (-4) e^1 e^2 e^3 e^4 e^5 (-1)(2) e^4 ]
   \nonumber\\
   &= C_b^{102} \int \left(
   \prod_{i=1}^3 dx_i \Phi_b(x_i)
   \right)
   \;
   e^{\pi i ( 3 x_1^2 + x_2^2 + 4 x_1 x_2 - 4 x_1 x_3 )}
\end{align}
The integral is the partition function of an $\CN=2$ $U(1)^3$ gauge theory coupled to three chiral multiplets with
\begin{align}
    K = \left(
    \begin{array}{ccc}
        4 & 2 & -2 \\
        2 & 2 & 0 \\
        -2 & 0 & 1
    \end{array}
    \right)
    \;\;,\;\;
    Q = \left(
    \begin{array}{ccc}
        1 & 0 & 0 \\
        0 & 1 & 0 \\
        0 & 0 & 1
    \end{array}
    \right)\ .
\end{align}
In this description, we have two gauge invariant half-BPS monopole operators
\begin{align}
    \phi_1^2 V_{(0,0,1)},\quad \phi_2^2 V_{(-1,0,-2)}\ .
    \label{eq: A8^4 monopole}
\end{align}
Deforming by the monopole superpotentials, we are left with
\be
A = U(1)_{\text{top}}^{(2)}\ .
\ee

\paragraph{Superconformal index}
Performing the maximization, the superconformal index at the fixed point reads
\begin{align}
    I_{\text{SCI}} = 1 - q - \left(\eta+\frac{1}{\eta}\right) q^{3/2} - 2 q^2 + \left(1+\eta^2 + \frac{1}{\eta^2}\right) q^3 + 2 \left(\eta+\frac{1}{\eta}\right)q^{7/2} + \left(3+\eta^2\right)q^4 + \cdots\ ,
\end{align}
which is strong evidence that the theory flows to an $\CN=4$ rank-zero fixed point.

\paragraph{Modular data ($\nu=-1$ twist)}
The modular data extracted from the partition function are
\begin{align}
    \left\{ | S_{0\a} | \right\} &= 
    \frac{2}{\sqrt{11}}
    \times
    \left\{
    \cos(5\pi/22),
    \cos(\pi/22),
    \sin(2\pi/11),
    \sin(\pi/11),
    \cos(3\pi/22)
    \right\}
    \nonumber\\
    \left\{T_{\a\a}\right\} &= 
    \left\{ 
    1,
    e^{2\pi i (\frac{6}{11})},
    e^{2\pi i (\frac{5}{11})},
    e^{2\pi i (\frac{8}{11})},
    e^{2\pi i (\frac{4}{11})}
    \right\}
\end{align}
They are complex conjugates of the modular data of $\nu=1$ twist in the $M=2$ case.

\paragraph{Modular data ($\nu=1$ twist)}
The modular data extracted from the partition function are
\begin{align}
    \left\{ | S_{0\a} | \right\} &= 
    \frac{2}{\sqrt{11}}
    \times
    \left\{
    \cos(3\pi/22),
    \cos(5\pi/22),
    \sin(\pi/11),
    \cos(\pi/22),
    \sin(2\pi/11)
    \right\}
    \nonumber\\
    \left\{T_{\a\a}\right\} &= 
    \left\{ 
    1,
    e^{2\pi i (\frac{8}{11})},
    e^{2\pi i (\frac{3}{11})},
    e^{2\pi i (\frac{7}{11})},
    e^{2\pi i (\frac{9}{11})}
    \right\}
\end{align}
They are complex conjugates of the modular data of $\nu=-1$ twist in the $M=2$ case.

\subsubsection{$M=5$}
The trace of the fifth power of the monodromy operator becomes
\begin{align}
   S_b^{(5)} &= \Tr \hat\Phi^5
   \nonumber\\
   &= C_b^{128} \Tr [ e^3 e^4 e^5 e^6 e^7 e^8 (e^1 e^2 e^3 e^4 e^5 e^6 e^7 e^8)^3 e^1 e^2 e^3 e^3 e^4 e^5 e^6 e^2 e^3 e^4 e^5 e^6 e^7 \nonumber\\
   &\qquad\qquad\qquad\qquad
   \times e^3 e^4 e^5 e^6 e^7 e^8 e^1 e^2 e^3 e^4 e^5 e^6 e^7 e^8 (2)(4)(6)(8) e^1 e^2 e^3 e^4 e^5 e^6 e^7 e^8 (2)(4)(6)(8)]
   \nonumber\\
   &= C_b^{128} \int \left(
   \prod_{i=1}^8 dx_i \; e^{-\pi i x_i^2} \Phi_b(x_i)
   \right)
   \;
   e^{4 \pi i ( x_1 x_3 + x_2 x_3 + x_2 x_5 - x_2 x_6 - x_3 x_4 - x_3 x_7 + x_4 x_6 - x_5 x_7 + x_6 x_7 + x_7 x_8 )}\ .
\end{align}
The integral is the partition function of an $\CN=2$ $U(1)^8$ gauge theory coupled to eight chiral multiplets with 
\begin{align}
    K = \left(
    \begin{array}{cccccccc}
        0 & 0 & 2 & 0 & 0 & 0 & 0 & 0 \\
        0 & 0 & 2 & 0 & 2 & -2 & 0 & 0 \\
        2 & 2 & 0 & -2 & 0 & 0 & -2 & 0 \\
        0 & 0 & -2 & 0 & 0 & 2 & 0 & 0 \\
        0 & 2 & 0 & 0 & 0 & 0 & -2 & 0 \\
        0 & -2 & 0 & 2 & 0 & 0 & 2 & 0 \\
        0 & 0 & -2 & 0 & -2 & 2 & 0 & 2 \\
        0 & 0 & 0 & 0 & 0 & 0 & 2 & 0
    \end{array}
    \right)
    \;\;,\;\;
    Q = \left(
    \begin{array}{cccccccc}
        1 & 0 & 0 & 0 & 0 & 0 & 0 & 0 \\
        0 & 1 & 0 & 0 & 0 & 0 & 0 & 0 \\
        0 & 0 & 1 & 0 & 0 & 0 & 0 & 0 \\
        0 & 0 & 0 & 1 & 0 & 0 & 0 & 0 \\
        0 & 0 & 0 & 0 & 1 & 0 & 0 & 0 \\
        0 & 0 & 0 & 0 & 0 & 1 & 0 & 0 \\
        0 & 0 & 0 & 0 & 0 & 0 & 1 & 0 \\
        0 & 0 & 0 & 0 & 0 & 0 & 0 & 1 \\
    \end{array}
    \right)
\end{align}
In this description, we have eight gauge invariant half-BPS monopole operators
\begin{align}
    &\phi_1^2 V_{(0,0,-1,0,0,-1,0,0)},\quad \phi_2^2 V_{(0,0,0,0,-1,0,0,-1)},\quad \phi_3^2 V_{(-1,0,0,0,0,0,0,0)},\quad \phi_4^2 V_{(0,0,0,0,-1,-1,0,0)} \notag\\
    &\phi_5^2 V_{(0,-1,0,-1,0,0,0,0)},\quad \phi_6^2 V_{(-1,0,0,-1,0,0,0,0)},\quad \phi_7^2 V_{(0,0,0,0,0,0,0,-1)},\quad \phi_8^2 V_{(0,-1,0,0,0,0,-1,0)}
    \label{eq: A8^5 monopole}
\end{align}
\paragraph{Superconformal index}
Upon the superpotential deformation by all of the monopole operators above, the superconformal index is
\begin{align}
    I_{\text{SCI}} = 1
\end{align}
which is a strong signal that the theory flows to a unitary topological field theory in the infrared.

\paragraph{Modular data}
The modular data extracted from the partition function are
\begin{align}
    \left\{ | S_{0\a} | \right\} &= 
    \frac{2}{\sqrt{11}}
    \times
    \left\{
    \sin(\pi/11),
    \sin(2\pi/11),
    \cos(5\pi/22),
    \cos(3\pi/22),
    \cos(\pi/22)
    \right\}
    \nonumber\\
    \left\{T_{\a\a}\right\} &= 
    \left\{ 
    1,
    e^{2\pi i (\frac{2}{11})},
    e^{2\pi i (\frac{9}{11})},
    e^{2\pi i (\frac{10}{11})},
    e^{2\pi i (\frac{5}{11})}
    \right\}
\end{align}
The data are compatible with the complex conjugate of the modular data of $(A_1,9)_{\frac12}$ \cite{Rowell:2007dge}
\bea
S &=\frac{2}{\sqrt{11}}\left(
\begin{array}{ccccc}
 \sin \left(\frac{\pi }{11}\right) & \sin \left(\frac{2 \pi }{11}\right) & \cos \left(\frac{5 \pi }{22}\right) & \cos \left(\frac{3 \pi }{22}\right) & \cos \left(\frac{\pi }{22}\right) \\
 \sin \left(\frac{2 \pi }{11}\right) & -\cos \left(\frac{3 \pi }{22}\right) & \cos \left(\frac{\pi }{22}\right) & -\cos \left(\frac{5 \pi }{22}\right) & \sin \left(\frac{\pi }{11}\right) \\
 \cos \left(\frac{5 \pi }{22}\right) & \cos \left(\frac{\pi }{22}\right) & \sin \left(\frac{2 \pi }{11}\right) & -\sin \left(\frac{\pi }{11}\right) & -\cos \left(\frac{3 \pi }{22}\right) \\
 \cos \left(\frac{3 \pi }{22}\right) & -\cos \left(\frac{5 \pi }{22}\right) & -\sin \left(\frac{\pi }{11}\right) & \cos \left(\frac{\pi }{22}\right) & -\sin \left(\frac{2 \pi }{11}\right) \\
 \cos \left(\frac{\pi }{22}\right) & \sin \left(\frac{\pi }{11}\right) & -\cos \left(\frac{3 \pi }{22}\right) & -\sin \left(\frac{2 \pi }{11}\right) & \cos \left(\frac{5 \pi }{22}\right) \\
\end{array}
\right) \ ,\\ 
T &=e^{2\pi i (2/33)}\left(
\begin{array}{ccccc}
 1 & 0 & 0 & 0 & 0 \\
 0 & e^{2\pi i (2/11)} & 0 & 0 & 0 \\
 0 & 0 & e^{2\pi i (9/11)} & 0 & 0 \\
 0 & 0 & 0 & e^{2\pi i (10/11)} & 0 \\
 0 & 0 & 0 & 0 & e^{2\pi i (5/11)} \\
\end{array}
\right) \ .
\eea
See appendix \ref{app:modular data} for the conventions. 
The bulk topological theory is dual to the orientation reversal of the pure CS theory $(SU(2)_{9}\times U(1)_{-2})/\mathbb{Z}^{[1]}_2$.

\subsubsection{$M=11$}\label{sec: A8^11}
Finally, one can check that 
\begin{align}
    \hat\Phi^{11} = C_b^{288} (e^1 e^2 e^3 e^4 e^5 e^6 e^7 e^8)^{18}\ .
\end{align}
In Appendix \ref{app: identity}, we show the operator $(e^1 e^2 e^3 e^4 e^5 e^6 e^7 e^8)^{18}$ is the identity operator in the four-particle Hilbert space
    \begin{equation}
        (e^1 e^2 e^3 e^4 e^5 e^6 e^7 e^8)^{18} = \mathbf{1}.
    \end{equation}
It then implies $\hat\Phi^{11} = C_b^{288}\bf{1}$ as expected.

\newpage

\section{$A_{2N+1}$}
\begin{figure}[!h]
  \centering
  \begin{tikzpicture}
			\node[W] (1) at (0,0){$\gamma_1$};
		 	\node[W] (2) at (2,0) {$\gamma_2$};
            \node[W] (3) at (4,0) {$\gamma_3$};
            \node[W] (4) at (6,0) {$\cdots$};
            \node[W] (5) at (8,0) {$\gamma_{2N}$};
            \node[W] (6) at (10,0) {$\gamma_{2N+1}$};
			\draw[->] (1)--(2);
            \draw[<-] (2)--(3);
            \draw[->] (3)--(4);
            \draw[->] (4)--(5);
            \draw[->] (6)--(5);
  \end{tikzpicture}
\caption{ \label{A2Np1 quiver}
BPS quiver for the $A_{2N+1}$ theory in the canonical chamber}
\end{figure}

Let us now consider the trace formula for the $A_{2N+1}$ theory, whose BPS spectrum in the canonical chamber is described by the quiver in Figure \ref{A2Np1 quiver}. The theory has a rank one Higgs branch, which can be identified with the orbifold $\mathbb{C}^2/\mathbb{Z}_{N+1}$. The following lattice vector
\be
\gamma_f = \gamma_1 - \gamma_3 + \cdots (-1)^N \gamma_{2N+1}
\ee
represents the corresponding $U(1)$ Higgs branch flavor symmetry for $N>1$.
For $N=1$, this symmetry enhances to $SU(2)$. The Coulomb branch is parametrized by the operators of dimensions $1+\frac{l}{N+2}$, where $l=1,\cdots, N$.

In this section, we compute
\be\label{trace formula A2N+1}
S_b^{(M)} = \text{Tr}~\hat\Phi^M\ ,
\ee
for $N<4$ and write down the 3d $\CN=2$ gauge theory descriptions that compute the modular data of corresponding VOAs for each $M$. The monodromy operator $\hat\Phi$ is constructed in the same way as \eqref{eq: A2n monodromy} and we will check that it satisfies the relation 
\be
\hat\Phi^{N+2} = C_b^{4(N+1)(2N+1)}~e^{2\pi i m^2}{\bf 1}\ ,
\ee
on the Hilbert space of the $N$-particle quantum mechanics,
as expected from the dimensions of the Coulomb branch operators.

The twisted compactification of $A_{2N+1}$ with $M=1$ completely lifts the Coulomb branch and generates a 3d $\CN=4$ SCFTs with a rank one Higgs branch. The topological $A$-twist of the 3d SCFT is expected to support the logarithmic $\CB_{N+2}$ algebra \cite{Creutzig:2017qyf,Auger:2019gts} with central charge $c_N = 2-6(N+1)^2/(N+2)$.

Using the wall-crossing identities, we simplify the trace formula \eqref{trace formula A2N+1} and identify various 3d $\CN=2$ UV gauge theories related by basic 3d dualities. For $M=1$, while these theories are expected to flow in the IR to 3d $\CN=4$ SCFTs with non-trivial Higgs branches, sufficient wall-crossing manipulations instead yield a gauge theory that flows directly to a unitary topological field theory. A similar phenomenon is observed in \cite{ArabiArdehali:2024vli}, where it is explained in detail that these unitary TFTs can naturally arise from deforming the original $\CN=4$ SCFTs by moment map operators associated with the Higgs branch symmetries. The two 3d topological theories related by such a deformation have identical supersymmetric partition functions on Seifert manifolds.

As a consistency check, we compute the partition functions of the unitary topological theories obtained in this way and show that they are indeed compatible with the modular data of the logarithmic VOAs that appear in the original SCFT/VOA correspondence for this model. 

For an integer $M$ co-prime to $N+2$, we again find that the modular data obtained from the corresponding topological field theory are related to those for $M=1$ by Galois transformations. For $M$ not co-prime to $N+2$, the corresponding 3d gauge theory has runaway vacua with ill-defined partition functions and indices.

\subsection{$A_3$}\label{sec: A3}
We start from the simplest example with the trace formula,
\begin{align}
    S_b^{(M)} = \Tr \hat\Phi^M = \Tr [ (1)(3)(2)(-1)(-3)(-2) ]^M\ .
\end{align}
In section \ref{sec: A3^3}, we show that
\begin{align}
    \hat\Phi^3 = C_b^{24} e^{2\pi i m^2}\, \bf{1}\ ,
\end{align}
which implies that the theories extracted from $S_b^{(M)}$ and $S_b^{(3-M)}$ are related by the orientation reversal up to background couplings.  Therefore, we will only consider $M=1$ below.

\subsubsection{$M=1$}
This example is discussed in \cite{Gaiotto:2024ioj}. We can write 
\begin{align}
    S_b^{(1)} &= \Tr \hat\Phi = C_b^8\, \Tr e^1 e^2 e^3 e^3 
    \nonumber\\
    &= C_b^8\,
    \int dx~ e^{\pi i ( 3 x^2 + 4 m x + 2 m^2 )}
    \nonumber\\
    &= \frac{1}{\sqrt{3}} C_b^8 \, e^{\frac{2\pi i m^2}{3}}
\end{align}
where $m$ is a fugacity for the $SU(2)$ flavor symmetry. While the last line implies that the bulk theory behaves in some ways as a pure $SU(2)$ CS theory at level $4/3$, we can also consider the expression in the next to the last line to extract a well-defined gauge theory description. This expression can be considered as a $U(1)\times U(1)_f$ pure CS theory at integer levels
\begin{align}
    K = \left(
    \begin{array}{c|c}
        3 & 2 \\
        \hline
        2 & 2
    \end{array}
    \right)
\end{align}
where we distinguish the levels associated with the dynamical $U(1)$ and the background $U(1)_f$ by horizontal and vertical lines in the matrix. We claim the unitary topological field theory is infrared dual to the theory obtained by deforming the original 3d $\CN=4$ theory with rank-one Higgs branch by the moment map operators of the flavor symmetry. \cite{ArabiArdehali:2024vli}

\paragraph{Modular data}
The modular data extracted from the supersymmetric partition function of the $U(1)\times U(1)_f$ CS theory read
\begin{align}
    |S_{00}| = |S_{01}| = |S_{02}| = \frac{1}{\sqrt{3}}
    \;\;,\;\;
    \left\{T_{\a\a}\right\} = 
    \left\{
    1
    ,
    e^{2\pi i (\frac{2}{3})}
    ,
    e^{2\pi i (\frac{2}{3})}
    \right\}\ .
\end{align}
They are compatible with
\be
S =\frac{1}{\sqrt{3}}\left(
\begin{array}{ccc}
 -1 & 1 & -1 \\
 1 & e^{2\pi i (1/6)} & e^{2\pi i (1/3)} \\
 -1 & e^{2\pi i (1/3)} & e^{2\pi i (1/6)} \\
\end{array}
\right) \ , \quad T =e^{2\pi i (1/4)}\left(
    \begin{array}{ccc}
        1 & 0 & 0 \\
        0 & e^{2\pi i (2/3)} & 0 \\
        0 & 0 & e^{2\pi i (2/3)}
    \end{array}
    \right) \ ,
\ee
which are the modular data of a logarithmic VOA, the affine $\rm{su}(2)$ at level $-4/3$ with central charge $c=-6$.

\subsubsection{$M=3$} \label{sec: A3^3}
The third power of the monodromy operator simplifies to
\begin{align}
    \hat\Phi^3 = 
    C_b^{24}\,(e^1 e^2 e^3)^4\ .
\end{align}
In the appendix \ref{app: identity}, we show the operator $(e^1 e^2 e^3)^4$ is the identity operator in the single-particle Hilbert space
    \begin{equation}
        (e^1 e^2 e^3)^4 = e^{2\pi i m^2} \mathbf{1},
    \end{equation}
where we keep track of the dependence of $m$ by a direct computation. It then implies $\hat\Phi^3 = C_b^{24} e^{2\pi i m^2}\, \bf{1}$ as expected.

\subsection{$A_5$}
The next example is the $A_5$ theory with the trace formula
\begin{align}
    S_b^{(M)} = \Tr \hat\Phi^M = \Tr [(1)(3)(5)(2)(4)(-1)(-3)(-5)(-2)(-4)]^M\ .
\end{align}
We show in section \ref{sec: A5^4} that
\begin{align}
    \hat\Phi^4 = C_b^{60}\,(e^1 e^2 e^3 e^4 e^5)^6
    =
    C_b^{60}e^{2\pi i m^2}\,\bf{1}\ ,
\end{align}
which implies that the theories extracted from $S_b^{(M)}$ and $S_b^{(4-M)}$ are related by the orientation reversal up to background couplings, which allows us to consider $M=1$ and $2$.

\subsubsection{$M=1$}
Consider the trace of the first power of the monodromy operator,
\begin{align}
    S_b^{(1)} &= \Tr \hat\Phi
    = C_b^{14}\, \Tr e^1 e^2 e^3 e^4 e^5 e^4 e^3 (4)
    \nonumber\\
    &= C_b^{14} \int dx_1 dx_2 dx_3\,
    e^{\pi i ( -x_1^2 + 2 x_3^2 + m^2 + 2 x_1 x_3 + 4 x_1 x_2 + 2 m x_3 )}
    \Phi_b(x_1)
    \nonumber\\
    &= \frac{1}{\sqrt{8}} C_b^{15} e^{\frac{\pi i m^2}{2}}
\end{align}
We can consider the integral expression in the second line, which can be interpreted as an $\CN=2$ $U(1)^3\times U(1)_f$ gauge theory coupled to a single chiral multiplet with 
\begin{align}
    K = \left(
    \begin{array}{ccc|c}
        0 & 2 & 1 & 0 \\
        2 & 0 & 0 & 0 \\
        1 & 0 & 2 & 1 \\
        \hline
        0 & 0 & 1 & 1
    \end{array}
    \right)
    \;\;,\;\;
    Q = \left(
    \begin{array}{c}
        1 \\
        0 \\
        0 \\
        \hline
        0
    \end{array}
    \right)
\end{align}
This description admits one half-BPS monopole operator
\be
(\phi_1)^2V_{(0,-1,0)}\ .
\ee
Turning on the monopole superpotential, all the flavor symmetries in the theory are lifted, and the theory is expected to flow to a unitary TFT.

\paragraph{Modular data}
The modular data extracted from the supersymmetric partition function are \footnote{Computation of the modular data via the Bethe vacua formalism is numerically very subtle in this example. To obtain the result below, one has to slightly turn on the fugacity for $U(1)_\text{top}^{(2)}$ and send it back to 1 at the end of the calculation.}
\begin{align} \label{Modular data of (A1,A5)}
    &| S_{0,\a} | = \frac{1}{\sqrt{8}}\;\;\text{for}\;\;
    \a=0,\cdots,7
    \nonumber\\
    &\left\{T_{\a\a} \right\} = \left\{
    1,
    1,
    e^{2\pi i(\frac{1}{4})},
    e^{2\pi i(\frac{1}{4})},
    e^{2\pi i(\frac{3}{4})},
    e^{2\pi i(\frac{3}{4})},
    e^{2\pi i(\frac{3}{4})},
    e^{2\pi i(\frac{3}{4})}
    \right\}
\end{align}

\subsubsection{$M=2$}
Consider the trace of the second power of the monodromy operator
\begin{align}
    S_b^{(2)} &= \Tr \hat\Phi^2  = C_b^{30} \Tr e^1 e^2 e^3 e^4 e^5 e^1 e^2 e^3 e^4 e^1 e^2 e^3 e^1 e^2 e^1
    \nonumber\\
    &= \frac{1}{3} C_b^{30} \int dx_1 dx_2 dx_3
    \,e^{\pi i ( x_1^2 + x_2^2 + x_3^2 + m^2 + 2 x_1 x_2 + 2 x_2 x_3 + 2 x_1 x_3 )}
\end{align}
Note that this is an example where $\text{gcd}(M,N+2)\neq 1$. The integral in the last line can be thought of as a partition function of an $U(1)^3 \times U(1)_f$ pure CS theory with the level matrix
\begin{align}
    K = \left(
    \begin{array}{ccc|c}
        1 & 1 & 1 & 0 \\
        1 & 1 & 1 & 0 \\
        1 & 1 & 1 & 0 \\
        \hline
        0 & 0 & 0 & 1
    \end{array}
    \right)\ .
\end{align}
This theory has runaway vacua and the supersymmetric indices and partition functions are ill-defined.
Indeed, we can check that a change of integration variables leads to the expression
\begin{align}
    S_b^{(2)} = \frac{e^{\pi i m^2}}{3} C_b^{30} \int dx_1 dx_2 dx_3 \,
    e^{\pi i x_3^2}\ .
\end{align}

\subsubsection{$M=4$}\label{sec: A5^4}
Finally, let us consider the fourth power of the monodromy operator
\begin{align}
    \hat\Phi^4 = C_b^{60}\, (e^1 e^2 e^3 e^4 e^5)^6\ .
\end{align}
In the appendix \ref{app: identity} we show the operator $ (e^1 e^2 e^3 e^4 e^5)^6$ is the identity operator in the Hilbert space of a two-particle quantum mechanics
    \begin{equation}
         (e^1 e^2 e^3 e^4 e^5)^6 = e^{2\pi i m^2} \mathbf{1},
    \end{equation}
where we keep track of the dependence of $m$ by a direct computation. It then implies $\hat\Phi^4 = C_b^{60} e^{2\pi i m^2}\bf{1}$ as expected.

\subsection{$A_7$}
The last example in this section is the $A_7$ theory with the trace formula
\begin{align}
    S_b^{(M)} = \Tr \hat\Phi^M
    = \Tr [ (1)(3)(5)(7)(2)(4)(6)(-1)(-3)(-5)(-7)(-2)(-4)(-6) ]^M\ .
\end{align}
We show in section \ref{sec: A7^5} that
\begin{align}
    \hat\Phi^5 = C_b^{112}e^{2\pi i m^2}\,\bf{1}\ ,
\end{align}
which implies that the theories obtained by $S_b^{(M)}$ and $S_b^{(5-M)}$ are related by the orientation reversal. so we only consider $M=1,2$ below.

\subsubsection{$M=1$}
Consider the trace of the first power of the monodromy operator, which simplifies to
\begin{align}
    S_b^{(1)} &= C_b^{20} \Tr \hat\Phi = \Tr e^1 e^2 e^3 e^4 e^5 e^6 e^7 (-2) e^4 e^3 e^2 (4)
    \nonumber\\
    &= C_b^{20} \int dx_1 dx_2 dx_3 dx_4\,
    e^{\pi i ( 3x_1^2 + 2 x_3^2 + 3 x_4^2 + m^2 - 4 x_1 x_2 - 2 x_1 x_3 + 6 x_1 x_4 - 2 x_3 x_4 + 2 m x_3 )}
    \Phi_b(x_1)\Phi_b(x_2)\ .
\end{align}
The integral is the partition function of an $\CN=2$ $U(1)^4\times U(1)_f$ theory coupled to two chiral multiplets with
\begin{align}
    K = \left(
    \begin{array}{cccc|c}
        4 & -2 & -1 & 3 & 0 \\
        -2 & 1 & 0 & 0 & 0 \\
        -1 & 0 & 2 & -1 & 1 \\
        3 & 0 & -1 & 3 & 0 \\
        \hline
        0 & 0 & 1 & 0 & 1
    \end{array}
    \right)
    \;\;,\;\;
    Q = \left(
    \begin{array}{cc}
        1 & 0  \\
        0 & 1  \\
        0 & 0  \\
        0 & 0  \\
        \hline
        0 & 0 
    \end{array}
    \right)\ .
\end{align}
In this description, there are two gauge invariant half-BPS monopole operators
\begin{align}
    \phi_1^2 V_{(0,1,0,0)}
    \;\;,\;\;
    \phi_2^2 V_{(1,0,0,-1)}
    \;\;.
    \label{eq: A7^1 monopole}
\end{align}
Deforming by the two monopole operators lifts all the flavor symmetries.
\paragraph{Superconformal index}
Upon turning on the monopole operators, we find
\begin{align}
    I_{\text{SCI}} = 1
\end{align}
which is a strong signal that the theory flows directly to a topological field theory in the infrared.

\paragraph{Modular data}
The modular data extracted from the supersymmetric partition function are
\begin{align}
    &\qquad\qquad\qquad | S_{00} | = | S_{01} | = | S_{02} | = | S_{03} | = | S_{04} | = \frac{2}{5}\sin\left(\frac{\pi}{5}\right)\;,
    \nonumber\\
    &\qquad\qquad\qquad | S_{05} | = | S_{06} | = | S_{07} | = | S_{08} | = | S_{09} | = \frac{2}{5}\sin\left(\frac{2\pi}{5}\right)\;,
    \nonumber\\
    \left\{T_{\a\a}\right\} & =  \left\{
    1,
    e^{2\pi i (\frac{4}{5})},
    e^{2\pi i (\frac{4}{5})},
    e^{2\pi i (\frac{1}{5})},
    e^{2\pi i (\frac{1}{5})},
    e^{2\pi i (\frac{2}{5})},
    e^{2\pi i (\frac{3}{5})},
    e^{2\pi i (\frac{3}{5})},
    e^{2\pi i (\frac{1}{5})},
    e^{2\pi i (\frac{1}{5})}
    \right\}
\end{align}

\subsubsection{$M=2$}
Consider the trace of the second power of the monodromy operator
\begin{align}
    S_b^{(2)} &= \Tr \hat\Phi^2 = C_b^{44}\,\Tr[ (e^1 e^2 e^3 e^4 e^5 e^6 e^7)^2 (1) e^1 e^2 e^3 e^4 e^5 e^6 e^1 e^2 ]
    \nonumber\\
    & = C_b^{44} \int dx_1 dx_2\,
    e^{\pi i ( x_1^2 + 5 x_2^2 + m^2 + 2 m x_2 )} \Phi_b(x_1)\ ,
\end{align}
where the integral is the partition function of an $\CN=2$ $U(1)^2 \times U(1)_f$ theory coupled to a single chiral multiplet with
\begin{align}
    K = \left(
    \begin{array}{cc|c}
        2 & 0 & 0 \\ 
        0 & 5 & 1 \\
        \hline
        0 & 1 & 1
    \end{array}
    \right)
    \;\;,\;\;
    Q = \left(
    \begin{array}{c}
        1 \\
        0 \\
        \hline
        0 
    \end{array}
    \right)\ .
\end{align}
This theory is a direct product of two $\CN=2$ gauge theories, from which we claim that the IR theory is a direct product of $\CT_{\text{min}}$ and a pure CS theory $\left(U(1)\times U(1)_f\right)_{K'}$ with the level matrix $K'$ given by the last $2\times 2$ block in $K$. The latter TFT factor will appear in section \ref{D5 M1} as a theory that captures the modular data of $D_5$ theory.
\paragraph{Superconformal index}
The superconformal index of the theory reads
\begin{align}
    I_{\text{SCI}} = 
    1 - q - \left(\eta+\frac{1}{\eta}\right) q^{3/2} - 2 q^2 - \left(\eta + \frac{1}{\eta}\right)q^{5/2} - 2 q^3- \left(\eta + \frac{1}{\eta}\right)q^{7/2} - 2 q^4 +\cdots
\end{align}
which coincides with that of $\CT_{\text{min}}$, in agreement with the claim that the theory is a direct product of $\CT_{\text{min}}$ and a topological theory.

\paragraph{Modular data ($\nu=-1$ twist)}
The modular data extracted from the supersymmetric partition function for $\nu=-1$ twist is
\begin{align}
    &\qquad\qquad\qquad | S_{00} | = | S_{01} | = | S_{02} | = | S_{03} | = | S_{04} | = \frac{2}{5}\sin\left(\frac{2\pi}{5}\right)
    \;,
    \nonumber\\
    &\qquad\qquad\qquad | S_{05} | = | S_{06} | = | S_{07} | = | S_{08} | = | S_{09} | = \frac{2}{5}\sin\left(\frac{\pi}{5}\right)\;,
    \nonumber\\
    \left\{T_{\a\a}\right\} &= \left\{
    1,
    e^{2\pi i (\frac{3}{5})},
    e^{2\pi i (\frac{3}{5})},
    e^{2\pi i (\frac{2}{5})},
    e^{2\pi i (\frac{2}{5})},
    e^{2\pi i (\frac{4}{5})},
    e^{2\pi i (\frac{1}{5})},
    e^{2\pi i (\frac{1}{5})},
    e^{2\pi i (\frac{2}{5})},
    e^{2\pi i (\frac{2}{5})}
    \right\}
\end{align}

\paragraph{Modular data ($\nu=1$ twist)}
The modular data extracted from the supersymmetric partition function for $\nu=1$ twist is
\begin{align}
    &\qquad\qquad\qquad | S_{00} | = | S_{01} | = | S_{02} | = | S_{03} | = | S_{04} | = \frac{2}{5}\sin\left(\frac{2\pi}{5}\right)
    \;,
    \nonumber\\
    &\qquad\qquad\qquad | S_{05} | = | S_{06} | = | S_{07} | = | S_{08} | = | S_{09} | = \frac{2}{5}\sin\left(\frac{\pi}{5}\right)\;,
    \nonumber\\
    \left\{T_{\a\a}\right\} &= \left\{
    1,
    e^{2\pi i (\frac{2}{5})},
    e^{2\pi i (\frac{2}{5})},
    e^{2\pi i (\frac{3}{5})},
    e^{2\pi i (\frac{3}{5})},
    e^{2\pi i (\frac{1}{5})},
    e^{2\pi i (\frac{4}{5})},
    e^{2\pi i (\frac{4}{5})},
    e^{2\pi i (\frac{3}{5})},
    e^{2\pi i (\frac{3}{5})}
    \right\}
\end{align}

\subsubsection{$M=5$}\label{sec: A7^5}
Finally, we compute the fifth power of the monodromy operator
\begin{align}
    \hat\Phi^5 = C_b^{112}\,(e^1 e^2 e^3 e^4 e^5 e^6 e^7)^8\ .
\end{align}
In the appendix \ref{app: identity}, we show the operator $(e^1 e^2 e^3 e^4 e^5 e^6 e^7)^8$ is the identity operator in the three-particle Hilbert space
    \begin{equation}
        (e^1 e^2 e^3 e^4 e^5 e^6 e^7)^8 = e^{2\pi i m^2} \mathbf{1},
    \end{equation}
where we keep track of the dependence of $m$ by a direct computation. It then implies  $\hat\Phi^5 = C_b^{112}e^{2\pi i m^2}\, \bf{1}$  as expected.

\newpage

\section{$D_{2N}$}
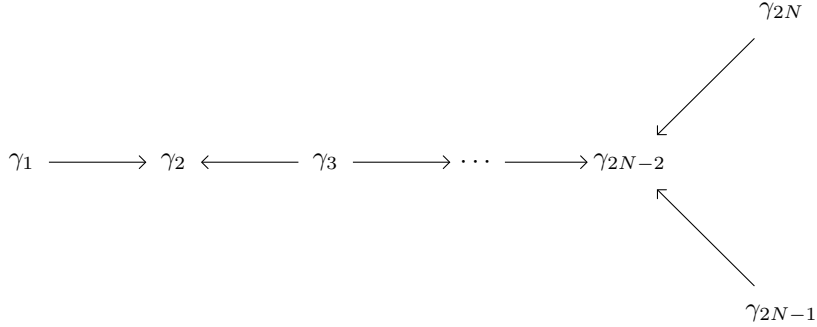
\begin{figure}[!h]
  \centering
  \begin{tikzpicture}
			\node[W] (1) at (0,0){$\gamma_1$};
		 	\node[W] (2) at (2,0) {$\gamma_2$};
            \node[W] (3) at (4,0) {$\gamma_3$};
            \node[W] (4) at (6,0) {$\cdots$};
            \node[W] (5) at (8,0) {$\gamma_{2N-2}$};
            \node[W] (6) at (10,2) {$\gamma_{2N}$};
            \node[W] (7) at (10,-2) {$\gamma_{2N-1}$};            
			\draw[->] (1)--(2);
            \draw[<-] (2)--(3);
            \draw[->] (3)--(4);
            \draw[->] (4)--(5);
            \draw[<-] (5)--(7);
            \draw[<-] (5)--(6);
  \end{tikzpicture}
\caption{ \label{D2N quiver}
BPS quiver for the $D_{2N}$ theory in the canonical chamber}
\end{figure}

Next, we consider the trace formula for the $D_{2N}$ theory for $N>1$, whose BPS spectrum in the canonical chamber is described by the quiver in Figure \ref{D2N quiver}. The theory has $SU(2)\times U(1)$ flavor symmetry, which corresponds to the vectors
\be
\gamma_{f_1} = \gamma_1-\gamma_3 + \cdots + (-1)^{N-1}\gamma_{2N-1}\ ,\quad  \gamma_{f_2} = \gamma_{2N} - \gamma_{2N-1}\ .
\ee
For $N=2$, the flavor symmetry enhances to $SU(3)$. The Coulomb branch is parametrized by the operators of dimensions $1+\frac{l}{N}$ for $l=1,\cdots, N-1$.

In this section, we compute
\be
S_b^{(M)} = \text{Tr}~\hat\Phi^M\ ,
\ee
for $N\leq 4$ and write down the 3d $\CN=2$ gauge theory descriptions that compute the modular data of corresponding VOAs for each $M$. The monodromy operator is written as
\be
\hat\Phi =
\bigg(
\prod_{i=1}^{N}\Phi_b(x_{\gamma_{2i-1}})
\bigg)
\Phi_b(x_{\gamma_{2N}})
\prod_{i=1}^{N-1}\Phi_b(x_{\gamma_{2i}})
\bigg(
\prod_{i=1}^{N}\Phi_b(-x_{\gamma_{2i-1}})
\bigg)
\Phi_b(-x_{\gamma_{2N}})
\prod_{i=1}^{N-1}\Phi_b(-x_{\gamma_{2i}})\ .
\label{eq: D2n monodromy}
\ee
We will check that this satisfies the relation
\be
\hat\Phi^{N}=C_b^{4N(2N-1)}f(m,m_1,m_2){\bf 1}\ ,
\ee
where the prefactor $f(m,m_1,m_2)$ is $e^{2\pi i m^2 + (2N-1)\pi i (m_1^2 + m_2^2)}$ for $N>2$.\footnote{Here $m,m_1$ and $m_2$ are normalized so that $m+(-1)^{N-1}m_1$ is the fugacity that corresponds to $\gamma_{f_1}$. And $m_1$, $m_2$ are the $SU(2)$ fugacity with $m_1+m_2=0$, which rotates the last two nodes.} As in the previous section, the twisted compactification of $D_{2N}$ theory with $M=1$ completely lifts the Coulomb branch and generates a 3d $\CN=4$ SCFTs with a two-dimensional Higgs branch. The topological $A$-twist of the SCFT is expected to support the logarithmic $\CW_{N}$ algebra \cite{Creutzig:2017qyf} with central charge $c_N = 4-6N$.

We again find that, after sufficient wall-crossing manipulations of the trace formula, we instead obtain a gauge theory description that flows directly to a unitary topological field theory. We expect that these unitary topological theories are related to the original 3d SCFTs by moment map deformations as discussed in the previous section, and therefore have identical supersymmetric partition functions on Seifert manifolds. For all examples in this section, we find that the trace formula after enough wall-crossing manipulations gives rise to a pure CS description of the TFT.

\subsection{$D_4$}
The first example in this section is the $D_4$ theory with the trace formula
\begin{align}
    S_b^{(M)} = \Tr \hat\Phi^M = \Tr [(1)(3)(4)(2)(-1)(-3)(-4)(-2)]^M\ .
\end{align}
We show in section \ref{sec: D4^2} that
\begin{align}
    \hat\Phi^2 = C_b^{24}e^{3\pi i (m_1^2 + m_2^2 + m_3^2)}\,\bf{1}\ .
\end{align}
Therefore we only consider $M=1$. 
\subsubsection{$M=1$}
We can simplify the trace formula to
\begin{align}
    S_b^{(1)} & = \Tr \hat\Phi = C_b^{12}\, \Tr[ e^1 e^2 e^3 e^4 e^5 e^6 e^2 e^3 ]
    \nonumber\\
    & = C_b^{12}\int dx_1 dx_2\,
    e^{\pi i ( 2m_1^2 + 3 m_2^2 + 4 m_1 m_2 + 2 m_1 x_1 - 2 m_2 x_1 + 2 m_1 x_2 - 4 x_1 x_2 )}
    \nonumber\\
    & = \frac{1}{2}C_b^{12} e^{\frac{3\pi i}{2}(m_1^2 + m_2^2 + m_3^2)}\ ,
\end{align}
where $m_1$, $m_2$, $m_3$ are the fugacities for the $SU(3)$ flavor symmetry satisfying $m_1+m_2+m_3=0$. The last expression implies that the theory behaves in some ways as a pure $SU(3)$ CS theory at level $\frac{3}{2}$. The integral in the second line gives rise to a well-defined $U(1)^2\times U(1)_{f_1} \times U(1)_{f_2}$ pure CS theory with
\begin{align}
    K = \left(
    \begin{array}{cc|cc}
        0 & -2 & 1 & -1 \\
        -2 & 0 & 1 & 0 \\
        \hline
        1 & 1 & 2 & 2 \\
        -1 & 0 & 2 & 3
    \end{array}
    \right)\ .
\end{align}
\paragraph{Modular data}
The modular data extracted from the partition functions are
\begin{align}
    |S_{00}| = |S_{01}| = |S_{02}| = |S_{03}| = \frac{1}{2}
    \;\;,\;\;
    \left\{T_{\a\a} \right\} = \left\{ 1 , e^{2\pi i (\frac{1}{2})} , e^{2\pi i (\frac{1}{2})} , e^{2\pi i (\frac{1}{2})} \right\}\ .
\end{align}
They are compatible with the modular data of affine $\mathrm{su}(3)$ at level $-3/2$ \cite{DiFrancesco:1997nk}
\begin{equation}
    S= \left(
\begin{array}{cccc}
 -\frac{1}{2} & -\frac{1}{2} & -\frac{1}{2} & \frac{1}{2} \\
 -\frac{1}{2} & -\frac{1}{2} & \frac{1}{2} & -\frac{1}{2} \\
 -\frac{1}{2} & \frac{1}{2} & -\frac{1}{2} & -\frac{1}{2} \\
 \frac{1}{2} & -\frac{1}{2} & -\frac{1}{2} & -\frac{1}{2} \\
\end{array}
\right),\quad T=\left( \begin{array}{cccc}
     e^{2\pi i(2/6)}&0&0&0\\
     0&e^{2\pi i (5/6)}&0&0\\
     0&0&e^{2\pi i (5/6)}&0\\
     0&0&0&e^{2\pi i (5/6)}
\end{array}\right).
\end{equation}
with central charge $c=-8$.

\subsubsection{$M=2$}\label{sec: D4^2}
The second power gives
\begin{align}
    \hat\Phi^2 = C_b^{24}\,(e^1 e^2 e^3 e^4)^3\ ,
\end{align}
In the appendix \ref{app: identity}, we show the operator $(e^1 e^2 e^3 e^4)^3$ is the identity operator in the single-particle Hilbert space
    \begin{equation}
        (e^1 e^2 e^3 e^4)^3 = e^{3\pi i (m_1^2 + m_2^2 + m_3^2)} \mathbf{1} ,
    \end{equation}
where we keep track of the dependence of $m_1,m_2,m_3$ by a direct computation. It then implies $\hat\Phi^2 = C_b^{24}e^{3\pi i(m_1^2 + m_2^2 + m_3^2)} \,\bf{1}$ as expected.

\subsection{$D_6$}
Let us move on to the $D_6$ theory with the trace formula
\begin{align}
    S_b^{(M)} = \Tr \hat\Phi^M = \Tr [(1)(3)(5)(2)(6)(4)(-1)(-3)(-5)(-2)(-6)(-4)]^M\ .
\end{align}
In section \ref{sec: D6^3} we show that 
\begin{align}
    \hat\Phi^3 = C_b^{60}e^{\pi i (2m^2 + 5m_1^2 + 5m_2^2)}\, \bf{1}\ ,
\end{align}
that implies that the modular data extracted from $S_b^{(M)}$ and $S_b^{(3-M)}$ are related by the complex conjugation. We will only consider $M=1$ below.

\subsubsection{$M=1$}
Consider the trace of the first power of the monodromy operator
\begin{align}
    S_b^{(1)} &= \Tr \hat\Phi = C_b^{20}\, \Tr[e^1 e^2 e^3 e^4 e^5 e^6 e^4 e^3 e^4 e^5]
    \nonumber\\
    &= C_b^{20} \int dx_1 dx_2\,
    e^{\pi i ( - 3 x_1^2 + m^2 + 3 m_1^2 - 6 x_1 x_2 - 2 m_1 x_1 - 2 m_1 x_2 )}
    \nonumber\\
    & =\frac{1}{3} C_b^{20} \, e^{\frac{2\pi i}{3} m^2 + \frac{5 \pi i}{3} (m_1^2 + m_2^2)}
\end{align}
where $m$, $m_1$ and $m_2$ are the fugacities for the $U(1)$ and $SU(2)$ flavor symmetries with $m_1+m_2=0$. Let us consider the integral in the second line, which can be thought of as an $U(1)^2\times U(1)_f\times U(1)_{f_1}$ pure CS theory with
\begin{align}
    K = \left(\begin{array}{cc|cc}
        -3 & -3 & 0 & -1 \\
        -3 & 0 & 0 & -1 \\
        \hline
        0 & 0 & 1 & 0 \\
        -1 & -1 & 0 & 3
    \end{array}
    \right)\ .
\end{align}

\paragraph{Modular data} 
The modular data extracted from the partition functions of the pure CS theory are
\begin{align}
    &\qquad\qquad |S_{0\a}| = \frac{1}{3}
    \;\;,\;\;\text{for}\;\;\a=0,1,\cdots,8
    \nonumber\\
    &\left\{T_{\a\a} \right\} =\left\{
    1,1,1,1,1,
    e^{2\pi i (\frac{1}{3})},
    e^{2\pi i (\frac{1}{3})},
    e^{2\pi i (\frac{2}{3})},
    e^{2\pi i (\frac{2}{3})}
    \right\}
\end{align}

\subsubsection{$M=3$}\label{sec: D6^3}
Now we consider the third power of the operator
\begin{align}
    \hat\Phi^3 = C_b^{60} \, (e^1 e^2 e^3 e^4 e^5 e^6)^5\ ,
\end{align}
In the appendix \ref{app: identity}, we show the operator $ (e^1 e^2 e^3 e^4 e^5 e^6)^5$ is the identity operator in the two-particle Hilbert space
    \begin{equation}
         (e^1 e^2 e^3 e^4 e^5 e^6)^5 = e^{\pi i (2m^2 + 5m_1^2 + 5m_2^2)} \mathbf{1},
    \end{equation}
where we keep track of the dependence of $m_1,m_2$ by a direct computation. It then implies $\hat\Phi^3 = C_b^{60}e^{\pi i (2m^2 + 5m_1^2 + 5m_2^2)}\,\bf{1}$ as expected. 

\subsection{$D_8$}
The last example in this section is the $D_8$ theory with the trace formula
\begin{align}
    S_b^{(M)} = \Tr \hat\Phi^M
    =
    \Tr[(1)(3)(5)(7)(2)(4)(8)(6)(-1)(-3)(-5)(-7)(-2)(-4)(-8)(-6)]^M\ .
\end{align}
We show in section \ref{sec: D8^4} that
\begin{align}
    \hat\Phi^4 = C_b^{112}e^{\pi i (2m^2 + 7m_1^2 + 7m_2^2)}\, \bf{1}
\end{align}
which implies that the modular data extracted from $S_b^{(M)}$ and $S_b^{(4-M)}$ are related by the complex conjugation.

\subsubsection{$M=1$}
Consider the trace of the first power of the monodromy operator
\begin{align}
    S_b^{(1)} &= \Tr \hat\Phi = C_b^{28} \Tr[ e^1 e^2 e^3 e^4 e^5 e^6 e^7 e^8 e^2 e^3 e^4 e^5 e^6 e^7 ]
    \nonumber\\
    &= C_b^{28} \int dx_1 dx_2\,
    e^{\pi i ( -8 x_1^2 + m^2 + 3 m_1^2 - 8 x_1 x_2 + 2 m x_2 + 4 m_1 x_1 + 2 m_1 x_2 )}
    \nonumber\\
    &= \frac{1}{4} C_b^{28}\, e^{\frac{\pi i}{2}m^2 + \frac{7\pi i}{4}(m_1^2 + m_2^2)}
\end{align}

We consider the integral in the second line, which can be thought of as a partition function of an $U(1)\times U(1)_f\times U(1)_{f_1}$ pure CS theory with 
\begin{align}
    K = \left(
    \begin{array}{cc|cc}
        -8 & -4 & 0 & 2 \\
        -4 & 0 & 1 & 1 \\
        \hline
        0 & 1 & 1 & 0 \\
        2 & 1 & 0 & 3
    \end{array}
    \right)
\end{align}
\paragraph{Modular data}
The modular data extracted from the partition functions are
    \begin{equation}
        \begin{gathered}
        |S_{0\alpha}| = \frac{1}{4}
\;\;,\;\;\text{for}\;\;\a=0,1,\cdots,15
    \nonumber\\
    \left\{T_{\a\a}\right\} = \left\{
    1,1,1,1,1,1,1,1,
    e^{2\pi i (\frac{1}{2})},
    e^{2\pi i (\frac{1}{2})},
    e^{2\pi i (\frac{1}{2})},
    e^{2\pi i (\frac{1}{2})},e^{2\pi i (\frac{1}{4})},
    e^{2\pi i (\frac{1}{4})},
    e^{2\pi i (\frac{3}{4})},
    e^{2\pi i (\frac{3}{4})}
    \right\}
    \end{gathered}
    \end{equation}

\subsubsection{$M=2$}
Next, we consider
\begin{align}
    S_b^{(2)} & = \Tr \hat\Phi^2 = C_b^{56}\, \Tr[ (e^1 e^2 e^3 e^4 e^5 e^6 e^7 e^8)^3 e^4 e^5 e^6 e^7 ]
    \nonumber\\
    &=C_b^{56}\int dx_1 dx_2 dx_3 dx_4\,
    e^{\pi i ( x^{T} K x  )}
\end{align}
where
\begin{align}
    K = \left(
    \begin{array}{cccc|cc}
        0 & 0 & 2 & 2 & 1 & -1 \\
        0 & 0 & -2 & -2 & -1 & 1 \\
        2 & -2 & -16 & -10 & -4 & 8 \\
        2 & -2 & -10 & -4 & -1 & 5 \\
        \hline
        1 & -1 & -4 & -1 & 1 & 2 \\
        -1 & 1 & 8 & 5 & 2 & 3
    \end{array}
    \right)\ .
\end{align}
Note that this is an example where gcd$(M,N)\neq 1$. As expected, the 3d gauge theory has runaway vacua, which can be inferred from the relation det$K=0$. The second line can be further simplified to 
\be   S_b^{(2)} = \frac{1}{6} C_b^{56} \int dx_1 dx_2 \,e^{\pi i m^2 + \frac{7\pi i}{2}(m_1^2 + m_2^2)}\ ,
\ee
by a simple change of variables, which makes it clear that the indices and partition functions are ill-defined.

\subsubsection{$M=4$}\label{sec: D8^4}
Finally, we consider the fourth power of the monodromy operator, which simplifies to
\begin{align}
    \hat\Phi^4 = C_b^{112}\,(e^1 e^2 e^3 e^4 e^5 e^6 e^7 e^8)^7\ .
\end{align}
In the appendix \ref{app: identity}, we show the operator $(e^1 e^2 e^3 e^4 e^5 e^6 e^7 e^8)^7$ is the identity operator in the three-particle Hilbert space
    \begin{equation}
         (e^1 e^2 e^3 e^4 e^5 e^6 e^7 e^8)^7 = e^{\pi i (2m^2 + 7m_1^2 + 7m_2^2)} \mathbf{1},
    \end{equation}
where we keep track of the dependence of $m_1,m_2$ by a direct computation. It then implies $\hat\Phi^4 = C_b^{112}e^{\pi i (2m^2 + 7m_1^2 + 7m_2^2)} \,\bf{1}$ as expected.

\newpage

\section{$D_{2N+1}$}

\begin{figure}[h]
  \centering
  \begin{tikzpicture}
			\node[W] (1) at (0,0){$\gamma_1$};
		 	\node[W] (2) at (2,0) {$\gamma_2$};
            \node[W] (3) at (4,0) {$\gamma_3$};
            \node[W] (4) at (6,0) {$\cdots$};
            \node[W] (5) at (8,0) {$\gamma_{2N-1}$};
            \node[W] (6) at (10,2) {$\gamma_{2N+1}$};
            \node[W] (7) at (10,-2) {$\gamma_{2N}$};            
			\draw[->] (1)--(2);
            \draw[<-] (2)--(3);
            \draw[->] (3)--(4);
            \draw[<-] (4)--(5);
            \draw[->] (5)--(7);
            \draw[->] (5)--(6);
  \end{tikzpicture}
\caption{ \label{D2Np1 quiver}
BPS quiver for the $D_{2N+1}$ theory in the canonical chamber}
\end{figure}
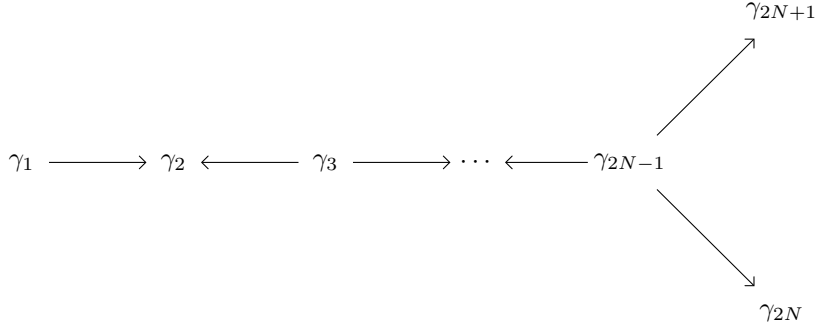

Let us now consider the trace formula for the $D_{2N+1}$ theory, where the BPS spectrum in the canonical chamber is described by the quiver in Figure \ref{D2Np1 quiver}. The theory has a $SU(2)_f$ Higgs branch flavor symmetry, which corresponds to
\be
\gamma_f = \gamma_{2N+1}-\gamma_{2N}\ .
\ee
The Coulomb branch of the theory is parametrized by the operators of dimensions $1+\frac{2l+1}{2N+1}$ for $l=0,\cdots N-1$.

In this section, we consider
\be
S_b^{(M)} = \text{Tr} ~\hat\Phi^M\ ,
\ee
for $M<4$ and write down the gauge theory description that computes the modular data of the corresponding VOAs for each $M$. The monodromy operator is written in the same way as \eqref{eq: A2n monodromy} and we will check that it satisfies the relation
\be
\hat\Phi^{2N+1} = C_b^{8N(2N+1)} e^{2\pi i\, N m^2}{\bf 1}\ ,
\ee
as expected. As in the previous two sections, the twisted compactifications of $D_{2N+1}$ theory with $M=1$ completely lift the Coulomb branch and generate a 3d $\CN=4$ SCFTs with the same Higgs branch. The topological A-twist of the SCFT is expected to support the affine $\rm{su}(2)$ algebra at level $-4N/(2N+1)$ with central charge $c_N=-6N$. Note that $D_3$ is equivalent to $A_3$ discussed in section \ref{sec: A3} so we do not consider it in this section.

We again find that, after sufficient wall-crossing manipulations of the trace formula, we obtain a pure abelian CS theory. We expect that this theory is related to the original 3d SCFT by a moment map deformation as discussed in earlier sections. For all examples in this section, we find that modular data extracted from these abelian CS theories are compatible with that of the affine $\rm{su}(2)$ VOAs at admissible levels. The modular data we obtain for $M>1$ are related to $M=1$ by a Galois transformation.

\subsection{$D_5$}
Let us consider $D_5$ theory with the trace formula
\begin{align}
    S_b^{(M)} = \Tr \hat\Phi^M = \Tr[(1)(3)(5)(2)(4)(-1)(-3)(-5)(-2)(-4)]^M\ .
\end{align}
In section \ref{sec: D5^5} we show that
\begin{align}
    \hat\Phi^5 = C_b^{80}e^{4\pi i m^2}\,\bf{1}\ ,
\end{align}
which implies that the modular data extracted from $S_b^{(M)}$ and $S_b^{(5-M)}$ are related by the complex conjugation. Therefore, we only consider $M=1,2$ below. 

\subsubsection{$M=1$} \label{D5 M1}
We have
\begin{align}
    S_b^{(1)} &= \Tr \hat\Phi = C_b^{16}\, \Tr [ e^1 e^2 e^3 e^4 e^3 e^5 e^2 e^3 ]
    \nonumber\\
    & = C_b^{16} \int dx\, e^{\pi i ( 5 x^2 + 2 m x + m^2 )}
    \nonumber\\
    & = \frac{1}{\sqrt{5}} C_b^{16}\, e^{\frac{4\pi i m^2}{5}}
\end{align}
where $m$ is the fugacity for the $SU(2)$ flavor symmetry. As in the $A_3$ example, the theory behaves in some way as a $SU(2)$ CS theory at level $\frac{8}{5}$. The integral in the second line provides a well-defined gauge theory description in terms of a $U(1)\times U(1)_f$ pure CS theory with the level matrix
\begin{align}
    K = \left(
    \begin{array}{c|c}
        5 & 1 \\
        \hline
        1 & 1
    \end{array}
    \right)\,.
\end{align}

\paragraph{Modular data}
The modular data extracted from the partition function of the abelian CS theory are
\begin{align}
    &\qquad\qquad |S_{0\a}| = \frac{1}{\sqrt{5}}
    \;\;
    \text{for}\;\;
    \a = 0,\cdots,4
    \nonumber\\
    &\left\{T_{\a\a}\right\} = 
    \left\{
    1,
    e^{2\pi i (\frac{3}{5})},
    e^{2\pi i (\frac{2}{5})},
    e^{2\pi i (\frac{2}{5})},
    e^{2\pi i (\frac{3}{5})}
    \right\}
\end{align}
These are compatible with
\bea
S &=\frac{1}{\sqrt{5}}\left(
\begin{array}{ccccc}
 1 & -1 & 1 & -1 & 1 \\
 -1 & e^{2\pi i (8/10)} & e^{2\pi i (1/10)} & e^{2\pi i (4/10)} & e^{2\pi i (7/10)} \\
 1 & e^{2\pi i (1/10)} & e^{2\pi i (2/10)} & e^{2\pi i (3/10)} & e^{2\pi i (4/10)} \\
 -1 & e^{2\pi i (4/10)} & e^{2\pi i (3/10)} & e^{2\pi i (2/10)} & e^{2\pi i (1/10)} \\
 1 & e^{2\pi i (7/10)} & e^{2\pi i (4/10)} & e^{2\pi i (1/10)} & e^{2\pi i (8/10)} \\
\end{array}
\right) \ ,\\ 
T &=\left(
\begin{array}{ccccc}
 e^{2\pi i (5/10)} & 0 & 0 & 0 & 0 \\
 0 & e^{2\pi i (1/10)} & 0 & 0 & 0 \\
 0 & 0 & e^{2\pi i (9/10)} & 0 & 0 \\
 0 & 0 & 0 & e^{2\pi i (9/10)} & 0 \\
 0 & 0 & 0 & 0 & e^{2\pi i (1/10)} \\
\end{array}
\right) \ ,
\eea
which are the modular data of affine $\mathrm{su}(2)$ at level $-8/5$ with central charge $c=-12$.

\subsubsection{$M=2$}
Consider the second power of the monodromy operator
\begin{align}
    S_b^{(2)} &= \Tr \hat\Phi^2 = C_b^{32}\,\Tr [ (e^1 e^2 e^3 e^4 e^5)^3 e^4 ]
    \nonumber\\
    &=C_b^{32}\int dx_1 dx_2\,
    e^{\pi i ( 3 x_1^2 + 7 x_2^2 + 3 m^2 - 8 x_1 x_2 - 4 m x_1 + 6 m x_2 )}
    \nonumber\\
    &= \frac{1}{\sqrt{5}} C_b^{32}\,e^{\frac{8\pi i}{5} m^2}
\end{align}

The integral in the second line can be thought of as the partition function of a $U(1)^2\times U(1)_f$ pure CS theory with
\begin{align}
    K = \left(
    \begin{array}{cc|c}
        3 & -4 & -2 \\
        -4 & 7 & 3 \\
        \hline
        -2 & 3 & 3
    \end{array}
    \right)\,.
\end{align}

\paragraph{Modular data}
The modular data extracted from the abelian CS theory description are
\begin{align}
    &\qquad\qquad |S_{0\a}| = \frac{1}{\sqrt{5}}
    \;\;
    \text{for}\;\;
    \a = 0,\cdots,4
    \nonumber\\
    &\left\{T_{\a\a}\right\} = 
    \left\{
    1,
    e^{2\pi i (\frac{1}{5})},
    e^{2\pi i (\frac{4}{5})},
    e^{2\pi i (\frac{4}{5})},
    e^{2\pi i (\frac{1}{5})}
    \right\}.
\end{align}

\subsubsection{$M=5$}\label{sec: D5^5}
Finally, the fifth power of the monodromy operator simplifies to
\begin{align}
    \hat\Phi^5 = C_b^{80}\,(e^1 e^2 e^3 e^4 e^5)^8\ ,
\end{align}
In the appendix \ref{app: identity}, we show the operator $(e^1 e^2 e^3 e^4 e^5)^8$ is the identity operator in the two-particle Hilbert space
    \begin{equation}
         (e^1 e^2 e^3 e^4 e^5)^8 = e^{4\pi i m^2} \mathbf{1},
    \end{equation}
where we keep track of the dependence of $m$ by a direct computation. It then implies $\hat\Phi^5 = C_b^{80} e^{4\pi i m^2}\,\bf{1}$ as expected.

\subsection{$D_7$}
The next example is the $D_7$ theory with the trace formula
\begin{align}
    S_b^{(M)} = \Tr \hat\Phi^M = \Tr[(1)(3)(5)(7)(2)(4)(6)(-1)(-3)(-5)(-7)(-2)(-4)(-6)]^M\ .
\end{align}
We show in section \ref{sec: D7^7} that
\begin{align}
    \hat\Phi^7 = C_b^{168}e^{6\pi i m^2}\,\bf{1},
\end{align}
which implies that the modular data extracted from $S_b^{(M)}$ and $S_b^{(7-M)}$ are related by the complex conjugation, which allows us to consider $M=1,2$ and $3$.
\subsubsection{$M=1$}
We have
\begin{align}
    S_b^{(1)} & = \Tr \hat\Phi = C_b^{24} \Tr [ e^1 e^2 e^3 e^4 e^5 e^6 e^7 e^2 e^3 e^4 e^5 e^7 ]
    \nonumber\\
    & = C_b^{24} \int dx e^{\pi i ( 7 x^2 + m^2 + 2 m x )}
    \nonumber\\
    & = \frac{1}{\sqrt{7}} C_b^{24}\, e^{\frac{6\pi i}{7} m^2},
\end{align}

The integral in the second line can be thought of as the partition function of the $U(1)\times U(1)_f$ pure CS theory with the CS level matrix
\begin{align}
    K=\left(
    \begin{array}{c|c}
        7 & 1 \\
        \hline
        1 & 1
    \end{array}
    \right)\ .
\end{align}

\paragraph{Modular data}
The modular data extracted from the partition function of the abelian CS theory are
\begin{align}
    &\qquad\qquad\qquad\qquad |S_{0\a}| = \frac{1}{\sqrt{7}}
    \;\;
    \text{for}\;\;
    \a = 0,\cdots,6
    \nonumber\\
    &\left\{T_{\a\a}\right\} = 
    \left\{
    1,
    e^{2\pi i (\frac{4}{7})},
    e^{2\pi i (\frac{2}{7})},
    e^{2\pi i (\frac{1}{7})},
    e^{2\pi i (\frac{1}{7})},
    e^{2\pi i (\frac{2}{7})},
    e^{2\pi i (\frac{4}{7})}
    \right\}\ .
\end{align}
They are compatible with
\bea
S &=\frac{1}{\sqrt{7}}\left(
\begin{array}{ccccccc}
 -1 & 1 & -1 & 1 & -1 & 1 & -1 \\
 
 1 & e^{2\pi i (5/14)} & e^{2\pi i (10/14)} & e^{2\pi i (1/14)} & e^{2\pi i (6/14)} & e^{2\pi i (11/14)} & e^{2\pi i (2/14)} \\
 
 -1 & e^{2\pi i (10/14)} & e^{2\pi i (13/14)} & e^{2\pi i (2/14)} & e^{2\pi i (5/14)} & e^{2\pi i (8/14)} & e^{2\pi i (11/14)} \\
 
 1 & e^{2\pi i (1/14)} & e^{2\pi i (2/14)} & e^{2\pi i (3/14)} & e^{2\pi i (4/14)} & e^{2\pi i (5/14)} & e^{2\pi i (6/14)} \\
 
 -1 & e^{2\pi i (6/14)} & e^{2\pi i (5/14)} & e^{2\pi i (4/14)} & e^{2\pi i (3/14)} & e^{2\pi i (2/14)} & e^{2\pi i (1/14)} \\
 
 1 & e^{2\pi i (11/14)} & e^{2\pi i (8/14)} & e^{2\pi i (5/14)} & e^{2\pi i (2/14)} & e^{2\pi i (13/14)} & e^{2\pi i (10/14)} \\
 
 -1 & e^{2\pi i (2/14)} & e^{2\pi i (11/14)} & e^{2\pi i (6/14)} & e^{2\pi i (1/14)} & e^{2\pi i (10/14)} & e^{2\pi i (5/14)} \\
\end{array}
\right) \ ,\\ 
T &=\left(
\begin{array}{ccccccc}
 e^{2\pi i (21/28)} & 0 & 0 & 0 & 0 & 0 & 0 \\
 0 & e^{2\pi i (9/28)} & 0 & 0 & 0 & 0 & 0 \\
 0 & 0 & e^{2\pi i (1/28)} & 0 & 0 & 0 & 0 \\
 0 & 0 & 0 & e^{2\pi i (25/28)} & 0 & 0 & 0 \\
 0 & 0 & 0 & 0 & e^{2\pi i (25/28)} & 0 & 0 \\
 0 & 0 & 0 & 0 & 0 & e^{2\pi i (1/28)} & 0 \\
 0 & 0 & 0 & 0 & 0 & 0 & e^{2\pi i (9/28)} \\
\end{array}
\right) \ ,
\eea
which are the modular data of affine $\mathrm{su}(2)$ at level $-12/7$ with central charge $c=-18$.

\subsubsection{$M=2$}
Now we consider the second power, which gives
\begin{align}
    S_b^{(2)} & = \Tr \hat\Phi^2 = C_b^{48}\,\Tr[ ( e^1 e^2 e^3 e^4 e^5 e^6 e^7 e^3 e^3 e^4 e^5 e^6 )^2 ]
    \nonumber\\
    & = C_b^{48} \int dx\,
    e^{\pi i ( 7 x^2 + 6 m x + 3 m^2 )}
    \nonumber\\
    &= \frac{1}{\sqrt{7}} C_b^{48}\, e^{\frac{12\pi i}{7} m^2}
\end{align}

The integral in the second line can be thought of as the partition function of an $U(1)\times U(1)_f$ CS theory with
\begin{align}
    K = \left(
    \begin{array}{c|c}
        7 & 3 \\
        \hline
        3 & 3
    \end{array}
    \right)\,.
\end{align}

\paragraph{Modular data}
The modular data extracted from the partition function of the abelian CS theory are
\begin{align}
    &\qquad\qquad\qquad\qquad |S_{0\a}| = \frac{1}{\sqrt{7}}
    \;\;
    \text{for}\;\;
    \a = 0,\cdots,6
    \nonumber\\
    &\left\{T_{\a\a}\right\} = 
    \left\{
    1,
    e^{2\pi i (\frac{1}{7})},
    e^{2\pi i (\frac{4}{7})},
    e^{2\pi i (\frac{2}{7})},
    e^{2\pi i (\frac{2}{7})},
    e^{2\pi i (\frac{4}{7})},
    e^{2\pi i (\frac{1}{7})}
    \right\}
\end{align}
which are compatible with the modular data of affine $\mathrm{su}(2)$ at level $-12/7$ as discussed in the previous example.

\subsubsection{$M=3$}
Consider the third power, which gives
\begin{align}
    S_b^{(3)} & = \Tr \hat\Phi^3 = C_b^{72} \, \Tr [ (e^1 e^2 e^3 e^4 e^5 e^6 e^7)^2 e^6 ]
    \nonumber\\
    & = C_b^{72} \int dx_1 dx_2 dx_3 e^{\pi i ( 2 x_1^2 + 7 x_2^2 + 11 x_3^2 + 5 m^2 + 6 x_1 x_2 + 8 x_1 x_3 + 16 x_2 x_3 + 4 m x_1 + 8 m x_2 + 10 m x_3 )}
    \nonumber\\
    & = \frac{1}{\sqrt{7}} C_b^{72}\, e^{\frac{18 \pi i }{7} m^2}
\end{align}

The integral in the second line can be thought of as an $U(1)^3\times U(1)_f$ CS theory with
\begin{align}
    K = \left(
    \begin{array}{ccc|c}
        2 & 3 & 4 & 2 \\
        3 & 7 & 8 & 4 \\
        4 & 8 & 11 & 5 \\
        \hline
        2 & 4 & 5 & 5
    \end{array}
    \right)\,.
\end{align}

\paragraph{Modular data}
The modular data extracted from the partition function of the abelian CS theory are
\begin{align}
    &\qquad\qquad\qquad\qquad |S_{0\a}| = \frac{1}{\sqrt{7}}
    \;\;
    \text{for}\;\;
    \a = 0,\cdots,6
    \nonumber\\
    &\{T_{\a\a}\} = 
    \left\{
    1,
    e^{2\pi i (\frac{5}{7})},
    e^{2\pi i (\frac{6}{7})},
    e^{2\pi i (\frac{3}{7})},
    e^{2\pi i (\frac{3}{7})},
    e^{2\pi i (\frac{6}{7})},
    e^{2\pi i (\frac{5}{7})}
    \right\},
\end{align}
which are compatible with the complex conjugate of the modular data of affine $\mathrm{su}(2)$ at level $-12/7$.

\subsubsection{$M=7$}\label{sec: D7^7}
Finally, the seventh power of the monodromy operator reads,
\begin{align}
    \hat\Phi^7 = C_b^{168}\, (e^1 e^2 e^3 e^4 e^5 e^6 e^7)^{12}
\end{align}
In the appendix \ref{app: identity}, we show the operator $(e^1 e^2 e^3 e^4 e^5 e^6 e^7)^{12}$ is the identity operator in the three-particle Hilbert space
    \begin{equation}
         (e^1 e^2 e^3 e^4 e^5 e^6 e^7)^{12} = e^{6\pi im^2} \mathbf{1},
    \end{equation}
where we keep track of the dependence of $m$ by a direct computation. It then implies $\hat\Phi^7 = C_b^{168} e^{6\pi im^2}\, \bf{1}$ as expected.

\newpage

\section{$E_N$} \label{sec:EN}

\subsection{$E_6$}
\begin{figure}[!h]
\centering
\begin{tikzpicture}
		\node[W] (1) at (0,0){$\gamma_1$};
		  \node[W] (2) at (2,0) {$\gamma_2$};
            \node[W] (3) at (4,0) {$\gamma_3$};
            \node[W] (4) at (6,0) {$\gamma_4$};
            \node[W] (5) at (8,0) {$\gamma_5$};
            \node[W] (6) at (4,2) {$\gamma_6$};
		\draw[->] (1)--(2);
            \draw[<-] (2)--(3);
            \draw[->] (3)--(4);
            \draw[<-] (4)--(5);
            \draw[->] (3)--(6);
\end{tikzpicture}
\caption{ \label{E6 quiver}
BPS quiver for the $E_6$ theory in the canonical chamber}
\end{figure}
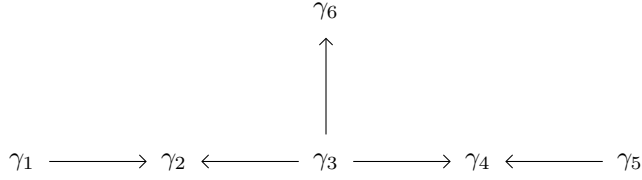
Let us consider the trace for the $E_6$ theory, whose BPS spectrum in the canonical chamber is described by the quiver in Figure \ref{E6 quiver}. The monodromy operator can be written as
\begin{align}
    \hat\Phi = (1)(3)(5)(2)(4)(6)(-1)(-3)(-5)(-2)(-4)(-6)\ .
\end{align}
As we will show in section \ref{sec: E6^7}, the operator satisfies
\begin{align}
    \hat\Phi^7 = C_b^{144} \bf{1}\ ,
\end{align}
which implies that the theories obtained from $S_b^{(M)}$ and $S_b^{(7-M)}$ are related by the orientation reversal up to a gravitational coupling. Therefore, we will only consider the cases $M=1,2,$ and $3$ in the following.

The twisted compactification for $M=1$ completely lifts the Coulomb branch and generates a 3d $\CN=4$ rank-zero SCFT. The topological A-twist of the SCFT is expected to support the W-algebra minimal model $W_3(3,7)$ with central charge $c=-114/7$.

\subsubsection{$M=1$}
The trace formula gives
\begin{align}
    S_b^{(1)} &= \Tr \hat\Phi
    \nonumber\\
    &= C_b^{18} \Tr [ e^1 e^2 e^3 e^4 e^5 (-6)(2) e^3 (4) e^2 e^3 e^6 ]
    \nonumber\\
    &= C_b^{18} \int dx_1 dx_2 dx_3\;
    e^{\pi i ( x_1^2 + x_3^2 + 2 x_1 x_2 + 2 x_2 x_3 - 2 x_1 x_3 )} \Phi_b(x_1) \Phi_b(x_2) \Phi_b(x_3)\ .
\end{align}
The integral is the partition function of an $\CN=2$ $U(1)^3$ gauge theory coupled to three chiral multiplets with 
\begin{align}
    K = \left(
    \begin{array}{ccc}
        2 & 1 & -1 \\
        1 & 1 & 1 \\
        -1 & 1 & 2
    \end{array}
    \right)
    \;\;,\;\;
    Q = \left(
    \begin{array}{ccc}
        1 & 0 & 0 \\
        0 & 1 & 0 \\
        0 & 0 & 1
    \end{array}
    \right)
\end{align}
This description has two gauge invariant half-BPS monopole operators
\begin{align}
    \phi_1^2 V_{(0,-1,1)}
    \;\;,\;\;
    \phi_3^2 V_{(1,-1,0)}
    \;\;.
    \label{eq: E6^1 monopole}
\end{align}
Deforming by these operators, the theory has an unbroken $U(1)$ flavor symmetry
\be
A = U(1)_{\text{top}}^{(1)}+U(1)_{\text{top}}^{(2)}+U(1)_{\text{top}}^{(3)}\ ,
\ee
which can be identified with the axial $U(1)$ R-symmetry in the IR $\CN=4$ SCFT.

\paragraph{Superconformal index}
Performing the F-maximization, we find that the superconformal index at the fixed point reads
\bea\label{sci w37}
    I_{\text{SCI}} =
    1\!-\!q\!-\!\eta\,  q^{3/2}+\left(\frac{1}{\eta ^2}\!-\!1\right) q^2\!+\!\frac{2 q^{5/2}}{\eta }+\left(\eta ^2\!+\!\frac{2}{\eta ^2}\!+\!3\right) q^3+\left(3 \eta \!+\!\frac{5}{\eta }\right) q^{7/2}\!+\!\left(\eta ^2\!+\!7\right) q^4+\cdots
\eea
which is strong evidence that the theory flows to a rank-zero fixed point.\footnote{From the superconformal index calculation, we claim that this theory is dual to the theories in class 8 in \cite{Gang:2024loa}. This is an example whose superconformal index does not contain the $-(\eta+\eta^{-1})q^{3/2}$ term, yet we still anticipate a supersymmetry enhancement. See \emph{loc.cit.} or the discussion in the appendix \ref{app: conventions}.}
\paragraph{Modular data ($\nu=-1$ twist)}
The modular data extracted from the partition function are
\begin{align}
    \left\{ |S_{0\a}| \right\} = \frac{2}{\sqrt{7}}\times\left\{ \sin\frac{3\pi}{14} , \sin\frac{\pi}{6} , \sin\frac{\pi}{14} , \sin\frac{5\pi}{14} ,\sin\frac{\pi}{6} \right\}
    \;\;,\;\;
    \left\{T_{\a\a}\right\} = \left\{
    1 , e^{2\pi i (\frac{4}{7})}
    , e^{2\pi i (\frac{2}{7})}
    , e^{2\pi i (\frac{3}{7})}
    , e^{2\pi i (\frac{4}{7})}
    \right\}.
\end{align}
They are compatible with the modular data of $W_3(3,7)$. See the appendix \ref{app:modular data} for the details.

\paragraph{Modular data ($\nu=1$ twist)}
The modular data extracted from the partition functions are 
\begin{align}
    \left\{ |S_{0\a}| \right\} = \frac{2}{\sqrt{7}}\times\left\{ \sin\frac{5\pi}{14} , 
    \sin\frac{\pi}{6} , 
    \sin\frac{\pi}{14} , 
    \sin\frac{3\pi}{14} ,
    \sin\frac{\pi}{6} \right\}
    \;\;,\;\;
    \left\{T_{\a\a}\right\} = \left\{
    1 , e^{2\pi i (\frac{2}{7})}
    , e^{2\pi i (\frac{5}{7})}
    , e^{2\pi i (\frac{1}{7})}
    , e^{2\pi i (\frac{2}{7})}
    \right\}.
\end{align}

\subsubsection{$M=2$}
The trace of the second power simplifies to
\begin{align}
    S_b^{(2)} &= \Tr \hat\Phi^2
    \nonumber\\
    &= C_b^{36} \Tr [ e^1 e^2 e^3 e^4 e^5 (-6) e^3 e^6 e^4 e^2 e^3 (4)(5)(-6) e^1 e^2 e^3 e^4 e^5 (-6) e^3 e^6 (-2) e^3 ]
    \nonumber\\
    &= C_b^{36} \int \prod_{i=1}^6 \left(dx_i\Phi_b(x_i)\right)\; \times
    \nonumber\\
    &e^{\pi i ( \!-\! x_1^2 \!-\! x_2^2 \!-\! 2 x_3^2 \!-\! 2 x_4^2 \!-\! x_5^2 \!-\! x_6^2 \!-\! 2 x_1 x_2 \!-\! 2 x_1 x_3 \!-\! 2 x_2 x_3 \!-\! 2 x_4 x_5 \!-\! 2 x_4 x_6 \!-\! 2 x_5 x_6 \!+\! 4 x_1 x_4 \!+\! 4 x_2 x_4 \!+\! 4 x_3 x_4 \!+\! 4 x_1 x_5 \!+\! 4 x_3 x_5 \!+\! 4 x_2 x_6 \!+\! 4 x_3 x_6 ) }
\end{align}
The integral is the partition function of an $\CN=2$ $U(1)^6$ gauge theory coupled to three chiral multiplets with 
\begin{align}
    K = \left(
    \begin{array}{cccccc}
        0 & -1 & -1 & 2 & 2 & 0 \\
        -1 & 0 & -1 & 2 & 0 & 2 \\
        -1 & -1 & -1 & 2 & 2 & 2 \\
        2 & 2 & 2 & -1 & -1 & -1 \\
        2 & 0 & 2 & -1 & 0 & -1 \\
        0 & 2 & 2 & -1 & -1 & 0 
    \end{array}
    \right)
    \;\;,\;\;
    Q = \left(
    \begin{array}{cccccc}
        1 & 0 & 0 & 0 & 0 & 0 \\
        0 & 1 & 0 & 0 & 0 & 0 \\
        0 & 0 & 1 & 0 & 0 & 0 \\
        0 & 0 & 0 & 1 & 0 & 0 \\
        0 & 0 & 0 & 0 & 1 & 0 \\
        0 & 0 & 0 & 0 & 0 & 1
    \end{array}
    \right)
\end{align}
This description admits six gauge invariant half-BPS monopole operators
\bea
&\phi_1^2V_{(0,0,0,-1,0,1)},\; \phi_2^2V_{(0,0,0,-1,1,0)},\; \phi_3^2V_{(0,0,0,1,-1,-1)}, \\ 
&\phi_4^2V_{(-1,-1,1,0,0,0)},\; \phi_5^2V_{(0,1,-1,0,0,0)},\; \phi_6^2V_{(1,0,-1,0,0,0)}
\;\;.
\label{eq: E6^2 monopole}
\eea
\paragraph{Superconformal index}
Upon a superpotential deformation with the monopole operators in \eqref{eq: E6^2 monopole}, all the global symmetries are lifted and the superconformal index reduces to
\begin{align}
    I_{\text{SCI}} = 1,
\end{align}
which implies that the theory directly flows to a unitary topological field theory in the infrared.
\paragraph{Modular data}
The modular data from the partition function calculation are
\begin{align}
    \left\{ |S_{0\a}| \right\} = \frac{2}{\sqrt{7}}\times\left\{ 
    \sin\frac{\pi}{14} , 
    \sin\frac{\pi}{6} , 
    \sin\frac{5\pi}{14} , 
    \sin\frac{3\pi}{14} ,
    \sin\frac{\pi}{6} \right\}
    \;\;,\;\;
    \left\{T_{\a\a}\right\} = \left\{
    1 , e^{2\pi i (\frac{1}{7})}
    , e^{2\pi i (\frac{4}{7})}
    , e^{2\pi i (\frac{6}{7})}
    , e^{2\pi i (\frac{1}{7})}
    \right\}
\end{align}

\subsubsection{$M=3$}
The trace of the third power gives
\begin{align}
    S_b^{(3)} &= \Tr \hat\Phi^3
    \nonumber\\
    &= C_b^{60} \Tr [ e^1 e^2 e^3 e^4 e^5 (6) (4) (e^2 e^3 e^4 e^5 e^6)^2 e^1 e^2 e^3 e^4 e^5 e^6 e^1 e^2 e^3 e^4 e^5 (-6) e^1 e^2 e^3 e^6 (4) ]
    \nonumber\\
    &= C_b^{60} \int 
    \left(\prod_{i=1}^4 dx_i \Phi_b(x_i) \right)
    e^{\pi i ( -x_3^2 - x_4^2 - 2 x_1 x_2 + 2 x_2 x_3 - 2 x_1 x_3 + 4 x_1 x_4 )}
\end{align}
The integral is the partition function of an $\CN=2$ $U(1)^4$ gauge theory coupled to four chiral multiplets with 
\begin{align}
    K = \left(
    \begin{array}{cccc}
        1 & -1 & -1 & 2 \\
        -1 & 1 & 1 & 0 \\
        -1 & 1 & 0 & 0 \\
        2 & 0 & 0 & 0 \\
    \end{array}
    \right)
    \;\;,\;\;
    Q = \left(
    \begin{array}{cccc}
        1 & 0 & 0 & 0 \\
        0 & 1 & 0 & 0 \\
        0 & 0 & 1 & 0 \\
        0 & 0 & 0 & 1 \\
    \end{array}
    \right)\ .
\end{align}
This description admits three gauge invariant half-BPS monopole operators
\begin{align}
    \phi_1^2 V_{(0,0,0,-1)}
    \;\;,\;\;
    \phi_2^2 V_{(0,0,-2,-1)}
    \;\;,\;\;
    \phi_4^2 V_{(-1,-1,0,0)}
    \;\;.
    \label{eq: E6^3 monopole}
\end{align}
Deforming by the monopole operators, we are left with
\be
A = -U(1)_{\text{top}}^{(1)} + U(1)_{\text{top}}^{(2)}\ .
\ee
\paragraph{Superconformal index}
Performing the F-maximization, the superconformal index at the fixed point reads \footnote{Note that this is equivalent to \eqref{sci w37} upon replacing $\eta\rightarrow \eta^{-1}$, which is evidence that it flows to a $\CN=4$ rank-zero theory mirror to that theory.}
\bea
    I_{\text{SCI}}
    =
    1\!-\!q\!-\!\frac{q^{3/2}}{\eta}+\left(\eta ^2\!-\!1\right) q^2+2\eta\,  q^{5/2}+\left(2\eta ^2\!+\!\frac{1}{\eta ^2}\!+\!3\right) q^3\!+\!\left(5 \eta \!+\!\frac{3}{\eta }\right) q^{7/2}\!+\left(\frac{1}{\eta ^2}\!+\!7\right) q^4 + \cdots\ .
\eea
\paragraph{Modular data ($\nu=-1$ twist)}
The modular data for the $\nu=-1$ twist extracted from the supersymmetric partition function is \footnote{The numerical computations of the partition functions in this example turn out to be very subtle. The two solutions to the Bethe equations corresponding to $\alpha=1,4$ are numerically unstable, and give rise to diverging answers. The entries marked with $\star$ are computed with the fugacity $a$ for $U(1)_{\text{top}}^{(3)}-2U(1)_{\text{top}}^{(4)}$ turned on, and taken the limit $a\rightarrow 1$ from the right, at the end of the computation. The expected values for $|S_{0\alpha}|$ for $\alpha=1,4$ (from the Galois transformation) are both 
$\sin\pi/6$.}
\begin{align}
    \{|S_{0\a}|\}
    =
    \frac{2}{\sqrt{7}}\times\left\{
    \sin\frac{5\pi}{14}
    ,
    0^\star
    ,
    \sin \frac{3\pi}{14}
    ,
    \sin \frac{\pi}{14}
    ,
    \sin \frac{\pi}{4}^\star
    \right\}
    \;\;,\;\;
    \{T_{\a\a}\}
    =
    \{ 
    1,
    \left.e^{2\pi i (\frac{21}{25})}\right. ^\star,
    e^{2\pi i (\frac{6}{7})},
    e^{2\pi i (\frac{2}{7})},
    \left.e^{2\pi i (\frac{5}{7})}\right.^\star
    \}
\end{align}

\paragraph{Modular data ($\nu=1$ twist)}
The modular data for the $\nu=1$ twist extracted from the supersymmetric partition function is
\begin{align}
    \{|S_{0\a}|\}
    =
    \frac{2}{\sqrt{7}}\times\left\{
    \sin\frac{3\pi}{14}
    ,
    0^\star
    ,
    \sin \frac{\pi}{14}
    ,
    \sin \frac{5\pi}{14}
    ,
    \sin \frac{\pi}{4}^\star
    \right\}
    \;\;,\;\;
    \{T_{\a\a}\}
    =
    \{ 
    1,
    \left.e^{2\pi i (\frac{111}{200})}\right.^\star,
    e^{2\pi i (\frac{5}{7})},
    e^{2\pi i (\frac{4}{7})},
    \left.e^{2\pi i (\frac{3}{7})}\right.^\star
    \}
\end{align}

\subsubsection{$M=7$}\label{sec: E6^7}
Finally, the seventh power of the monodromy operator simplifies to
\begin{align}
    \hat\Phi^7 = C_b^{144} (e^1 e^2 e^3 e^4 e^5 e^6)^{12}\ .
\end{align}
The operator $(e^1 e^2 e^3 e^4 e^5 e^6)^{12}$ can be evaluated by Gauss integrals and is the identity operator in the three-particle Hilbert space 
    \begin{equation}
         (e^1 e^2 e^3 e^4 e^5 e^6)^{12} = \mathbf{1}.
    \end{equation}
It then implies $\hat\Phi^7 = C_b^{144} \bf{1}$ as expected.

\subsection{$E_7$}
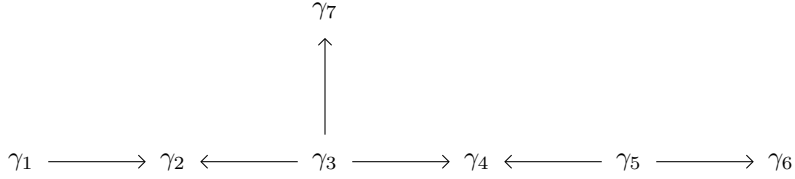
\begin{figure}[h]
  \centering
  \begin{tikzpicture}
			\node[W] (1) at (0,0){$\gamma_1$};
		 	\node[W] (2) at (2,0) {$\gamma_2$};
            \node[W] (3) at (4,0) {$\gamma_3$};
            \node[W] (4) at (6,0) {$\gamma_4$};
            \node[W] (5) at (8,0) {$\gamma_5$};
            \node[W] (6) at (10,0) {$\gamma_6$};
            \node[W] (7) at (4,2) {$\gamma_7$};
		\draw[->] (1)--(2);
            \draw[<-] (2)--(3);
            \draw[->] (3)--(4);
            \draw[<-] (4)--(5);
            \draw[->] (5)--(6);
            \draw[->] (3)--(7);
  \end{tikzpicture}
\caption{ \label{E7 quiver}
BPS quiver for the $E_7$ theory in the canonical chamber}
\end{figure}

Let us consider the trace formula for the $E_7$ theory, whose BPS spectrum in the canonical chamber is described by the quiver in Figure \ref{E7 quiver}. The monodromy operator is now
\begin{align}
    \hat\Phi = (1)(3)(5)(7)(2)(4)(6)(-1)(-3)(-5)(-7)(-2)(-4)(-6)\ .
\end{align}
The theory has a $U(1)$ flavor symmetry associated with the vector
\be
\gamma_f = -\gamma_4+\gamma_6+\gamma_7\ .
\ee

We will show in section \ref{sec: E7^5} that
\begin{align}
    \hat\Phi^5 = C_b^{126} e^{3\pi i m^2}\, \bf{1}
\end{align}
This implies that the theories extracted from $S_b^{(M)}$ and $S_b^{(5-M)}$ are related by the orientation reversal up to background couplings. We will only consider the cases $M=1$ and $2$ in the following.
\subsubsection{$M=1$}
We have
\begin{align}
    S_b^{(1)} & = \Tr \hat\Phi = C_b^{24}\, \Tr [ e^1 e^2 e^3 e^4 e^5 e^6 (2) e^3 e^4 e^1 e^2 e^3 e^7 ] 
    \nonumber\\
    & = C_b^{24}\int dx_1 dx_2 dx_3\,
    e^{\pi i ( 3 x_2^2 + 3 x_3^2 + m^2 - 2 x_1 x_2 + 2 x_1 x_3 + 4 x_2 x_3 + 2 m x_2 + 2 m x_3 )}\Phi_b(x_1)\ ,
\end{align}
which is the partition function of an $\CN=2$ $U(1)^3\times U(1)_f$ theory coupled to a single chiral multiplet with
\begin{align}
    K = \left(
    \begin{array}{ccc|c}
        1 & -1 & 1 & 0 \\
        -1 & 3 & 2 & 1 \\
        1 & 2 & 3 & 1 \\
        \hline
        0 & 1 & 1 & 1
    \end{array}
    \right)
    \;\;,\;\;
    Q = \left(
    \begin{array}{c}
        1 \\
        0 \\
        0 \\
        \hline
        0
    \end{array}
    \right)
\end{align}
This description does not admit any gauge invariant half-BPS monopole operator.

\paragraph{Superconformal index}
The superconformal index of the theory reads
\begin{align}
    I_{\text{SCI}} = 1 - q - \left(\eta+\frac{1}{\eta}\right)q^{3/2} - 2 q^2 - \left(\eta+\frac{1}{\eta}\right)q^{5/2} - 2 q^3- \left(\eta+\frac{1}{\eta}\right)q^{7/2} - 2 q^4 + \cdots
\end{align}
which is strong evidence that the theory flows to $\CN=4$ rank-zero theory in the infrared. Notice that this expansion coincides with the superconformal index of $\CT_{\text{min}}$, which implies that the theory is dual to the $\CT_{\text{min}}$ multiplied by a unitary TFT.

\paragraph{Modular data ($\nu=-1$ twist)}
The modular data extracted from the supersymmetric partition function for $\nu=-1$ twist is 
\bea
    \qquad & |S_{00}| = |S_{01}| = |S_{02}| = |S_{03}| =  |S_{04}| = \frac{2}{5}\sin\left(\frac{2\pi}{5}\right)\,,\\
    \qquad & |S_{05}| = |S_{06}| = |S_{07}| = |S_{08}| =  |S_{09}| = \frac{2}{5}\sin\left(\frac{\pi}{5}\right)\,,\\
    \left\{T_{\a\a}\right\} =& 
    \left\{
    1, 
    e^{2\pi i (\frac{1}{5})},
    e^{2\pi i (\frac{1}{5})},
    e^{2\pi i (\frac{4}{5})},
    e^{2\pi i (\frac{4}{5})},
    e^{2\pi i (\frac{1}{5})},
    1,
    1,
    e^{2\pi i (\frac{2}{5})},
    e^{2\pi i (\frac{2}{5})}
    \right\}\,.
\eea

\paragraph{Modular data ($\nu=1$ twist)}
The modular data extracted from the supersymmetric partition function for $\nu=1$ twist is 
\bea
    \qquad & |S_{00}| = |S_{01}| = |S_{02}| = |S_{03}| =  |S_{04}| = \frac{2}{5}\sin\left(\frac{2\pi}{5}\right)\,,\\
    \qquad & |S_{05}| = |S_{06}| = |S_{07}| = |S_{08}| =  |S_{09}| = \frac{2}{5}\sin\left(\frac{\pi}{5}\right)\,,\\
    \left\{T_{\a\a}\right\} =& 
    \left\{
    1, 
    e^{2\pi i (\frac{4}{5})},
    e^{2\pi i (\frac{4}{5})},
    e^{2\pi i (\frac{1}{5})},
    e^{2\pi i (\frac{1}{5})},
    e^{2\pi i (\frac{4}{5})},
    1,
    1,
    e^{2\pi i (\frac{3}{5})},
    e^{2\pi i (\frac{3}{5})}
    \right\}\,.
\eea

\subsubsection{$M=2$}
Now the second power gives
\begin{align}
    S_b^{(2)} & = \Tr \hat\Phi = C_b^{48}\, \Tr [ ( e^1 e^2 e^3 e^1 e^5 e^7 (3) e^2 e^4 e^6 e^5 e^3 e^4 )^2) ] 
    \nonumber\\
    & = C_b^{48}\int dx_1 dx_2 dx_3\,
    e^{\pi i ( -5 x_1^2 - 5 x_3^2 + m^2 - 4 x_1 x_2 - 10 x_1 x_3 - 2 m x_1 - 2 m x_3 )}\Phi_b(x_1) \Phi_b(x_2)
\end{align}
which is the partition function of an $\CN=2$ $U(1)^3\times U(1)_f$ theory coupled to two chiral multiplets with
\begin{align}
    K = \left(
    \begin{array}{ccc|c}
        -4 & -2 & -5 & -1 \\
        -2 & 1 & 0 & 0 \\
        -5 & 0 & -5 & -1 \\
        \hline
        -1 & 0 & -1 & 1
    \end{array}
    \right)
    \;\;,\;\;
    Q = \left(
    \begin{array}{cc}
        1 & 0 \\
        0 & 1 \\
        0 & 0 \\
        \hline
        0 & 0
    \end{array}
    \right)
\end{align}
This description admits two gauge invariant half-BPS monopole operators
\begin{align}
    \phi_1^2 V_{(0,1,0)}\;\;,\;\;
    \phi_2^2 V_{(1,0,-1)}\;.
    \label{eq: E7^2 monopole}
\end{align}
Deforming by the monopole operators does not leave any global symmetry that acts non-trivially on the IR theory.

\paragraph{Superconformal index}
The superconformal index with the monopole deformation reads
\begin{align}
    I_{\text{SCI}} = 1\ ,
\end{align}
which is a signal that the theory flows to a unitary topological field theory.

\paragraph{Modular data}
The modular data extracted from the supersymmetric partition function are 
\bea
    \qquad & |S_{00}| = |S_{01}| = |S_{02}| = |S_{03}| =  |S_{04}| = \frac{2}{5}\sin\left(\frac{\pi}{5}\right)\,,\\
    \qquad & |S_{05}| = |S_{06}| = |S_{07}| = |S_{08}| =  |S_{09}| = \frac{2}{5}\sin\left(\frac{2\pi}{5}\right)\,,\\
    \left\{T_{\a\a}\right\} =& 
    \left\{
    1, 
    e^{2\pi i (\frac{2}{5})},
    e^{2\pi i (\frac{2}{5})},
    e^{2\pi i (\frac{3}{5})},
    e^{2\pi i (\frac{3}{5})},
    e^{2\pi i (\frac{2}{5})},
    1,
    1,
    e^{2\pi i (\frac{4}{5})},
    e^{2\pi i (\frac{4}{5})}
    \right\}\,.
\eea

\subsubsection{$M=5$}\label{sec: E7^5}
The fifth power of the monodromy operator simplifies to
\begin{align}
    \hat\Phi^5 = C_b^{126}\,(e^1 e^2 e^3 e^4 e^5 e^6 e^7)^9\ ,
\end{align}
The operator $(e^1 e^2 e^3 e^4 e^5 e^6 e^7)^9$ can be evaluated by Gauss integrals and is the identity operator in the three-particle Hilbert space 
    \begin{equation}
         (e^1 e^2 e^3 e^4 e^5 e^6 e^7)^9 = e^{3\pi i m^2} \mathbf{1}.
    \end{equation}
It then implies $\hat\Phi^5 = C_b^{126}\,e^{3\pi i m^2} \bf{1}$ as expected.

\subsection{$E_8$}
\begin{figure}[!h]
  \centering
  \begin{tikzpicture}
			\node[W] (1) at (0,0){$\gamma_1$};
		 	\node[W] (2) at (2,0) {$\gamma_2$};
            \node[W] (3) at (4,0) {$\gamma_3$};
            \node[W] (4) at (6,0) {$\gamma_4$};
            \node[W] (5) at (8,0) {$\gamma_5$};
            \node[W] (6) at (10,0) {$\gamma_6$};
            \node[W] (7) at (12,0) {$\gamma_7$};
            \node[W] (8) at (4,2) {$\gamma_8$};
		\draw[->] (1)--(2);
            \draw[<-] (2)--(3);
            \draw[->] (3)--(4);
            \draw[<-] (4)--(5);
            \draw[->] (5)--(6);
            \draw[<-] (6)--(7);
            \draw[->] (3)--(8);
\end{tikzpicture}
\caption{ \label{E8 quiver}
BPS quiver for the $E_8$ theory in the canonical chamber}
\end{figure}
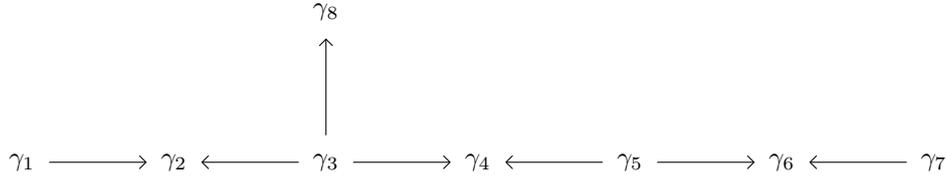
Let us consider the traces for the $E_8$ theory, whose BPS spectrum in the canonical chamber is described by the quiver in Figure \ref{E8 quiver}. The monodromy operator reads
\begin{align}
    \hat\Phi = (1)(3)(5)(7)(2)(4)(6)(8)(-1)(-3)(-5)(-7)(-2)(-4)(-6)(-8)\ .
\end{align}
We will show in section \ref{sec: E8^8} that it satisfies
\begin{align}
    \hat\Phi^8 = C_b^{280} \bf{1}\ ,
\end{align}
which implies the theories obtained from $S_b^{(M)}$ and $S_b^{(8-M)}$ are related by the orientation reversal. We consider the cases $M=1,2,3,$ and 4 in the following.

\subsubsection{$M=1$}
The trace of the first power of the monodromy operator reads
\begin{align}
    S_b^{(1)} &= \Tr \hat\Phi 
    \nonumber\\
    &= C_b^{26}
    \Tr [ e^1 e^2 e^3 e^{4+5} (-6) e^5 e^7 e^6 e^8 e^3 (4) e^5 (7) (-6) e^2 e^3 e^4 ]
    \nonumber\\
    &= C_b^{26}
    \int 
    \left(
    \prod_{i=1}^4
    dx_i \Phi_b(x_i)
    \right)dx_5
    \;
    e^{\pi i ( \!-\! 2 x_1^2 \!-\! 2 x_2^2 \!-\! x_3^2 \!-\! x_4^2 \!+\! 4 x_1 x_2 \!+\! 2 x_1 x_3 \!-\! 2 x_1 x_4 \!+\! 2 x_2 x_4 \!-\! 4 x_1 x_5 \!+\! 2 x_2 x_5 \!+\! 2 x_3 x_5 \!-\! 4 x_4 x_5 ) }
\end{align}
which is the partition function of an $\CN=2$ $U(1)^5$ gauge theory with four chiral multiplets with
\begin{align}
    K=
    \left(
    \begin{array}{ccccc}
        -1 & 2 & 1 & -1 & -2 \\
        2 & -1 & 0 & 1 & 1 \\
        1 & 0 & 0 & 0 & 1 \\
        -1 & 1 & 0 & 0 & -2 \\
        -2 & 1 & 1 & -2 & 0
    \end{array}
    \right)
    \;\;,\;\;
    Q=
    \left(
    \begin{array}{cccc}
        1 & 0 & 0 & 0 \\
        0 & 1 & 0 & 0 \\
        0 & 0 & 1 & 0 \\
        0 & 0 & 0 & 1 \\
        0 & 0 & 0 & 0
    \end{array}
    \right)
\end{align}
This description admits three gauge invariant half-BPS monopole operators
\begin{align}
    \phi_1\phi_2 V_{(0,0,-2,-1,0)}
    \;\;,\;\;
    \phi_2\phi_3 V_{(-1,0,0,1,0)}
    \;\;,\;\;
    \phi_1\phi_3 V_{(0,-2,0,-1,-1)}
    \;\;.
    \label{eq: E8^1 monopole}
\end{align}
Deforming by the monopole superpotentials, we are left with two independent $U(1)$ flavor symmetries. We identify
\be
A = U(1)_{\text{top}}^{(2)} -2 U(1)_{\text{top}}^{(5)}\ .
\ee
The other $U(1)$ flavor symmetry decouples in the infrared, as one can check from the superconformal index calculation below.

\paragraph{Superconformal index}
Upon a superpotential deformation with the monopole operators in \eqref{eq: E8^1 monopole}, the theory flows to $\CN=4$ rank-zero theory with the superconformal index
\bea
    I_{\text{SCI}} =& 
    1-q+\left(\eta^2+\frac{1}{\eta^2}+1\right) q^2+\left(3 \eta+\frac{3}{\eta}\right) q^{5/2}+\left(\eta^2+\frac{1}{\eta^2}+7\right) q^3 \\
    &+\left(-\eta^3-\frac{1}{\eta^3}+6 \eta+\frac{6}{\eta}\right) q^{7/2}+\left(-2 \eta^2-\frac{2}{\eta^2}+10\right) q^4+\cdots
\eea
which is evidence that the theory reduces to an $\CN=4$ rank-zero fixed point. \footnote{This is another example whose superconformal index does not contain the $-(\eta+\eta^{-1})q^{3/2}$ term, yet we still anticipate a supersymmetry enhancement. See the discussion in the appendix \ref{app: conventions}}
\paragraph{Modular data ($\nu=-1$ twist)}
The modular data extracted from the partition functions are 
\bea
    \left\{ |S_{0\a}| \right\} &= \left\{ \frac{1}{4} \left(\sqrt{2}+1\right),\frac{1}{2\sqrt{2}},\frac{1}{2},\frac{1}{4} \left(\sqrt{2}-1\right),\frac{1}{2\sqrt{2}},\frac{1}{4},\frac{1}{4} \right\}
    \;\;,\;\; \\
    \left\{T_{\a\a}\right\} &= \left\{
    1 , e^{2\pi i (\frac{2}{8})}, e^{2\pi i (\frac{1}{8})}, e^{2\pi i (0)}, e^{2\pi i (\frac{2}{8})}, e^{2\pi i (\frac{4}{8})}, e^{2\pi i (\frac{4}{8})}
    \right\}\ .
\eea
These are compatible with the modular data of $W_3(3,8)$. See the appendix \ref{app:modular data} for details.

\paragraph{Modular data ($\nu=1$ twist)}
The modular data from the partition functions are
\bea
    \left\{ |S_{0\a}| \right\} &= \left\{ \frac{1}{4} \left(\sqrt{2}+1\right),\frac{1}{2\sqrt{2}},\frac{1}{2},\frac{1}{4} \left(\sqrt{2}-1\right),\frac{1}{2\sqrt{2}},\frac{1}{4},\frac{1}{4} \right\}
    \;\;,\;\; \\
    \left\{T_{\a\a}\right\} &= \left\{
    1 , e^{2\pi i (\frac{2}{8})}, e^{2\pi i (\frac{1}{8})}, e^{2\pi i (0)}, e^{2\pi i (\frac{2}{8})}, e^{2\pi i (\frac{4}{8})}, e^{2\pi i (\frac{4}{8})}
    \right\}\ .
\eea
These are again compatible with $W_3(3,8)$ discussed above, which is a strong signal that the 3d $\CN=4$ SCFT is self-mirror.

\subsubsection{$M=2$}
The second power simplifies to
\begin{align}
    S_b^{(2)} &= \Tr \hat\Phi^2 = C_b^{60}\, \Tr [ (e^1 e^2 e^3 e^4 e^5 e^6 e^7 e^8)^3 e^1 e^2 e^3 e^4 e^5 e^8 ]
    \nonumber\\
    &= C_b^{60} \int dx_1 dx_2 dx_3 dx_4 dx_5\,
    e^{\pi i ( x_4^2 - 7 x_5^2 - 2 x_1 x_2 - 4 x_1 x_3 - 2 x_2 x_3 - 2 x_1 x_4 + 2 x_3 x_4 + 8 x_1 x_5 + 2 x_2 x_5 + 8 x_3 x_5 )}
\end{align}
which can be thought of as the partition function of an $U(1)^5$ CS theory with level matrix
\begin{align}
    K = \left(
    \begin{array}{ccccc}
        0 & -1 & -2 & -1 & 4 \\
        -1 & 0 & -1 & 0 & 1 \\
        -2 & -1 & 0 & 1 & 4 \\
        -1 & 0 & 1 & 1 & 0 \\
        4 & 1 & 4 & 0 & -7
    \end{array}
    \right)
\end{align}
This theory has runaway vacua, which can be inferred from the relation det$K=0$. In fact, this is an example with gcd$(M,8)\neq 1$.

By a change of variables, we can simplify the partition function to
\begin{align}
    S_b^{(2)} = C_b^{60} \int dx_1 dx_2 dx_3 dx_4 dx_5 \,
    e^{\pi i ( x_2^2 + x_4^2 - x_5^2 )}\ .
\end{align}

\subsubsection{$M=3$}
The third power gives
\begin{align}
    S_b^{(3)} &= \Tr \hat\Phi^3 = C_b^{78}\, \Tr \left[ \left( e^1 e^2 e^3 e^4 e^5 e^6 e^7 e^8 (2+3) e^4 e^5 e^3 e^4 (-8) (2) e^3 (-8) \right)^3 \right]
    \nonumber\\
    &=C_b^{78}\int \prod_{i=1}^{12}\left(dx_i \Phi_b(x_i) \right) dx_{13}\,
    e^{\pi i\, {\bf{x}}^{t} A {\bf{x}}}
\end{align}
where $A$ is a rank 13 square matrix with 
\begin{align}
    K = \left(
    \begin{array}{ccccccccccccc}
        2 & -1 & -1 & 1 & -1 & 1 & 1 & -1 & 0 & 0 & 0 & -1 & -1 \\
        -1 & 1 & 0 & 0 & 0 & 0 & 1 & 0 & 1 & 0 & 0 & 1 & 0 \\
        -1 & 0 & 1 & -1 & 1 & 0 & 0 & 1 & 0 & 1 & 1 & 0 & 1 \\
        1 & 0 & -1 & 2 & -1 & 1 & 0 & 0 & -1 & 0 & 1 & 0 & 0 \\
        -1 & 0 & 1 & -1 & 2 & -1 & -1 & 1 & -1 & 1 & 1 & -1 & 1 \\
        1 & 0 & 0 & 1 & -1 & 1 & 0 & 0 & 0 & 0 & 1 & 0 & 0 \\
        1 & 1 & 0 & 0 & -1 & 0 & 1 & -1 & 1 & 0 & 0 & 1 & -1 \\
        -1 & 0 & 1 & 0 & 1 & 0 & -1 & 2 & -1 & 1 & 0 & 0 & 0 \\
        0 & 1 & 0 & -1 & -1 & 0 & 1 & -1 & 2 & -1 & -1 & 1 & -1 \\
        0 & 0 & 1 & 0 & 1 & 0 & 0 & 1 & -1 & 1 & 0 & 0 & 0 \\
        0 & 0 & 1 & 1 & 1 & 1 & 0 & 0 & -1 & 0 & 1 & -1 & 1 \\
        -1 & 1 & 0 & 0 & -1 & 0 & 1 & 0 & 1 & 0 & -1 & 2 & 0 \\
        -1 & 0 & 1 & 0 & 1 & 0 & -1 & 0 & -1 & 0 & 1 & 0 & 2
    \end{array}
    \right)
    \;\;,\;\;
    Q = \left(
    \begin{array}{cccccccccccc}
        1 & 0 & 0 & 0 & 0 & 0 & 0 & 0 & 0 & 0 & 0 & 0 \\
        0 & 1 & 0 & 0 & 0 & 0 & 0 & 0 & 0 & 0 & 0 & 0 \\
        0 & 0 & 1 & 0 & 0 & 0 & 0 & 0 & 0 & 0 & 0 & 0 \\
        0 & 0 & 0 & 1 & 0 & 0 & 0 & 0 & 0 & 0 & 0 & 0 \\
        0 & 0 & 0 & 0 & 1 & 0 & 0 & 0 & 0 & 0 & 0 & 0 \\
        0 & 0 & 0 & 0 & 0 & 1 & 0 & 0 & 0 & 0 & 0 & 0 \\
        0 & 0 & 0 & 0 & 0 & 0 & 1 & 0 & 0 & 0 & 0 & 0 \\
        0 & 0 & 0 & 0 & 0 & 0 & 0 & 1 & 0 & 0 & 0 & 0 \\
        0 & 0 & 0 & 0 & 0 & 0 & 0 & 0 & 1 & 0 & 0 & 0 \\
        0 & 0 & 0 & 0 & 0 & 0 & 0 & 0 & 0 & 1 & 0 & 0 \\
        0 & 0 & 0 & 0 & 0 & 0 & 0 & 0 & 0 & 0 & 1 & 0 \\
        0 & 0 & 0 & 0 & 0 & 0 & 0 & 0 & 0 & 0 & 0 & 1 \\
        0 & 0 & 0 & 0 & 0 & 0 & 0 & 0 & 0 & 0 & 0 & 0
    \end{array}
    \right)
\end{align}
Namely, the theory is an $\CN=2$ $U(1)^{13}$ gauge theory coupled to 12 chiral multiplets. We do not perform any further analysis of the theory.

\subsubsection{$M=4$}
The fourth power gives
\begin{align}
    S_b^{(4)} &= \Tr \hat\Phi^4 = C_b^{120}\,\Tr[ (e^1 e^2 e^3 e^4 e^5 e^6 e^7 e^8)^2 e^2 e^3 e^4 e^5 e^8 (e^1 e^2 e^3 e^4 e^5 e^6 e^7 e^8)^4 e^2 e^3 e^4 e^5 e^6 e^7 e^8 ]
    \nonumber\\
    &= C_b^{120}\int \left( \prod_{i=1}^{6}dx_i\right) \,
    e^{\pi i ( 5 x_2^2 + 2 x_3^2 + 2 x_4^2 + 2 x_5^2 + x_6^2 + 6 x_2 x_3 + 6 x_2 x_4 + 4 x_3 x_4 + 2 x_2 x_5 + 4 x_2 x_6 + 2 x_3 x_6 + 2 x_4 x_6 + 2 x_5 x_6 )}
\end{align}
which is a pure $U(1)^6$ CS theory with
\begin{align}
    K = \left(
    \begin{array}{cccccc}
        0 & 0 & 0 & 0 & 0 & 0 \\
        0 & 5 & 3 & 3 & 1 & 2 \\
        0 & 3 & 2 & 2 & 0 & 1 \\
        0 & 3 & 2 & 2 & 0 & 1 \\
        0 & 1 & 0 & 0 & 2 & 1 \\
        0 & 2 & 1 & 1 & 1 & 1
    \end{array}
    \right)
\end{align}
This theory again has runaway vacua, which we expected from the fact that gcd$(M,8)\neq 1$.
In fact, by a change of variable, we can rewrite the integral as

\begin{align}
    S_b^{(4)} = C_b^{120}\int dx_1 dx_2 dx_3 dx_4 dx_5 dx_6\,
    e^{\pi i (u_5^2 + u_6^2)}\ .
\end{align}

\subsubsection{$M=8$}\label{sec: E8^8}
Finally, the eighth power of the monodromy operator simplifies to
\begin{align}
    \hat\Phi^8 = C_b^{280}\,(e^1 e^2 e^3 e^4 e^5 e^6 e^7 e^8)^{15}\ ,
\end{align}
The operator $(e^1 e^2 e^3 e^4 e^5 e^6 e^7 e^8)^{15}$ can be evaluated by Gauss integrals and is the identity operator in the four-particle Hilbert space
    \begin{equation}
         (e^1 e^2 e^3 e^4 e^5 e^6 e^7 e^8)^{15} =  \mathbf{1}.
    \end{equation}
It then implies $\hat\Phi^8 = C_b^{280}\,\bf{1}$ as expected.

\newpage

\section*{Acknowledgements}

We thank Davide Gaiotto, Thomas Creutzig, Niklas Garner, Jaewon Song, Dongmin Gang, and Jin Chen for useful discussions and collaborations on related works. The work of BG, QJ, and HK is supported by the National Research Foundation of Korea (NRF) grant NRF2023R1A2C1004965 and RS-2024-00405629. The work of SK is supported by KIAS Individual Grant PG092101 at Korea Institute for Advanced Study. The work of HK is also supported by POSCO Science Fellowship of POSCO TJ Park Foundation. The work of QJ is also supported by Jang Young-Sil Fellow Program at the Korea Advanced Institute of Science and Technology.

\appendix

\section{Conventions on the twisted partition functions}\label{app: conventions}

In this appendix, we will review and summarize our conventions for the calculation of supersymmetric partition functions on a Seifert manifold $\CM_{g,p}$, which is a degree $p$ circle bundle over a genus $g$ closed Riemann surface. See e.g., \cite{Closset:2017zgf, Closset:2019hyt} for a review.

\paragraph{Building block of supersymmetric partition functions}

Let us consider an $\CN=2$ $U(1)^N$ gauge theory with an effective Chern-Simons level $K$ coupled to chiral multiplets with a matrix of gauge charges $Q$. We introduce the fugacities for the gauge symmetries and the $U(1)_{\text{top}}^N$ topological symmetries \footnote{The rank of the total $U(1)$ flavor symmetries for this class of theories is the same as the number of chiral multiplets.}

\begin{equation}
    x_a = e^{2\pi i u_a},\quad y_a = e^{2\pi i v_a}\ ,
\end{equation}
where $a=1,\cdots,N$. In the main text, we consider dressed monopole operators 
    \begin{equation}
        \mathcal{O}_{\mathbf{n},\mathbf{m}} := \left(\prod_{i=1}^{N_f} (\phi_i)^{n_i} \right) V_{\mathbf{m}},
    \end{equation}
where $V_{\mathbf{m}}$ is a bare monopole operator with magnetic charge $\mathbf{m}$ and $\phi_i$ is a scalar field in a chiral multiplet $\Phi_i$. This operator becomes half-BPS if it satisfies the relation 
\begin{equation} \mathbf{m}\cdot \mathbf{Q}_i = 0,\quad (i=1,\cdots,N_f)\ ,
\end{equation}
where $\mathbf{Q}_i = (Q_{i1},\cdots,Q_{iN})$ is the gauge charge vectors of $\phi_i$. The monopole operator carries gauge charge 
    \begin{equation}
        \sum_{b} K_{ab} m_b + \sum_{i|\mathbf{m}\cdot \mathbf{Q}_i = 0} \mathbf{n} \cdot \mathbf{Q}_i - \sum_{i|\mathbf{m}\cdot \mathbf{Q}_i > 0} (\mathbf{m}\cdot \mathbf{Q}_i) \mathbf{Q}_i = 0
    \end{equation}
which must vanish for it to be a gauge invariant operator.

The supersymmetric partition functions on $\CM_{g,p}$ are fully determined by the effective twisted superpotential $\mathcal{W}(u,v)$ and the effective dilaton $\Omega(u,v)$ coupling. The effective twisted superpotential $\mathcal{W}(u,v)$ for this class of theory reads 
\begin{equation}
    \mathcal{W} = \mathcal{W}_{\textrm{CS}} + \mathcal{W}_{\Phi}
\end{equation}
with
\begin{equation}
    \begin{split}
    \mathcal{W}_{\textrm{CS}} =& \frac{1}{2} \sum_{a,b} K_{ab} u_a u_b + \frac{1}{2} (1+2 v_R) \sum_a K_{aa} u_a + \sum_{a i} \delta_{a b} u_a v_b + \cdots \\
    \mathcal{W}_{\Phi} =& \frac{1}{(2\pi i)^2} \sum_{i} \textrm{Li}_2 (x(\mathbf{Q}_i))\ ,
    \end{split}.
\end{equation}
where we defined $x(\mathbf{Q}_i) = \prod_a (x_a)^{Q_{ia}}$.
The effective dilaton is 
\begin{equation}
    \Omega(u,v) = -\frac{1}{2\pi i} \sum_i (r_i - 1) \sum_{i} \log(1 - x(\mathbf{Q}_i))+\frac{1}{2}k_{RR}\ ,
\end{equation}
where $r_i$'s are the R-charge of $\phi_i$.\footnote{All the $\mathcal{H}(u_*,v)$ can be set to positive so that $\sum_\alpha S_{0\alpha}^2=1$ by properly choosing $k_{RR}$. It can be done for our examples by setting $k_{RR}$ to 0 or 1.} The building blocks of the supersymmetric partition functions are the handle gluing operators $\mathcal{H}(u,v)$, the fibering operators $\mathcal{F}(u,v)$ and the flux operators $\Pi^g_a(u,v),\Pi^f_i(u,v)$ defined as
    \begin{equation}
        \begin{split}
            \mathcal{H}(u,v) =& \exp\left(2\pi i \Omega(u,v) \right)\det \left(\frac{\partial^2 \mathcal{W}(u,v)}{\partial u_a \partial u_b}\right)\\
            \mathcal{F}(u,v) =&\exp \left(2 \pi i \left(\mathcal{W}(u,v) - u_a \frac{\partial \mathcal{W}(u,v)}{\partial u_a} -  v_a \frac{\partial \mathcal{W}(u,v)}{\partial v_a}\right)\right)\\
            \Pi^g_a(u,v) =& \exp\left(2\pi i \frac{\partial \mathcal{W}(u,v)}{\partial u_a} \right),\quad \Pi^f_a(u,v) = \exp\left(2\pi i \frac{\partial \mathcal{W}(u,v)}{\partial v_a} \right).
        \end{split}
    \end{equation}
The supersymmetric vacua of effective 2d theory on the Riemann surface are given by the solutions to the so-called Bethe equations
    \begin{equation}
        \mathcal{S}_{\textrm{BE}} = \left\{ (u_{1},\cdots,u_N) ~|~ \Pi^g_a(u,v) = 1, \forall a=1,\cdots,N\right\}.
    \end{equation}
The partition functions are then written as
    \begin{equation}
        \mathcal{Z}_{\mathcal{M}_{g,p}}(v,v_R,\mathfrak{n}_i) = \sum_{u_* \in \mathcal{S}_{\textrm{BE}}(v,v_R)} \mathcal{H}(u_*,v)^{g-1} \mathcal{F}(u_*,v)^{p} \prod_{i} \Pi^f_i(u_*,v)^{{\mathfrak n}_i}\ .
    \end{equation}

\paragraph{Monopole deformation and F-Maximization}

Let us denote $T_1,\cdots,T_N$ by the generators of the $U(1)_{\textrm{top}}^N$ topological symmetry and $R$ as the generator of the $U(1)$ R-symmetry of $\CN=2$ algebra. We define the mixing of the R-symmetry with the flavor symmetry by
    \begin{equation}
        R_{\mu} = R + \boldsymbol{\mu} \cdot \mathbf{T},
    \end{equation}
where $\boldsymbol{\mu}=(\mu_1,\cdots,\mu_N)$ is the mixing parameters. In this paper, we  consider superpotential deformations by $\frac{1}{2}$-BPS monopole operators
    \begin{equation}
        \Delta W = \sum_{I} \mathcal{O}^{(I)}_{\mathbf{n},\mathbf{m}}\ ,\quad I=1,\cdots,N_{\mathcal{O}}\ .
    \end{equation}
An important constraint is that the R-charge of each term must be 2, which requires that the mixing parameters satisfy the relation
\begin{equation} \label{a13}
        R_{\mu} (\mathcal{O}^{(I)}_{\mathbf{n},\mathbf{m}}) =  \mathbf{m} \cdot \boldsymbol{\mu} + \sum_{i|\mathbf{m}\cdot \mathbf{Q}_i = 0} n_i r_i + \sum_{i|\mathbf{m}\cdot \mathbf{Q}_i > 0} (\mathbf{m}\cdot \mathbf{Q}_i) (1-r_i) = 2\ ,\quad \forall I = 1,\cdots,N_{\mathcal{O}},
    \end{equation}
which leaves $N-N_{\mathcal{O}}$ unconstrained mixing parameters. If the theory flows to a rank-0 $\mathcal{N}=4$ theory in the IR, which leaves one independent mixing parameter, we can write\footnote{Throughout this paper, however, there are some cases where $N_{\mathcal{O}}$ is less than $N-1$ or equal to $N$, which results in performing F-maximization with several maximizing parameters for the former case and performing no F-maximization for the latter case. The former includes $A_5, A_7, E_7, E_8$. and the latter happens for various cases including $A_{\text{even}}$.}
    \begin{equation}
        \boldsymbol{\mu} = \boldsymbol{\mu}_0 + \nu \boldsymbol{a} ,\quad \nu \in \mathbb{R},
    \end{equation}
where $\boldsymbol{\mu}_0$ is determined by the equation \eqref{a13} together with the requirement that the F-maximization happens at $\nu=0$, and $\boldsymbol{a}$ is the normalized charge vector of the unbroken $U(1)_A$ symmetry.\footnote{$\boldsymbol{a}$ is normalized so that it is a primitive element in $\mathbb{Z}^{d}$ where $d$ is dimension of boundary VOA.} 
 
The mixing of R- and flavor current
    \begin{equation}
        j^{(R)}_{\mu} \rightarrow j^{(R)}_{\mu} + \boldsymbol{\mu} \cdot \mathbf{j}^{(F)}_{\mu},
    \end{equation}
implies that the flavor fugacities $\mathbf{v}$ and the fluxes $\mathbf{n} = (n_1,\cdots,n_N)$ also get shifted by
    \begin{equation}
        \mathbf{v} \rightarrow \mathbf{v} + \boldsymbol{\mu} v_R,\quad \mathbf{n} \rightarrow \mathbf{n} + \boldsymbol{\mu} n_R,
    \end{equation}
where $n_R$ is an integer that calculates the R-symmetry flux on $\Sigma_g$. The variable $v_R$ should satisfy
    \begin{equation}
        v_R \in \left\{ \begin{array}{l}
            \frac{1}{2} \mathbb{Z},\quad \  p\  \textrm{even} \\
            \mathbb{Z},\quad \quad p\  \textrm{odd}
        \end{array} \right. ,\quad \quad n_R = g-1+v_R p\ .
    \end{equation}
The mixing also shift the flavor fluxes, which can be equivalently described by the shift of the effective dilaton
    \begin{equation}
        \Omega \rightarrow \Omega + \left(\frac{n_R}{g-1}\right) \sum_a \mu_a \frac{\partial \mathcal{W}}{\partial v_a}\ .
    \end{equation}
For the background that we perform the F-maximization, we set $(g,p,v_R,n_R) = (0,1,1,0)$, namely \footnote{For a more detailed explanation on F-maximization, see Appendix A in \cite{Gang:2024loa}.}
    \begin{equation}
        F = -\log |\mathcal{Z}_{\mathcal{M}_{0,1}} (\mathbf{v} = \boldsymbol{\mu}, v_R=1, n_R=0)|\ .
    \end{equation}

\paragraph{Modular data}

The supersymmetric partition functions of a 3d $\CN=4$ rank-zero theory on $\mathcal{M}_{g,p}$ with the topological $A$-and $B$-twist can computed by taking $\nu=-1$ and $\nu=1$ respectively with $v_R=-1/2$
    \begin{equation}
    \begin{split}
        \mathcal{Z}^{A/B\textrm{-twisted}}_{\mathcal{M}_{g,p}} =& \left(\mathcal{Z}_{\mathcal{M}_{g,p}}(\mathbf{v},\mathbf{n};v_R)\Big|_{\mathbf{v}=v_R \boldsymbol{\mu} , \, \mathbf{n} = (g-1+v_R p)\boldsymbol{\mu}} \right)\Big|_{\boldsymbol{\mu}=\boldsymbol{\mu}_0+\nu \boldsymbol{a}}\\
        =& \left(\sum_{\mathbf{u}_*(\mathbf{v};v_R)} \mathcal{H}(\mathbf{u}_*,\mathbf{v})^{g-1} \mathcal{F}(\mathbf{u}_*,\mathbf{v})^p \prod_{a} \Pi^f_a (\mathbf{u}_*,\mathbf{v})^{n_a}\Big|_{\mathbf{v}=v_R \boldsymbol{\mu} , \, \mathbf{n} = (g-1+v_R p)\boldsymbol{\mu}} \right)\Big|_{\boldsymbol{\mu}=\boldsymbol{\mu}_0+\nu \boldsymbol{a}}
    \end{split}
    \end{equation}
On the other hand, the partition function of a finite semisimple 3D TQFT can be written as
    \begin{equation}
        Z^{A/B \textrm{-twisted}}_{\mathcal{M}_{g,p}} = \sum_{\alpha} (S_{0\alpha})^{2-2g} (T_{\alpha \alpha})^{-p}\ ,
    \end{equation}
where $S/T$ are the modular matrices of boundary rational VOA. Comparing the two expressions, we can identify
    \begin{equation}
        S^{-2}_{0\alpha} = \mathcal{H}(\mathbf{u}_{*\alpha},\mathbf{v}) \prod_a \left(\Pi_a^f\right)^{\mu_a},\quad T^{-1}_{\alpha \alpha} = \mathcal{F}(\mathbf{u}_{*\alpha},\mathbf{v}) \prod_a \left(\Pi_a^f\right)^{v_R \mu_a}
    \end{equation}
with $\mathbf{v}=v_R \boldsymbol{\mu} = v_R (\boldsymbol{\mu}_0 + \nu \boldsymbol{a})$.

\paragraph{Superconformal Index}
The superconformal index of 3d $\CN=2$ abelian Chern-Simons matter theory described above is \cite{Dimofte:2011py}
\bea
    \mathcal{I}_{\textrm{SCI}} (q,\mathbf{v},\mu) &= \sum_{\mathbf{m}\in \mathbb{Z}^N} \oint \prod_{a=1}^N \frac{d x_a}{2\pi i x_a} \prod_{a,b=1}^{N} x_a^{K_{ab} m_b} \prod_{a=1}^{N} \left(v_a(-q^{\frac{1}{2}})^{\mu_a} \right)^{m_a} \prod_{i} \mathcal{I}_{\Phi_i}(x,\mathbf{m},r_i) \\
    \mathcal{I}_{\Phi_i}(x,\mathbf{m},r_i) &= \frac{(x^{-1}(\mathbf{Q})q^{1-\frac{r_i}{2}-\frac{\mathbf{Q}\cdot\mathbf{m}}{2}};q)_\infty}{(x(\mathbf{Q})q^{\frac{r_i}{2}-\frac{\mathbf{Q}\cdot\mathbf{m}}{2}};q)_\infty}\; ,
\eea
where we use ${\bf q}$-Pochhammer symbol $(x; {\bf q})_\infty= \prod_{k=0}^\infty (1- x {\bf q}^k)$.

Let $m$ be the rank of the flavor symmetry of the abelian Chern-Simons matter theory. If the gauge theory admits $(m-1)$ gauge invariant half-BPS chiral primary operators, one can perform F-maximization with respect to this symmetry as described in the paragraph above. We then compute the superconformal index at the fixed point, and claim that the theory flows to a rank-zero theory if the following conditions hold:
\begin{itemize}
\item[i)] The superconformal R-charge $R_{{\mu_0}}$ satisfies the condition $R_{{\mu_0}}\in (\frac12 \mathbb{Z})^m$.
\item[ii)] The superconformal index satisfies
\be
\CI_{\text{SCI}} (q,\nu=0,\eta)\neq 1\ , \quad \text{and}\quad  \CI_{\text{SCI}} (q,\nu=\pm 1,\eta=1)= 1\ .
\ee
\end{itemize}
See \cite{Gang:2024loa} for more discussions on these conditions. Note that the existence of the term
\be
-\left(\eta+\eta^{-1}\right) q^{3/2}
\ee
at the superconformal fixed point is a strong signal of a supersymmetry enhancement, because it is a contribution from the extra supercurrent multiplet. However, the existence of this term in the superconformal index is not a necessary or sufficient condition, as there may exist other BPS operators that contribute to the same term possibly with the opposite sign.

\section{Numerical evaluation of the partition functions} \label{app:numerical evaluataion}
In this appendix, we numerically evaluate the absolute values of the ellipsoid partition functions $|S_b^{(M)}|$'s, which can be used to provide a non-trivial consistency check for our proposal. See table \ref{tab: numerical} for a summary of the result.

If the 3d theory that we extract from the trace formula flows to a SCFT with supersymmetry enhancement, it is natural to expect that the corresponding $S_b^{(M)}$ calculates the partition function of the topologically A-twisted theory. We also expect that it is independent of the squashing parameter $b$, except for an overall phase factor $C_b^{s}$ from the gravitational Chern-Simons couplings.
On the other hand, a component of the modular S-matrix $S_{\a\b}$ can be identified with the three-sphere expectation value of a Hopf link composed of the lines that correspond to $\a$ and $\b$ modules. Therefore we expect
\begin{align}
    |\, S_b^{(M)} \,|
    =
    |\, S_{00} \,|_{\text{A-twist}}
    \;,
\end{align}
for all the examples that flow to a superconformal point. As we summarize in table \ref{tab: numerical}, we find strong numerical evidence that this relation indeed holds at $b=1$ once we identify the $\nu=-1$ twist with the topological A-twist.  A slightly non-trivial consistency check is that the $T$-matrices computed for the A-twist (or the T-matrix of the unitary VOA if the 3d theory flows directly to a TFT point), satisfy the relation

\begin{align}
    T^{(M)} = \big( T^{(1)} \big)^M\ ,
\end{align}
in all the examples that we checked, which is motivated by the Galois transformation of the modular data of rational VOAs discussed briefly in section \ref{sec: review}.

\begin{table}[h]
\centering
\begin{tabular}{|c|c|c|c|}
\hline
 Theory & $M$ & $\qquad\big|\, S_{b=1}^{(M)} \, \big|\qquad$ & $| S_{00} |$  \\
\hline
\multirow{4}{*}{$A_2$}
 & 1 & 0.85130 & $\frac{2}{\sqrt{5}}\sin(2\pi/5)\approx$ 0.85065 \\
 \cline{2-4}
& \textcolor{blue}{2} & 0.52576  & $\frac{2}{\sqrt{5}}\sin(\pi/5)\approx$ 0.52573  \\
\cline{2-4}
& \textcolor{blue}{3} & 0.52577  & $\frac{2}{\sqrt{5}}\sin(\pi/5)\approx$ 0.52573  \\
\cline{2-4}
& 4 & 0.85140  & $\frac{2}{\sqrt{5}}\sin(2\pi/5)\approx$ 0.85065  \\
\hline
\multirow{2}{*}{$A_4$}& 1 & 0.58951 & $\frac{2}{\sqrt{7}}\cos(3\pi/14)\approx$ 0.59101 \\
 \cline{2-4}
& 2 & 0.73799 & $\frac{2}{\sqrt{7}}\cos(\pi/14)\approx$ 0.73698 \\
 \cline{2-4}
\hline
\multirow{2}{*}{$A_6$}& 1 & 0.42857 & $\frac{2}{3}\sin(2\pi/9)\approx$ 0.42853 \\
 \cline{2-4}
 & 2 & 0.65735 & $\frac{2}{3}\sin(\pi/9) + \frac{2}{3} \sin(2\pi/9)\approx$ 0.65654 \\
\hline
$A_8$& 4 & 0.45569 & $\frac{2}{\sqrt{11}}\cos(5\pi/22)\approx$ 0.45573 \\
\hline
\multirow{2}{*}{$A_7$}& \textcolor{blue}{1} & 0.52577 & $\frac{2}{\sqrt{5}}\sin(\pi/5)\approx$ 0.52573 \\
 \cline{2-4}
 & 2 & 0.85134 & $\frac{2}{\sqrt{5}}\sin(2\pi/5)\approx$ 0.85065 \\
\hline
$E_6$ & 1 & 0.47101 & $\frac{2}{\sqrt{7}}\sin(3\pi/14) \approx$ 0.47131 \\
\hline
\multirow{2}{*}{$E_7$}& 1 & 0.38076 & $\frac{2}{5}\sin(2\pi/5)\approx$ 0.38042 \\
 \cline{2-4}
 & \textcolor{blue}{2} & 0.23513 & $\frac{2}{5}\sin(\pi/5)\approx$ 0.23511 \\
\hline
\end{tabular}
\caption{\label{tab: numerical} Numerical evaluations of $|S_{b=1}^{(M)}|$ and $|S_{00}|$ of the A-twisted theory. We compute for all the cases that the integral involves equal or less than 3 integration variables. The blue colors denote the cases where the $3d$ UV theory directly flows to unitary TFT upon monopole superpotential deformations. The results for the $D_n$ and $A_5$ examples are spelled out in the main text.}
\end{table}

\section{Useful identities of Faddeev dilogarithm} \label{app: identity}

In this appendix, we will summarize the conventions of multi-particle quantum mechanics we used to evaluate the trace formula. We will also prove the interesting identities 
\be
\left(\prod_{i=1}^{2N} e^{i\pi x_{\gamma_i}^2} \right)^{4N+2} = {\bf 1} \quad \textrm{and} \quad \quad \left(\prod_{i=1}^{2N+1} e^{i\pi x_{\gamma_i}^2} \right)^{2N+2} = {\bf 1} \ 
\ee
separately for $(A_1,A_{2N})$ and $(A_1,A_{2N+1})$-theories up to some constant phases. The proof of identities for $(A_1,D_*)$-type can be done similarly and we will not repeat them here.

\subsection*{Single-particle quantum mechanics}
Let us begin with a review of the single-particle quantum mechanics and fix conventions. The canonical commutator between position operator $x$ and momentum operator $p$ is
    \begin{equation}
        [p,x] = \frac{1}{2\pi i},
    \end{equation}
and one can also identify
    \begin{equation}
        \quad p \leftrightarrow \frac{1}{2\pi i} \partial_{x}\quad \textrm{or} \quad x \leftrightarrow -\frac{1}{2\pi i} \partial_p,
    \end{equation}
when acting on the square-integrable functions in $L^2(x)$ or $L^2(p)$. The position and momentum eigenstates $|x\rangle$ and $|p\rangle$ are normalized according to
    \begin{equation}
        \langle x' | x\rangle = \delta(x'-x),\quad \langle p' | p\rangle = \delta(p'-p),\quad \langle x | p\rangle = e^{2\pi i p x},
    \end{equation}
and they are related by Fourier transformations
    \begin{equation}
        |x\rangle = \int dp |p\rangle \langle p|x\rangle = \int dp e^{-2\pi i p x} |p\rangle,\quad |p\rangle = \int dx |x\rangle \langle x|p\rangle = \int dx e^{2\pi i p x}|x\rangle,
    \end{equation}
where we used the completion relations
    \begin{equation}
         \int dx | x\rangle \langle x| = \int dp |p\rangle \langle p| = \mathbf{1}.
    \end{equation}

In the trace formula, we also encounter the exponential operators like $e^{i\pi x^2}$ and $e^{i\pi p^2}$. The following properties turn out to be useful in the computation
    \begin{equation}
        e^{i\pi x^2} p = (-x+p) e^{i\pi x^2},\quad  e^{i \pi p^2} x = (x + p) e^{i \pi p^2}.
    \end{equation}

\subsection*{Quantum mechanics system associated with Argyres-Douglas theories}
We then turn to the Argyres-Douglas theories $(A_1,G)$ with $G$ an ADE-type group. In the canonical chamber, the BPS quiver is labeled by the quiver nodes with charge $\gamma_1,\cdots,\gamma_{\textrm{rank}(G)}$. For each $\gamma_a$ we assign a quantum mechanics variable $x_{\gamma_a}$ and the canonical commutation relations are defined by
    \begin{equation}
        \left[ x_{\gamma_a} , x_{\gamma_b} \right] = \frac{\langle \gamma_a,\gamma_b \rangle}{2\pi i},\quad a,b=1,\cdots,\textrm{rank}(G).
    \end{equation}
and they satisfy
    \begin{equation}
        x_{\gamma+\gamma'} = x_{\gamma} + x_{\gamma'}.
    \end{equation}
It is convenient to change a basis and rewrite the variables $x_{\gamma}$ in terms of $\{x\}$ and $\{p\}$ operators in the multi-particle quantum mechanics system and we will illustrate the procedure with $(A_1,A_*)$ theory. Consider the $N$-particles quantum mechanics systems described by the position/momentum operators $x_1,\cdots,x_N$ and $p_1,\cdots,p_N$ with the canonical commutators
    \begin{equation}
        \left[ p_i , x_j \right] = \frac{\delta_{ij}}{2\pi i}.\quad (i,j=1,\cdots,N)
    \end{equation}
For $(A_1,A_{2N})$-theory, we can write the variables $x_{\gamma}$ in terms of $\{x\}$ and $\{p\}$ operators according to
    \begin{equation}
        x_{\gamma_{2i}} = x_i\quad (n=1,\cdots,N)
    \end{equation}
and 
    \begin{equation}
        x_{\gamma_{2i+1}} = p_i + p_{i-1}\quad (n=2,\cdots,N)\quad \textrm{and} \quad x_{\gamma_1} = p_1. 
    \end{equation}
As an example, for $(A_1,A_6)$ theory we have the following assignment.
\begin{figure}[!h]
  \centering
  \begin{tikzpicture}
			\node[W] (1) at (0,0){$\gamma_1$};
		 	\node[W] (2) at (2,0) {$\gamma_2$};
            \node[W] (3) at (4,0) {$\gamma_3$};
            \node[W] (4) at (6,0) {$\gamma_4$};
            \node[W] (5) at (8,0) {$\gamma_5$};
            \node[W] (6) at (10,0) {$\gamma_6$};
			\node[W] at (0,-1){$p_1$};
		 	\node[W] at (2,-1) {$x_1$};
            \node[W] at (4,-1) {$p_2 + p_1$};
            \node[W] at (6,-1) {$x_2$};
            \node[W] at (8,-1) {$p_3+p_2$};
            \node[W] at (10,-1) {$x_3$};
			\draw[->] (1)--(2);
            \draw[<-] (2)--(3);
            \draw[->] (3)--(4);
            \draw[<-] (4)--(5);
            \draw[->] (5)--(6);
  \end{tikzpicture}
  \end{figure}
  
For $(A_1,A_{2N+1})$-theory, we need to add the $(2N+1)$-th variables $x_{\gamma_{2N+1}}$ as
    \begin{equation}
        x_{\gamma_{2N+1}} = p_{N} + m,
    \end{equation}
where $m$ is a central charge that commutes with position and momentum operators.

In this paper, we use the following abbreviation
\be
(a) := \Phi_b(x_{\gamma_a})\ ,\quad e^a :=e^{i\pi x_{\gamma_a}^2} \ .
\ee
Based on the derivations in the single-particle case, we can check the following properties of the exponential operators
\begin{equation}
    e^a x_{\gamma} = x_{\gamma + \langle \gamma_a , \gamma\rangle \gamma_a} e^a,
\end{equation}
and in particular, one has
\begin{equation}
    e^a (b) = (b + \langle \gamma_a,\gamma_b \rangle a) e^a, \quad e^a e^b = e^{b + \langle \gamma_a,\gamma_b \rangle a} e^a,
\end{equation}
where $(b+\langle \gamma_a , \gamma_b \rangle a)$ and $e^{b+\langle \gamma_a , \gamma_b \rangle a}$ are understood as $\Phi_b(x_{\gamma_b+\langle \gamma_a , \gamma_b \rangle \gamma_a})$ and $e^{i\pi x^2_{\gamma_b+\langle \gamma_a , \gamma_b \rangle \gamma_a}}$.

\subsection*{Proof of the identity for $(A_1,A_{2N})$}
We will prove the following identity for the $(A_1,A_{2N})$ theory
    \begin{equation}
        \hat{P}_{N}^{4N+2} \propto \mathbf{1}, \quad \textrm{with}\quad \hat{P}_N := \prod_{a=1}^{2N} e^a
    \end{equation}
which is proportional to the identity operator acting on the $N$-particles Hilbert space. The proof is split into two steps. First, we will show $\hat{P}^{4N+2}_N$ commute with any operator $x_{\gamma_a}$, namely
    \begin{equation}
        \hat{P}^{4N+2}_N x_{\gamma_a} = x_{\gamma_a} \hat{P}^{4N+2}_N. \quad (a=1,\cdots,2N)
    \end{equation}
Second, we will argue this is enough to conclude $\hat{P}^{4N+2}_{N} \propto 1$ in the $N$-particle quantum mechanics system.

Consider a generic $\gamma = \sum_a v_a \gamma_a$ and the corresponding operator $\sum_{a} v_a x_{\gamma_a}$. Define the $2N\times 2N$ matrix $P_{N}$ as the representation of $\hat{P}_N$ in terms of the operator basis $x_{\gamma_a}$, namely
    \begin{equation}
        \hat{P}_N x_{\gamma_a} =  \sum_b (P_N)_{ab}x_{\gamma_b} \hat{P}_N.
    \end{equation}
Equivalently, it will map the vector $\mathbf{v} = (v_1,\dots,v_{2N})$ according to
    \begin{equation}
        \mathbf{v} \rightarrow \mathbf{v} \cdot P_N = (\sum_{a} (P_N)_{a,1} v_a,\cdots,\sum_{a} (P_N)_{a,2N} v_a).
    \end{equation}
For example, when $N=1$, $\hat{P}_1 = e^{i\pi p^2} e^{i\pi x^2}$ and it is easy to obtain
    \begin{equation}
        \hat{P}_1 p = - x \hat{P}_1, \quad \hat{P}_1 x = (p + x) \hat{P}_1,
    \end{equation}
and one has
    \begin{equation}
        P_1 = \left(\begin{array}{cc}
            0 & -1 \\
            1 & 1
        \end{array} \right) \quad \Rightarrow \quad P_1^{6} = 1.
    \end{equation}

Instead of showing $P_{N}^{4N+2} = \mathbf{1}_{(2N)\times(2N)}$ directly, we will prove the eigenvalue equation of $P_N$ is
    \begin{equation}\label{App:identity-eigenvalue-equations}
        \det \left(P_{N} - \lambda \mathbf{1}_{(2N)\times(2N)} \right) = \sum_{a=0}^{2N} (-1)^a \lambda^a = 0,
    \end{equation}
and the eigenvalues of $P_N$ are all different and are solved by
    \begin{equation}
        \lambda_n = e^{\frac{2\pi i}{2N+1} (n+\frac{1}{2}) },\quad n=0,1,\cdots,2N-1.
    \end{equation}
Therefore, one can diagonalize $P_N$ using an invertible matrix $S_N$
    \begin{equation}
        P_N = S^{-1}_N \Lambda_N S_N,
    \end{equation}
with $\Lambda_N = \textrm{diag}(\lambda_0,\cdots,\lambda_{2N-1})$. Then it is straightforward to show
    \begin{equation}
        P_N^{4N+2} = S_N^{-1} \Lambda_N^{4N+2} S_N = \mathbf{1}.
    \end{equation}

When $N=1$ one can check
    \begin{equation}
        \det(P_1 - \lambda \mathbf{1}_{2\times 2}) = \lambda^2 - \lambda +1.
    \end{equation}
For higher $N$, let us build the matrix $P_{N+1}$ using $P_N$. To do this, we split the operator $\hat{P}_{N+1}$ as
    \begin{equation}
        \hat{P}_{N+1} = \left(\sum_{a=1}^{2N} e^{a}\right) \times \left(e^{2N+1} e^{2N+2}\right),
    \end{equation}
and the $P_{N+1}$ matrix can be written as
    \begin{equation}
        P_{N+1} = \widetilde{P}_1 \widetilde{P}_{N}
    \end{equation}
where both $\widetilde{P}_{N}$ and $\widetilde{P}_1$ are $(2N+2)\times (2N+2)$ matrices built as
    \begin{equation}
        \widetilde{P}_{1} = \left(\begin{array}{c|cc}
            \mathbf{1}_{2N\times 2N}& \mathbf{\delta}^T_{2N} & \mathbf{0}^T_{2N} \\ \hline
            \mathbf{0}_{2N} & 0&-1\\
            \mathbf{0}_{2N} & 1 & 1
        \end{array} \right),\quad \widetilde{P}_N = \left(\begin{array}{c|cc}
            P_N & \mathbf{0}_{2N}^T & \mathbf{0}_{2N}^T  \\ \hline
             - \mathfrak{p}_{2N}& 1 & 0\\
             \mathbf{0}_{2N} & 0 & 1
        \end{array} \right).
    \end{equation}
Here we introduce the $2N$-vectors $\mathbf{0}_{2N} = (0,\cdots,0)$, $\delta_{2N} = (0,\cdots,0,1)$ and $\mathfrak{p}_{2N}$ is the last line in $P_N$
\begin{equation}
    \mathfrak{p}_{2N} = ((P_N)_{2N,1} , \cdots, (P_N)_{2N,2N}).
\end{equation}

Therefore $P_{N+1}$ can be written as
    \begin{equation}
        P_{N+1} = \left( \begin{array}{c|cc}
            P'_N & \delta^T_{2N} & \mathbf{0}^T_{2N}\\ \hline
            \mathbf{0}_{2N} & 0&-1\\
            -\mathfrak{p}_{2N} & 1&1
        \end{array} \right),\quad \textrm{with} \quad         P'_N = P_N - \left(\begin{array}{c}
            \mathbf{0}_{2N}\\
            \vdots\\
            \mathbf{0}_{2N}\\
            \mathfrak{p}_{2N}
        \end{array} \right),
    \end{equation}
where $P'_N$ is the $P_N$ matrix with the last line erased. To evaluate the eigenvalues of $P_{N+1}$, let us consider $\det (P_{N+1} - \lambda \mathbf{1}_{(2N+2) \times (2N+2)})$
    \begin{equation}
        \det \left( \begin{array}{c|cc}
            P'_N - \lambda \mathbf{1}_{2N \times 2N} & \delta^T_{2N} & \mathbf{0}^T_{2N}\\ \hline
            \mathbf{0}_{2N} & -\lambda&-1\\
            -\mathfrak{p}_{2N} & 1&1-\lambda
        \end{array} \right) = \det \left( \begin{array}{c|cc}
            P'_N - \lambda \mathbf{1}_{2N \times 2N} & \delta^T_{2N} & \mathbf{0}^T_{2N}\\ \hline
            \mathbf{0}_{2N} & 0&-1\\
            -\mathfrak{p}_{2N} & 1-\lambda+\lambda^2&1-\lambda
        \end{array} \right).
    \end{equation}
It is then written as
    \begin{equation}
        \det \left(\begin{array}{c|c}       
             P'_N - \lambda \mathbf{1}_{2N \times 2N} & \delta^T_{2N}\\
              -\mathfrak{p}_{2N} & 1-\lambda+\lambda^2
        \end{array} \right) = (\lambda^2-\lambda+1) \det \left(  P'_N - \lambda \mathbf{1}_{2N \times 2N} \right) + \det\left(P_N - \lambda \mathbf{1}'_{2N \times 2N} \right)
    \end{equation}
where $\mathbf{1}'_{2N \times 2N} = \textrm{diag}(1,1,\cdots,1,0)$ is the identity matrix with the last element set to zero. Both factors can be calculated inductively. For the first factor, it is straightforward to check 
    \begin{equation}
        \det\left(P'_N - \lambda \mathbf{1}_{2N \times 2N} \right) = \lambda^2 \det\left(P'_{N-1} - \lambda \mathbf{1}_{2(N-1)\times 2(N-1)} \right).
    \end{equation}
Using the fact
    \begin{equation}
        P'_1 = \left(\begin{array}{cc}
            0 & -1 \\
            0 & 0
        \end{array} \right) \quad \Rightarrow \quad \det\left(P'_1 - \lambda \mathbf{1}_{2 \times 2} \right) = \lambda^2,
    \end{equation}
we conclude
    \begin{equation}
        \det\left(P'_N - \lambda \mathbf{1}_{2N \times 2N} \right) = \lambda^{2N}.
    \end{equation}
The second factor $\det\left(P_N - \lambda \mathbf{1}'_{2N \times 2N} \right)$ is
    \begin{equation}
        \det \left( \begin{array}{c|cc}
            P'_{N-1} - \lambda \mathbf{1}_{2(N-1) \times 2(N-1)} & \delta^T_{2(N-1)} & \mathbf{0}^T_{2(N-1)}\\ \hline
            \mathbf{0}_{2(N-1)} & -\lambda&-1\\
            -\mathfrak{p}_{2(N-1)} & 1&1
        \end{array} \right) = \det \left( \begin{array}{c|cc}
            P'_{N-1} - \lambda \mathbf{1}_{2(N-1) \times 2(N-1)} & \delta^T_{2(N-1)} & \mathbf{0}^T_{2(N-1)}\\ \hline
            \mathbf{0}_{2(N-1)} & 0&-1\\
            -\mathfrak{p}_{2(N-1)} & 1-\lambda&1
        \end{array} \right)
    \end{equation}
which can be further written as
    \begin{equation}
        \det \left(\begin{array}{c|c}       
             P'_{N-1} - \lambda \mathbf{1}_{2(N-1) \times 2(N-1)} & \delta^T_{2(N-1)}\\ \hline
              -\mathfrak{p}_{2(N-1)} & 1-\lambda
        \end{array} \right) = (1-\lambda) \det \left(  P'_{N-1} - \lambda \mathbf{1}_{2(N-1) \times 2(N-1)} \right) + \det\left(P_{N-1} - \lambda \mathbf{1}'_{2(N-1) \times 2(N-1)} \right)
    \end{equation}
and we obtain another recursion relation
    \begin{equation}
        \det\left(P_N - \lambda \mathbf{1}'_{2N \times 2N} \right) = \det\left(P_{N-1} - \lambda \mathbf{1}'_{2(N-1) \times 2(N-1)} \right) + \lambda^{2N-2} - \lambda^{2N-1}.
    \end{equation}
Using $\det\left(P_1 - \lambda \mathbf{1}'_{2 \times 2} \right) = 1-\lambda$, one has
    \begin{equation}
        \det\left(P_N - \lambda \mathbf{1}'_{2N \times 2N} \right) = \sum_{a=0}^{2N-1} (-1)^a \lambda^a.
    \end{equation}
Combined everything, we can prove
    \begin{equation}
        \det (P_{N+1} - \lambda \mathbf{1}_{(2N+2) \times (2N+2)}) = (\lambda^2-\lambda+1) \lambda^{2N} + \sum_{a=0}^{2N-1} (-1)^a \lambda^a = \sum_{a=0}^{2N+2} (-1)^a \lambda^{2N+2},
    \end{equation}
Which is the desired result.

In the next step, we will argue if an operator $\hat{\mathcal{O}}$ commute with all $x_{\gamma_a}$, or equivalently all position/momentum operators $\{x\}$ and $\{p\}$, it is then proportional to identity operator. It is enough to work with single-particle quantum mechanics and assume 
    \begin{equation}
        \hat{\mathcal{O}} x = x \hat{\mathcal{O}},\quad \hat{\mathcal{O}} p = p \hat{\mathcal{O}}.
    \end{equation}
Working with the eigenstate $|x\rangle$, it implies $\hat{\mathcal{O}}|x\rangle$ is still an eigenstate of $x$ and we can write
    \begin{equation}
        \hat{\mathcal{O}} |x\rangle = f(x) |x\rangle,
    \end{equation}
up to some function $f(x)$. Similarly, one has
    \begin{equation}
        \langle p| \hat{\mathcal{O}} = g(p) \langle p|
    \end{equation}
up to some function $g(p)$. They will imply
    \begin{equation}
        \langle p | \mathcal{\hat{O}} |x\rangle = f(x) \langle p|x\rangle = g(p) \langle p|x\rangle
    \end{equation}
which holds for all $x,p \in \mathbb{R}$. The only possibility is
    \begin{equation}
        f(x) = g(p) = c,
    \end{equation}
where $c$ is some constant. Therefore we can expand
    \begin{equation}
        \hat{\mathcal{O}} = \int dp dp' |p\rangle \langle p| \hat{\mathcal{O}} |p'\rangle \langle p'| = c \int dp |p\rangle \langle p| = c \mathbf{1},
    \end{equation}
which must be proportional to identity. The multi-particle case is a straightforward generalization.

\subsection*{Proof of the identity for $(A_1,A_{2N+1})$}
We then turn to the identity for the $(A_1,A_{2N+1})$ theory
    \begin{equation}
        \hat{P}_{N}^{2N+2} \propto \mathbf{1}, \quad \textrm{with}\quad \hat{P}_N := \prod_{a=1}^{2N+1} e^a
    \end{equation}
and it is also sufficient to show $\hat{P}^{2N+2}_N$ commutes with any operator $x_{\gamma_a}$, namely
    \begin{equation}
        \hat{P}^{2N+2}_N x_{\gamma_a} = x_{\gamma_a} \hat{P}^{2N+2}_N. \quad (a=1,\cdots,2N+1).
    \end{equation}
The proof is basically the same as the $(A_1,A_{2N})$ case. To begin with, when $N=1,\ \hat{P}_1 = e^{i\pi p^2} e^{i\pi x^2} e^{i \pi (p+m)^2}$ where $m$ is a central charge commuted with both $p$ and $x$. One can check
    \begin{equation}
        \hat{P}_1 p = - x \hat{P}_1,\quad \hat{P}_1 x = (p + m)\hat{P}_1,\quad \hat{P}_1 m = m\hat{P}_1,
    \end{equation}
so that one has
    \begin{equation}
        P_1 = \left(\begin{array}{ccc}
            0 & -1 & 0 \\
            1 & 0  & 1\\
            0 & 0 & 1
        \end{array} \right)\quad \Rightarrow \quad P_1^4 = 1,
    \end{equation}
with the basis $(p,x,m)$. For general $P_N$ we will prove the eigenvalue equations of $P_N$ is
    \begin{equation}\label{App:identity-eigenvalue-equations-2}
        \det \left(P_{N} - \lambda \mathbf{1}_{(2N+1)\times(2N+1)} \right) = \sum_{a=0}^{2N+1} (-1)^a \lambda^a = 0,
    \end{equation}
and the eigenvalues of $P_N$ are all different and are solved by
    \begin{equation}
        \lambda_n = e^{\frac{2\pi i n}{2N+2}},\quad n=0,1,\cdots,N-1,N+1,\cdots,2N+1.
    \end{equation}
Then we can use the same trick to show $P_N^{2N+2}=1$.

When $N=1$ one can check
    \begin{equation}
        \det(P_1 - \lambda \mathbf{1}_{2\times 2}) = -\lambda^3+\lambda^2 - \lambda +1.
    \end{equation}
For higher $N$ we split the operator $\hat{P}_{N+1}$ as
    \begin{equation}
        \hat{P}_{N+1} = \left(\sum_{a=1}^{2N} e^{a}\right) \times \left(e^{2N+1} e^{2N+2}e^{2N+3}\right),
    \end{equation}
and the $P_{N+1}$ matrix can be written accordingly as
    \begin{equation}
        P_{N+1} = \widetilde{P}_1 \widetilde{P}_{N}
    \end{equation}
where both $\widetilde{P}_{N}$ and $\widetilde{P}_1$ are $(2N+3)\times (2N+3)$ matrices built as
    \begin{equation}
        \widetilde{P}_{1} = \left(\begin{array}{c|ccc}
            \mathbf{1}_{2N\times 2N}& \mathbf{\delta}^T_{2N} & \mathbf{0}^T_{2N} & \mathbf{0}^T_{2N} \\ \hline
            \mathbf{0}_{2N} & 0&-1&0\\
            \mathbf{0}_{2N} & 1 & 0 & 1\\
            \mathbf{0}_{2N} & 0 & 0 & 1
        \end{array} \right),\quad \widetilde{P}_N = \left(\begin{array}{c|ccc}
            P_N & \mathbf{0}_{2N}^T & \mathbf{0}_{2N}^T &\mathbf{0}_{2N}^T \\ \hline
             - \mathfrak{p}_{2N}& 1 & 0 & 0\\
             \mathbf{0}_{2N} & 0 & 1 & 0\\
             \mathbf{0}_{2N} & 0 & 0 & 1
        \end{array} \right),
    \end{equation}
where the $2N$-vectors $\mathbf{0}_{2N},\delta_{2N}$ and $\mathfrak{p}_{2N}$ are defined as before. Therefore $P_{N+1}$ can be written as
    \begin{equation}
        P_{N+1} = \left( \begin{array}{c|ccc}
            P'_N & \delta^T_{2N} & \mathbf{0}^T_{2N} & \mathbf{0}^T_{2N}\\ \hline
            \mathbf{0}_{2N} & 0&-1&0\\
            -\mathfrak{p}_{2N} & 1&0&1\\
            \mathbf{0}_{2N}&0&0&1
        \end{array} \right),\quad \textrm{with} \quad         P'_N = P_N - \left(\begin{array}{c}
            \mathbf{0}_{2N}\\
            \vdots\\
            \mathbf{0}_{2N}\\
            \mathfrak{p}_{2N}
        \end{array} \right).
    \end{equation}
where $P'_N$ is the $P_N$ matrix with the last line erased. To evaluate the eigenvalues of $P_{N+1}$, let us consider $\det (P_{N+1} - \lambda \mathbf{1}_{(2N+3) \times (2N+3)})$
    \begin{equation}
        \det \left( \begin{array}{c|ccc}
            P'_N - \lambda \mathbf{1}_{2N \times 2N} & \delta^T_{2N} & \mathbf{0}^T_{2N}&\mathbf{0}^T_{2N} \\ \hline
            \mathbf{0}_{2N} & -\lambda&-1&0\\
            -\mathfrak{p}_{2N} & 1&-\lambda&1\\
            \mathbf{0}_{2N} & 0 & 0 & 1-\lambda
        \end{array} \right) = (1-\lambda)\det \left( \begin{array}{c|cc}
            P'_N - \lambda \mathbf{1}_{2N \times 2N} & \delta^T_{2N} & \mathbf{0}^T_{2N}\\ \hline
            \mathbf{0}_{2N} & 0&-1\\
            -\mathfrak{p}_{2N} & 1+\lambda^2&-\lambda
        \end{array} \right).
    \end{equation} 
which is then evaluated as
    \begin{equation}
        (1-\lambda)(1+\lambda^2) \det \left(  P'_N - \lambda \mathbf{1}_{2N \times 2N} \right) + \det\left(P_N - \lambda \mathbf{1}'_{2N \times 2N} \right),
    \end{equation}
where we have
    \begin{equation}
        \det\left(P'_N - \lambda \mathbf{1}_{2N \times 2N} \right) = \lambda^{2N},\quad \det\left(P_N - \lambda \mathbf{1}'_{2N \times 2N} \right) = \sum_{a=0}^{2N-1} (-1)^a \lambda^a,
    \end{equation}
form the discussion before. We then prove that
    \begin{equation}
        \det \left(P_{N+1} - \lambda \mathbf{1}_{(2N+3)\times(2N+3)} \right) = (1-\lambda)(1+\lambda^2) \lambda^{2N} + \sum_{a=0}^{2N-1} (-1)^a \lambda^a = \sum_{a=0}^{2N+3} (-1)^a \lambda^a.
    \end{equation}

\section{Summary of the modular data} \label{app:modular data}

In this appendix, we summarize the modular data of VOAs that appear in the main text.

\paragraph{$\bm{M(2,2k+3)}$} The first class of examples is the Virasoro minimal model $M(2,2k+3)$ with central charge
\be
c = -\frac{2 k (6 k+5)}{2 k+3}\ .
\ee
There are $k+1$ simple modules labeled by $\alpha=0,\cdots, k$, with conformal weights 
\be
h_\alpha = \frac{\a (\a - 2 k-1)}{4k +6 }\ .
\ee
The modular data are 
\begin{align}
\begin{split}
&S_{\alpha \beta} = \frac{2(-1)^{k+\alpha +\beta}}{\sqrt{2k+3}}  \sin \left( \frac{2 \pi (\a+1)(\b+1)}{2k+3} \right)\;,
\\
&T_{\alpha \beta} = \delta_{\alpha, \beta} \exp \left[2\pi i \left(h_\alpha - \frac{c}{24}\right)\right]\;.
\end{split} \label{modular}
\end{align}

\paragraph{Affine osp$\bm{(1|2)}$ at positive integer levels} The structure of simple modules and their modular data are summarized in \cite{creutzig2018representation, Creutzig:2018ogj, Ferrari:2023fez}. The central charge of affine $\mathrm{osp}(1|2)$ at level $k$ is
\be
c = \frac{2k}{2k+3}\ .
\ee
The modular data can be understood as a permutation of those of $M(2,2k+3)$ discussed above.
\be
	S = P S_{M(2,2k+3)} P^{-1} \qquad T = P T_{M(2,2k+3)} P^{-1} \qquad P = \begin{pmatrix}
		0 & 0 & \dots & (-1)^{k}\\
		\vdots & \vdots & \ddots & \vdots\\
		0 & -1 & \dots & 0\\
		1 & 0 & \dots & 0\\
	\end{pmatrix}\ .
\ee

\paragraph{Affine su(2) at admissible levels} The modular data for affine $\mathrm{su}(2)$ at level $k$ for
    \begin{equation}
        k = \frac{t}{u},\quad u \in \mathbb{N},\quad t\in \mathbb{Z}/\{0\},\quad \gcd(t,u)=1\ ,
    \end{equation}
are summarized in \cite{DiFrancesco:1997nk}.
The highest weight $\lambda$ can be decomposed into two integral weights $\lambda^I$ and $\lambda^F$
    \begin{equation}
        \lambda = \lambda^I - (k+2)\lambda^F
    \end{equation}
with level $k^I$ and $k^F$ separately
    \begin{equation}
        k^I = u(k+2),\quad k^F = u-1,\quad k^I,k^F\in \mathbb{N}.
    \end{equation}
We also require $t \geq 2-2u$ in order for $k^I$ to be non-negative. The admissible highest weight $\lambda$ is labeled by the $(\lambda^I,\lambda^F)$ pairs with
    \begin{equation}
        \lambda^I=0,1,\cdots ,k^I,\quad \lambda^F=0,1,\cdots ,k^F,
    \end{equation}
and there are $(k^I+1)(k^F+1)$ of them in total. The conformal weights and the central charge are given by
    \begin{equation}
        h_{\lambda} = \frac{\lambda(\lambda+2)}{4(k+2)},\quad c\left[\mathrm{su}(2)_k \right] = \frac{3k}{k+2}\ .
    \end{equation}
The modular data are given by
    \begin{equation}
        \begin{split}
        &S_{\lambda \mu} = \sqrt{\frac{2}{u^2(k+2)}} (-1)^{\mu^F(\lambda^I+1)+\lambda^F(\mu^I+1)} e^{-i \pi \lambda^F \mu^F (k+2)} \sin\left(\frac{\pi(\lambda^I+1)(\mu^I+1)}{k+2}\right)\nonumber\\
        &T_{\lambda \mu} = \delta_{\lambda,\mu} \exp\left(2\pi i(h_{\lambda} - \frac{c}{24}) \right)
        \end{split}
    \end{equation}
with $\lambda,\mu$ admissible highest weights given above.

\paragraph{$\bm{(A_1,k)_{\frac12}}$}  It is a class of modular data that appears in the $A_{2N}$ family. See \cite{Rowell:2007dge,Schoutens:2015uia,Harvey:2019qzs} for the notation. For the 3d bulk topological theory, we consider the pure $SU(2)_k$ CS theory for odd $k$, which has a $\mathbb{Z}^{[1]}_2$ anomalous one-form symmetry generated by the spin $k/4$ line. In order to cancel the anomaly, we consider a tensor product with $U(1)_{\pm2}$ and gauging the non-anomalous one-form symmetry. More precisely, we consider
\be
\left[SU(2)_k\otimes U(1)_2\right]/\mathbb{Z}_2^{[1]}
\ee
for $k=3$ (mod 4) or
\be
\left[SU(2)_k\otimes U(1)_{-2}\right]/\mathbb{Z}_2^{[1]}
\ee
for $k=1$ (mod 4). We denote $(A_1,k)_{\frac12}$ by the modular data associated with these unitary TFTs. Accordingly, the central charge of the boundary algebra is
\bea
c[(A_1,k)_{\frac12}]
&=c[(A_1,k)]+(-1)^{(k+1)/2}c[(A_1,1)] \\
&=\frac{3k}{k+2}+(-1)^{(k+1)/2} \ .
\eea
The resulting model does not have a twisted sector, and the modular data can be obtained by reducing the S-matrix of the tensor product theory $SU(2)_k\otimes U(1)_{\pm2}$. See e.g., \cite{Hsin:2018vcg,Gang:2021hrd}.
It is possible to show that the modular data of the reduced theory can be written as
\begin{equation}
    \begin{split}
        &S_{\alpha \beta} = 2(-1)^{\alpha \beta}\sqrt{\frac{1}{k+2}} \sin\left(\frac{\pi(\alpha+1)(\beta+1)}{k+2} \right)\\
        &T_{\alpha \beta} = \delta_{\alpha,\beta} \exp\left[ 2\pi i \left(\frac{\alpha(\alpha+2)}{4(k+2)}+(-1)^{\frac{k+1}{2}}\frac{\alpha^2}{4} - \frac{c[(A_1,k)_{\frac12}]}{24} \right)\right]
    \end{split}
\end{equation}
with $\alpha=0,\cdots,\frac{k-1}{2}$.

\paragraph{$\bm{W_3(3,7)}$} The module structure and the modular data of $W_N(p,q)$ can be found in \cite{Gannon:1992ty, Beltaos:2010ka}. For $W(3,7)$, we have \footnote{The $S$ and $T$ matrices we write in \eqref{Modular matrices of W_3(3,7)} and \eqref{Modular matrices of W_3(3,8)} are slightly different from those in the references, by an overall phase.}
\bea \label{Modular matrices of W_3(3,7)}
S &=\frac{2}{\sqrt{7}}\left(
\begin{array}{ccccc}
 -\sin\frac{3\pi}{14} & \sin\frac{\pi}{6} & \sin\frac{\pi}{14} &-\sin\frac{5\pi}{14} & \sin\frac{\pi}{6} \\
 
 \sin\frac{\pi}{6} & \frac{1}{4}(-1+i\sqrt{7}) &  \sin\frac{\pi}{6} &  -\sin\frac{\pi}{6} & -\frac{1}{4}(1+i\sqrt{7}) \\
 
 \sin\frac{\pi}{14} & \sin\frac{\pi}{6} & \sin\frac{5\pi}{14} & \sin\frac{3\pi}{14} & \sin\frac{\pi}{6} \\
 
 -\sin\frac{5\pi}{14} & -\sin\frac{\pi}{6} & \sin\frac{3\pi}{14} & \sin\frac{\pi}{14} & -\sin\frac{\pi}{6} \\
 
 \sin\frac{\pi}{6} & -\frac{1}{4}(1+i\sqrt{7}) & \sin\frac{\pi}{6} & -\sin\frac{\pi}{6} & \frac{1}{4}(-1+i\sqrt{7}) \\
 
\end{array}
\right) \ ,\\ 
T &=\left(
\begin{array}{ccccc}
 e^{2\pi i (19/28)} & 0 & 0 & 0 & 0 \\
 0 & e^{2\pi i (7/28)} & 0 & 0 & 0 \\
 0 & 0 & e^{2\pi i (27/28)} & 0 & 0 \\
 0 & 0 & 0 & e^{2\pi i (3/28)} & 0 \\
 0 & 0 & 0 & 0 & e^{2\pi i (7/28)} \\
\end{array}
\right) \ ,
\eea
with central charge $c=-114/7$.

\paragraph{$\bm{W_3(3,8)}$} The modular data of $W_3(3,8)$ is 
\bea \label{Modular matrices of W_3(3,8)}
S&=\frac{1}{4}\left(
\begin{array}{ccccccc}
 \sqrt{2}+1 & \sqrt{2} & -2 & \sqrt{2}-1 & \sqrt{2} & -1 & -1 \\
 \sqrt{2} & -\sqrt{2}+i \sqrt{2} & 0 & \sqrt{2} & -\sqrt{2}-i \sqrt{2} & -i \sqrt{2} & i \sqrt{2} \\
 -2 & 0 & 0 & 2 & 0 & -2 & -2 \\
 \sqrt{2}-1 & \sqrt{2} & 2 & \sqrt{2}+1 & \sqrt{2} & 1 & 1 \\
 \sqrt{2} & -\sqrt{2}-i \sqrt{2} & 0 & \sqrt{2} & -\sqrt{2}+i \sqrt{2} & i \sqrt{2} & -i \sqrt{2} \\
 -1 & -i \sqrt{2} & -2 & 1 & i \sqrt{2} & 1-i \sqrt{2} & 1+i \sqrt{2} \\
 -1 & i \sqrt{2} & -2 & 1 & -i \sqrt{2} & 1+i \sqrt{2} & 1-i \sqrt{2} \\
\end{array}
\right)\ , \\
T&=\left(
\begin{array}{ccccccc}
 e^{2\pi i (23/24)} & 0 & 0 & 0 & 0 & 0 & 0 \\
 0 & e^{2\pi i (5/24)} & 0 & 0 & 0 & 0 & 0 \\
 0 & 0 & e^{2\pi i (2/24)} & 0 & 0 & 0 & 0 \\
 0 & 0 & 0 & e^{2\pi i (23/24)} & 0 & 0 & 0 \\
 0 & 0 & 0 & 0 & e^{2\pi i (5/24)} & 0 & 0 \\
 0 & 0 & 0 & 0 & 0 & e^{2\pi i (11/24)} & 0 \\
 0 & 0 & 0 & 0 & 0 & 0 & e^{2\pi i (11/24)} \\
\end{array}
\right)\ .
\eea
with central charge $c=-23$.

\newpage
\bibliographystyle{nb}
\bibliography{jobname}

\end{document}